\documentclass[12pt, preprint]{aastex}

\slugcomment{Accepted For Publication in {\it The Astrophysical
    Journal}}

\shorttitle{Radiation Transfer and Internal Shock Model}
\shortauthors{Joshi \& B\"ottcher}

\begin{document}

\title{Time-dependent Radiation Transfer in the Internal Shock Model
  Scenario for Blazar Jets}

\author{M. Joshi\altaffilmark{1,2} and M. B\"ottcher\altaffilmark{1}}
\affil{Astrophysical Institute, Department of Physics and Astronomy,
\\Clippinger Labs, Ohio University, Athens, OH 45701, USA}
\affil{Institute for Astrophysical Research, Boston University,
\\725 Commonwealth Ave., Boston, MA 02215, USA}

\begin{abstract}
We describe the time-dependent radiation transfer in blazar jets,
within the internal shock model. We assume that the central engine,
which consists of a black hole and an accretion disk, spews out
relativistic shells of plasma with different velocity, mass, and
energy. We consider a single inelastic collision between a faster
(inner) and a slower (outer) moving shell. We study the dynamics of
the collision and evaluate the subsequent emission of radiation via
the synchrotron and synchrotron self Compton (SSC) processes after the
interaction between the two shells has begun. The collision results in
the formation of a forward shock (FS) and a reverse shock (RS) that
convert the ordered bulk kinetic energy of the shells into magnetic
field energy and accelerate the particles, which then radiate. We
assume a cylindrical geometry for the emission region of the jet. We
treat the self-consistent radiative transfer by taking into account
the inhomogeneity in the photon density throughout the region. In this
paper, we focus on understanding the effects of varying relevant input
parameters on the simulated spectral energy distribution (SED) and
spectral variability patterns.
\end{abstract}


\section{\label{intro}Introduction}

Blazars are a class of extreme active galactic nuclei (AGNs). They
consist of flat-spectrum radio quasars (FSRQs) and BL Lac objects, and
are known for their rapid variability, at all wavelengths, on
timescales of months to a few days and even to less than an hour in
some cases \citep[e.g.,][]{ga1996, cw1999, an2007, at2007}. Such
variability is indicative of highly violent phenomena and the presence
of ultra-relativistic particles in the jets of blazars. The SED of
blazars consists of two broad spectral bumps that are associated with
synchrotron radiation for the low-energy component and inverse
Comptonization for the high-energy one. The spectral variability
patterns and SEDs provide valuable information regarding the
acceleration of particles and the time-dependent interplay of various
radiation mechanisms responsible for the observed emission. It is,
therefore, imperative to develop a theoretical approach that is
capable of simulating these observables to gain a deeper understanding
of the physics of blazar jets.

Many authors have studied blazar jets in the past \citep[see,
  e.g.,][]{mgc1992, ds1993, bms1997, jb2007} in the framework of
one-zone models. In those works, a spherical emission region has been
considered and a power-law energy distribution of injected electrons
and positrons assumed. The evolution of the particle population and
the subsequent photon emission is solved numerically in a
time-dependent manner in order to generate theoretical light curves,
spectral variability patterns, SEDs, and other features to obtain a
satisfactory fit to the observed data of a given source. Such
theoretical efforts have successfully reproduced the observed SED and
the spectral variability patterns of various blazars. They have
provided information about the relevant physical parameters of the
emission region, but the location and the mode of particle
acceleration in the jets of such systems still need to be explored in
more detail. Most of the theoretical approaches have assumed a
spherical geometry for the emission region, continuous injection of
particles, and homogeneous population of both electrons and photons
throughout the emission region \citep[e.g.][]{sbr1994, bm1996,
  bb2000}. Such assumptions are good only to a first approximation and
do not completely account for the observations. Thus, a more
sophisticated approach is required to treat the characteristics of the
energized plasma, such as electron energy distribution and magnetic
field, not as free parameters but as products of global
parameters. This would enable us to link the observed emission
properties to the physics of energy transport inside a jet.

In the case of leptonic jet models, various authors have treated the
question of particle acceleration and their subsequent injection in
different ways. \cite{smm2004} consider the formation of a system of
standing oblique shock waves in the jet that accelerate particles. The
local pressure imbalance between the ambient medium and the jet gives
rise to standing oblique shock waves, which can terminate in a Mach
disk, near the axis, that lies perpendicular to the jet flow. As the
relativistic flow of plasma passes through the disk, the plasma
electrons accelerate to highly relativistic energies and radiate away
their energy via synchrotron and inverse Compton processes to produce
the observed emission.

\cite{sp2001} (hereafter S01) and \cite{mi2004}, consider the internal
shock model to accelerate the particles inside a jet. In this case,
shells of plasma with different mass, energy, and velocity are ejected
intermittently by the central engine. The faster moving shells catch
up with the slower moving ones to result in collisions that produce
shocks internal to the jet. These shocks accelerate the plasma
electrons to ultrarelativistic energies, which lose their energy via
various radiative processes to produce the observed radiation.

All of the above-mentioned approaches consider the formation of
reverse and forward shocks that lead to particle acceleration in a
region inside the jet. The injection of particles into the emission
region continues until the shocks cross the entire region. The
population of electrons and photons has been generally approximated to
be homogeneous in density throughout the emission region, while the
geometry of the region has been assumed to be cylindrical (cubic
geometry in case of \cite{cg1999}), as appropriate for internal
shocks.

\cite{gr2008} invoke a standing or propagating shock in a collimated
jet to accelerate particles. They adopt a multi-zone pipe geometry for
the emission region to simulate the resultant radiation and
variability of blazars via synchrotron and SSC processes. Their model
considers the inhomogeneity in the particle and photon population only
in the axial direction when calculating the non-local, time-delayed
SSC emission of the source. Their model does not take into account the
volume and angle-averaged photon escape timescale for a cylindrical
region. Other acceleration scenarios, such as isolated shocks
propagating along the jet \citep[]{mg1985, krm1998, si2001}, oblique
shocks propagating through a pre-existing jet \citep{ob2002}, or
particle acceleration in shear flows \citep{rd2004} have also been
considered as plausible acceleration mechanisms in blazar jets.

In a recent approach, \citet{bd2010} employ the internal shock model
to investigate its effects on the time-dependent spectral evolution
and cross-frequency correlation of the emission from the jets of
blazars. The model considers a one-zone cylindrical emission region
bounded by forward and reverse shocks that energize the background
plasma to very high energies. The resultant non-thermal distribution
of particles then decays in energy through synchrotron, SSC, and
external Compton (EC, Compton upscattering of photons external to the
jet) emission processes. The time-dependent synchrotron and EC spectra
are calculated analytically, while the SSC emission is computed
semi-analytically, with the calculations restricted to the Thomson
regime. As a result, their model is unable to reproduce TeV emission
from the high-frequency peaked BL Lac objects (HBLs).

In this paper, we consider the internal shock scenario to develop a
1-D multi-slice time-dependent leptonic jet model for simulating the
radiative transfer in relativistic blazar jets. We use this scenario
to explore the relevant physical parameters of a cylindrical emission
region that is responsible for the observed radiation. We take into
account the inhomogeneity in the photon and electron populations
throughout the region by dividing it into multiple slices and
considering time-dependent radiation transfer within and in between
each slice. The multi-slice scheme with radiation feedback
automatically lets us address the non-local, time-delayed SSC emission
of the source in a self-consistent manner along the length of the
emission region. It also takes into account the volume and
angle-averaged photon escape timescale. We calculate the
time-dependent synchrotron and SSC emission spectra without
restricting our calculations to the Thomson regime, hence making it
applicable to all classes of blazars. We also incorporate the
light-travel time delays to correctly calculate the observed spectra
and light curves in the observer's frame. We do not consider the
retarded-time photon field individually in the radial direction for
calculating the SSC emission neither do we evaluate the microphysics
of particle acceleration as that is beyond the scope of the current
work.

We assume a background thermal plasma in the jet that contains
non-relativistic electrons, positrons, and protons. In \S
\ref{collision}, we describe the interaction between two shells of
plasma, responsible for producing internal shocks, and the subsequent
treatment of the collision in the framework of relativistic
hydrodynamics. As the shocks propagate through the plasma, the
internal kinetic energy of the shocked fluid that is stored in the
baryons is transferred to electrons and positrons. This transfer of
energy accelerates electrons and positrons to relativistic and
non-thermal energies. In this section, we derive various emission
region parameters resulting from shock propagation and particle
acceleration. In \S \ref{timescale}, we present a semi-analytical
calculation of the volume and angle-averaged photon escape timescale
for a cylindrical geometry. In \S \ref{method}, we describe the
numerical approach that we use for calculating various radiative
energy loss rates as well as photon emissivities. Since positrons lose
the same amount of energy as the electrons via the same radiative loss
mechanisms, we do not distinguish between them throughout the
paper. We discuss our multi-slice radiation transfer scheme that
addresses the inhomogeneity in the density of the photon and electron
population throughout the emission region. \S \ref{delays} deals with
the relevant light-travel time delays that are required to accurately
register the radiation in the observer's frame of reference. The
effects of varying the relevant physical input parameters on the
simulated SED and lightcurves are discussed in \S \ref{study}. We
present our results of the parameter study in \S \ref{outcome}. We
summarize our results in \S \ref{conc}.

Appendices \ref{coeffeqns} and \ref{photesctime} delineate the details
of some of the cumbersome equations used in the numerical computation
of radiative output. In appendix \ref{synfit}, we discuss the
numerical approximation used to calculate the synchrotron photon
density in the simulations.

\section{\label{collision}Hydrodynamics of the collision}

In the internal shock scenario, the central engine ejects relativistic
shells of plasma with disparate mass and energy that travel with
different velocities. As a result, faster moving shells catch up with
slower moving ones, released at earlier times, resulting in inelastic
collisions. Each collision results in the formation of a FS that
propagates into the slower moving outer shell and a RS that propagates
into the faster moving inner shell (S01). The shocks are separated by
a contact discontinuity (CD) across which the pressure of the shocked
fluids is in equilibrium. As the shocks propagate into their
respective shells, they convert the ordered bulk kinetic energy of the
shocked plasma into magnetic field energy and internal energy of
electrons. As a result, electrons are accelerated to highly
relativistic energies and radiate to produce the observed emission
throughout the electromagnetic spectrum.

In the present paper, we consider a single inelastic collision between
a slower moving outer shell and a faster moving inner shell. We study
the dynamics of the collision and the subsequent photon emission from
non-thermal relativistic electrons after the interaction between the
two shells has started. The entire treatment of shell collision and
shock propagation is hydrodynamic and relativistic in nature. We do
not consider the expansion of individual shells or the merged shell
along or transverse to the direction of motion. The respective
expansions are neglected here because the jet remains well-collimated
at sub-pc and pc scales \citep{jo2005} and the collision lasts for a
short time during which the shell will have a nearly constant width
\citep{bd2010}.

\subsection{\label{colparam}Collision parameters}

We consider an ejection event of two shells lasting for time
$t_w$. The outer shell is ejected with rest mass $M_{\rm o}$, width
$\Delta_{\rm o}$ and a bulk Lorentz factor (BLF) $\Gamma_{\rm o}$. The
inner shell is ejected with rest mass $M_{\rm i}$, width $\Delta_{\rm
  i}$ and BLF $\Gamma_{\rm i}$. The ejection of the two shells is
separated by a time interval $t_{\rm e}$. The kinetic luminosity of
the entire ejection event is given by $L_w$ such that the sum of the
kinetic energies of the two shells is not more than the total
specified through $L_{\rm w}$ and $t_{\rm w}$ and satisfies the
condition:

\begin{equation}
\label{lweqn}
M_{\rm o}\Gamma_{\rm o} c^2 + M_{\rm i}\Gamma_{\rm i} c^2 = L_{\rm w}
t_{\rm w}~.
\end{equation}

The radii, R, of both shells are taken to be equal to each other and
to that of the collimated jet. The z-axis is along the axis of the jet
and the radius of the jet is perpendicular to this axis. At time $t =
0$, the outer shell is at a distance $z_{\rm o}$ and the inner shell
is at a distance $z_{\rm i}$ from the central engine. In this section,
all non-primed quantities refer to the rest frame of the AGN (lab
frame), all primed quantities refer to the comoving frame of the
shocked fluid behind the shock front, and all quantities with an
overline refer to the frame of the unshocked fluid. The time of
collision \citep{kps1997}, $\delta{t}$ can then be calculated by

\begin{equation}
\label{timeeqn}
\delta t = \frac{\left(z_o - z_i - \Delta_{\rm
o}\right)}{c\left(\beta_{\rm i} - \beta_{\rm o}\right)}
\end{equation}
and the position at which the two shells would collide, which also
defines the location of the CD, is given by

\begin{equation}
\label{poseqn}
z_{\rm c} = z_{\rm i} + c \beta_{\rm i} \delta t~.
\end{equation}

Considering the collision of these two shells to be an inelastic one,
the BLF, $\Gamma_{\rm m}$, and the total internal energy of the merged
shell, $E_{\rm int}$, can be obtained by using the conservation of
momentum and energy (S01). This yields

\begin{equation}
\label{gameqn}
\Gamma_{\rm m} = \sqrt{\frac{M_{\rm i} \Gamma_{\rm i} + M_{\rm o}
\Gamma_{\rm o}}{\frac{M_{\rm i}}{\Gamma_{\rm i}} + \frac{M_{\rm o}}
{\Gamma_{\rm o}}}}
\end{equation}

and

\begin{equation}
\label{eeqn}
E_{\rm int} = M_{\rm i} c^{2} \left(\Gamma_{\rm i} - \Gamma_{\rm
m}\right) + M_{\rm o} c^{2} \left(\Gamma_{\rm o} - \Gamma_{\rm
m}\right)~.
\end{equation}

The internal energy of both shells before the collision has been
assumed to be relatively negligible, so that essentially all of the
energy is stored as bulk kinetic energy. The efficiency of conversion
of this energy into the internal energy of the merged shell from a
single collision is given by

\begin{equation}
\label{effeqn}
\epsilon = 1 - \left[\left(M_{\rm i} + M_{\rm o}\right) \Gamma_{\rm m}
\over \left(M_{\rm i} \Gamma_{\rm i} + M_{\rm o} \Gamma_{\rm o}\right)
\right]~.
\end{equation}

\begin{figure}
\plotone{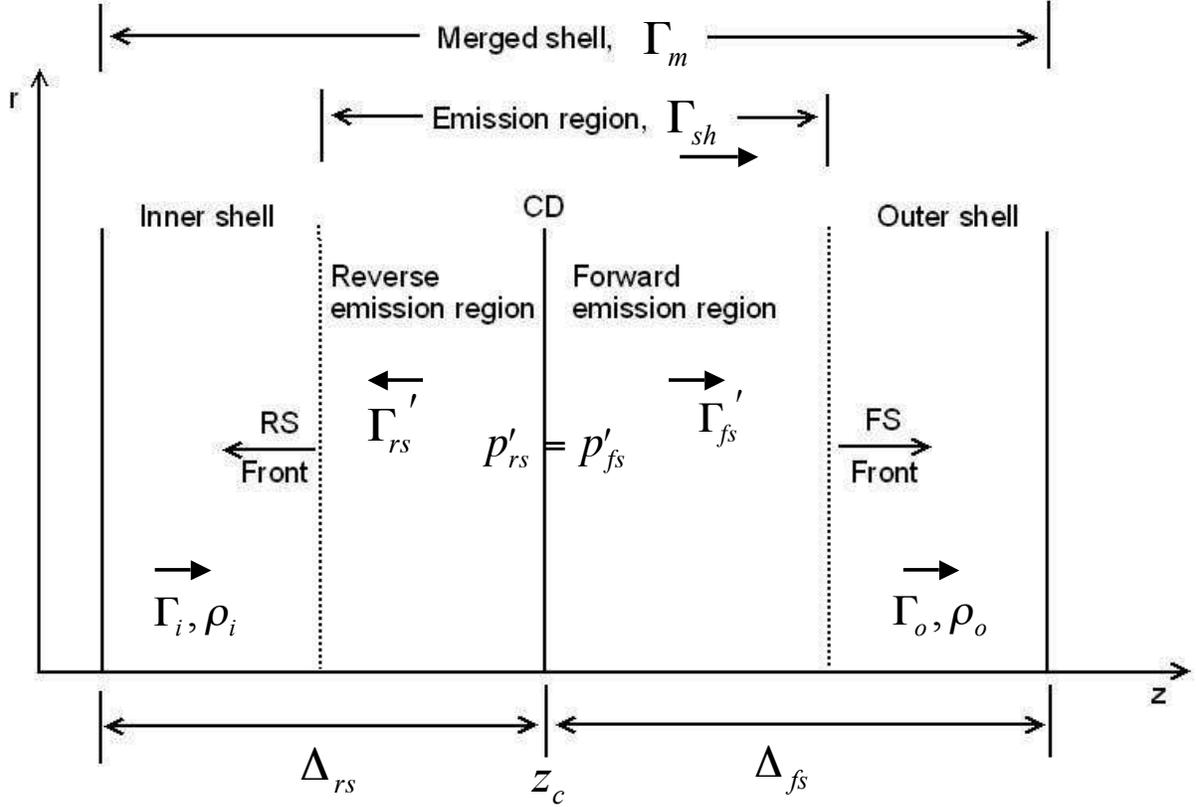}
\caption{Schematic diagram of the two-shell inelastic collision
  resulting in a merged shell of bulk Lorentz factor (BLF)
  $\Gamma_{m}$. The forward shock (FS) and reverse shock (RS) fronts
  start to propagate in the outer and inner shells, respectively,
  creating the forward and reverse emission regions. In the comoving
  frame, the shock fronts move with BLFs $\Gamma_{\rm fs}^\prime$ and
  $\Gamma_{\rm rs}^\prime$. The contact discontinuity (CD) is located
  at a distance $z_{c}$ from the central engine. The comoving
  pressures $p_{\rm fs}^\prime$ and $p_{\rm rs}^\prime$ of the shocked
  fluids across the CD are equal. Ultra-relativistic electrons are
  only present in the forward and reverse emission regions. These
  emission regions keep expanding until the shocks reach the
  respective boundaries of their regions.}
\label{emission_region}
\end{figure}

In the comoving frame of the shocked (emission) region, the two shocks
appear to move in opposite directions, away from the CD, with Lorentz
factors $\Gamma_{\rm fs(rs)}^\prime$, respectively. In the lab frame,
the entire shocked material continues to move forward with BLF
$\Gamma_{\rm sh}$. As a result of shock propagation, the emission
region keeps increasing until the shocks reach the respective
boundaries of the merged shell, as shown in Figure
\ref{emission_region}.

The BLF of the entire emission region in the lab frame, $\Gamma_{\rm
  sh}$, is calculated by using the relativistic hydrodynamic
Rankine-Hugoniot jump conditions for energy and particle density
across both shock fronts, and the condition of equal pressure of the
shocked fluids across the CD (S01). The jump equations
\citep[]{bm1976, pm1999} for the kinetic energy density $U_{\rm
  fs}^\prime$ and the matter density of baryons $\rm \rho_{fs}^\prime$
in the comoving frame of the shocked-fluid frame are given by

\begin{equation}
\label{releeqn}
{U_{\rm fs(rs)}^\prime \over \rho_{\rm fs(rs)}^\prime} = c^{2}
\left(\Gamma_{\rm fs(rs)}^\prime - 1\right)
\end{equation}

and

\begin{equation}
\label{reldeneqn}
\frac{\rho_{\rm fs(rs)}^\prime}{\overline{\rho}_{\rm o(i)}} =
\frac{\left({\hat{\gamma} \Gamma_{\rm fs(rs)}^\prime} +
  1\right)}{\left(\hat{\gamma} - 1\right)}~,
\end{equation}
where $\overline{\rho}_{\rm o(i)} = \frac{M_{\rm o(i)}}{\Gamma_{\rm
    o(i)} \pi R^2 \Delta_{\rm o(i)}}$ is the matter baryon density of
the preshocked fluid in the comoving frame of the unshocked fluid. The
quantity $\hat{\gamma}$ is the adiabatic index, whose value is assumed
to be 4/3 (for an ultra-relativistic plasma).

The pressure of the shocked fluids is obtained from equations
\ref{releeqn} \& \ref{reldeneqn}, and is given by

\begin{equation}
\label{peqn}
p_{\rm fs(rs)}^\prime = \left(\Gamma_{\rm fs(rs)}^\prime - 1\right)
\left({\hat{\gamma} \Gamma_{\rm fs(rs)}^\prime} + 1\right)
\overline{\rho}_{\rm o(i)} c^{2}~.
\end{equation}

The BLF of each shell relative to the shocked fluids in their
respective comoving frames can be written as

\begin{equation}
\label{gamfreqn}
\Gamma^{\prime}_{\rm fs(rs)} = \Gamma_{\rm o(i)} \Gamma_{\rm sh}
\left(1 - \beta_{\rm o(i)} \beta_{\rm sh}\right)~.
\end{equation}

Substituting Eqn. \ref{gamfreqn} in Eqn. \ref{peqn} and applying the
condition of equal pressures yields a quartic equation in
$\Gamma_{\rm sh}$

\begin{equation}
\label{gamsheqn}
a\Gamma^4_{\rm sh} + b\Gamma^3_{\rm sh} + c\Gamma^2_{\rm sh} +
d\Gamma_{\rm sh} + e = 0~.
\end{equation}

The values of each of these coefficients are given in Appendix
\ref{coeffeqns}.

As the shocks cross through their respective regions, they compress
the plasma present in those regions. As a result, the width of the
shocked-fluid region gets compressed. This is determined using the
conservation of rest mass of the shocked and unshocked fluids, which
is given by

\begin{equation}
\label{masseqn}
\rho_{\rm fs(rs)}^{\prime} V_{\rm fs(rs)}^{\prime} =
\overline{\rho}_{\rm o(i)} \overline{V}_{\rm o(i)}.
\end{equation}

Equation \ref{masseqn} provides the width of the shocked-fluid regions
in their respective comoving frames and is given by

\begin{equation}
\label{widtheqn}
\Delta_{\rm fs(rs)}^{\prime} = \overline{\Delta}_{\rm o(i)}
\left(\frac{\overline{\rho}_{\rm o(i)}}{\rho_{\rm
    fs(rs)}^{\prime}}\right)~.
\end{equation}

The time of crossing the forward or reverse region by the FS or RS, in
the shocked-fluid frame, can then be obtained by

\begin{equation}
\label{tcreqn}
t^{\prime}_{\rm cr,fs(rs)} = \frac{\Delta^{\prime}_{\rm fs(rs)}}
{c\beta^{\prime}_{\rm fs(rs)}}~.
\end{equation}

\subsection{\label{emitparam}Emission Region Parameters}

The calculation of the emitted radiation is dependent on the values of
the magnetic field ($B^{\prime}$), minimum ($\gamma^{\prime}_{\rm
  min}$) and maximum ($\gamma^{\prime}_{\rm max}$) electron Lorentz
factors (ELF), and the normalization factor $Q^{\prime~ \rm inj}_{\rm
  0}$~($\rm cm^{-3}s^{-1}$) of the electron injection function
$Q^{\prime~ \rm inj}_{\rm e}(\gamma, \rm {t}) (\rm
cm^{-3}s^{-1})$. The power-law distribution of the injected
relativistic electrons is given by

\begin{eqnarray}
\label{Qinjeqn}
Q^{\prime~ \rm inj}_{\rm e, fs(rs)}(\gamma^\prime) = Q^{\prime~ \rm
  inj}_{0, fs(rs)} \gamma^{\prime -q^{\prime}} & \textrm{for
  $\gamma^{\prime}_{\rm min, fs(rs)} \leq
  \gamma^{\prime}\leq\gamma^{\prime}_{\rm max, fs(rs)}$}
\end{eqnarray}
where $q^{\prime}$ is the electron injection power-law index.

In order to proceed with the calculation of emission, we first derive
the values of these parameters from shock dynamics (see below) for
each of the shocked regions. We calculate the radiative energy loss
rates of electrons and corresponding photon emissivities for both
regions, separately, and add their respective contributions to the
observed spectra.

Some fraction of the bulk kinetic energy density of the shocked fluid
is converted into the magnetic and electron energy density by the
resulting shocks. We define a magnetic field partition parameter
$\varepsilon^{\prime}_{\rm B}$ such that

\begin{equation}
\label{ubeqn}
\varepsilon^{\prime}_{\rm B} = \frac{U^{\prime}_{\rm B,
    fs(rs)}}{U^{\prime}_{\rm fs(rs)}}~,
\end{equation}
where $U^{\prime}_{\rm fs(rs)}$ is the kinetic energy density of the
shocked fluid. This is obtained by using Eqns. \ref{releeqn} \&
\ref{reldeneqn}, and is given by

\begin{equation}
\label{ufsrseqn}
U^{\prime}_{\rm fs(rs)} = \overline{\rho}_{\rm o(i)} c^2
\left(\Gamma^{\prime}_{\rm fs(rs)} - 1\right) \frac{\hat{\gamma}
  \Gamma^{\prime}_{\rm fs(rs)} + 1}{\hat{\gamma} - 1}
\end{equation}

The quantity $U^{\prime}_{\rm B, fs(rs)} = B^{\prime 2}_{\rm
  fs(rs)}/8\pi$ is the magnetic energy density. Equation \ref{ubeqn}
yields the value of the magnetic field, which is assumed to be
randomly oriented in space and tangled in the jet plasma. The maximum
ELF of the injected electrons is obtained by balancing the power
gained from the acceleration, i.e. from gyro-resonant processes, with
the synchrotron losses \citep[]{dh1992, cd1999}. This condition yields

\begin{equation}
\label{gmaxeqn}
\gamma^{\prime}_{\rm max, fs(rs)} = 4.6 \times 10^{7}
\sqrt{\frac{\alpha^{\prime}} {B^{\prime}_{\rm fs(rs)}}}~,
\end{equation}
where $\alpha^{\prime} \leq 1$ is the electron acceleration rate
parameter, defined through $\alpha^{\prime} = t^{\prime}_{\rm
  gyr}/t^{\prime}_{\rm acc}$. The quantities $t^{\prime}_{\rm gyr}$
and $t^{\prime}_{\rm acc}$ are, respectively, the gyration and
acceleration timescales of an electron. The quantity $B^{\prime}_{\rm
  fs(rs)}$ is in units of Gauss.

The minimum ELF is obtained by using the equation

\begin{equation}
\label{ueeqn}
U^{\prime}_{\rm e, fs(rs)} = \varepsilon^{\prime}_{\rm
  e}U^{\prime}_{\rm fs(rs)}~.
\end{equation}
Here, $U^{\prime}_{\rm e, fs(rs)} = n^{\prime}_{\rm e, fs(rs)}
\langle{\gamma^{\prime}_{\rm fs(rs)}}\rangle m_{e}c^{2}$ is the
electron energy density, $n^{\prime}_{\rm e, fs(rs)} =
\zeta^{\prime}_{e} \frac{\rho^{\prime}_{\rm fs(rs)}}{m_{p}}$ is the
number density of non-thermal electrons injected by shocks into their
respective emission regions, $\zeta^{\prime}_{e}$ is the fraction of
electrons that is accelerated behind the shock fronts into the
power-law distribution given by Eqn. \ref{Qinjeqn}, and
$\varepsilon^{\prime}_{\rm e}$ is the electron energy partition
parameter. The average electron energy $\langle{\gamma^{\prime}_{\rm
    fs(rs)}}\rangle$ for the energy range $\gamma^{\prime}_{\rm min,
  fs(rs)} < \gamma^{\prime}_{\rm fs(rs)} < \gamma^{\prime}_{\rm max,
  fs(rs)}$ is given by

\begin{equation}
\label{avgeqn}
\langle{\gamma^{\prime}_{\rm fs(rs)}}\rangle =
\frac{\int\limits_{\gamma^{\prime}_{\rm min,
      fs(rs)}}^{\gamma^{\prime}_{\rm max, fs(rs)}} n^{\prime}_{\rm e,
    fs(rs)} (\gamma^{\prime}_{\rm fs(rs)}) \gamma^{\prime}_{\rm
    fs(rs)} d\gamma^{\prime}_{\rm
    fs(rs)}}{\int\limits_{\gamma^{\prime}_{\rm min, fs(rs)}}
  ^{\gamma^{\prime}_{\rm max, fs(rs)}} n^{\prime}_{\rm e,
    fs(rs)}(\gamma^{\prime}_{\rm fs(rs)}) d\gamma^{\prime}_{\rm
    fs(rs)}}~.
\end{equation}
Here $n^{\prime}_{\rm e, fs(rs)}(\gamma^{\prime}_{\rm fs(rs)}) =
n^{\prime}_{o} \gamma^{\prime -q}_{\rm fs(rs)}$ is the number density
per unit $\gamma^{\prime}_{\rm fs(rs)}$ and $n^{\prime}_{o}$ is the
initial electron number density. As most of the internal kinetic
energy of the shocked fluid is stored in the baryons, implying that
$\varepsilon^{\prime}_{\rm e} << 1$, we can write $U^{\prime}_{\rm
  fs(rs)} = n^{\prime}_{\rm p, fs(rs)} (\Gamma^{\prime}_{\rm fs(rs)} -
1) m_{\rm p} c^{2}$. Substituting all of the above expressions along
with equation \ref{gmaxeqn} into equation \ref{ueeqn}, we obtain the
following expressions for $\gamma^{\prime}_{\rm min, fs(rs)}$:

\begin{eqnarray}
\label{gmineqns}
C \ln(\gamma^{\prime}_{\rm min, fs(rs)}) - \gamma^{\prime}_{\rm min,
  fs(rs)} + \gamma^{\prime}_{\rm max, fs(rs)} - C
\ln(\gamma^{\prime}_{\rm max, fs(rs)}) = 0 & \textrm{if $q^{\prime} =
  1$} \nonumber\\
\ln(\gamma^{\prime}_{\rm min, fs(rs)}) + \frac{C}{\gamma^{\prime}_{\rm
    min, fs(rs)}} - \ln(\gamma^{\prime}_{\rm max, fs(rs)}) -
\frac{C}{\gamma^{\prime}_{\rm max, fs(rs)}} = 0 & \textrm{if
  $q^{\prime} = 2$} \nonumber\\
\gamma^{\prime (1 - q^{\prime})}_{\rm min, fs(rs)}
(\gamma^{\prime}_{\rm min, fs(rs)} - C_q) - \gamma^{\prime (1 -
  q^{\prime})}_{\rm max, fs(rs)} (\gamma^{\prime}_{\rm max, fs(rs)} -
C_q) = 0 & \textrm{if $q^{\prime} \neq$ 1 or 2}~,
\end{eqnarray} 
where $C = 1837 \frac{\varepsilon^{\prime}_{\rm
    e}}{\zeta^{\prime}_{\rm e}} (\Gamma^{\prime}_{\rm fs(rs)} - 1)$,
and $C_q = 1837 \frac{q^{\prime} - 2}{q^{\prime} - 1}
\frac{\varepsilon^{\prime}_{\rm e}}{\zeta^{\prime}_{\rm
    e}}(\Gamma^{\prime}_{\rm fs(rs)} - 1)$. Equation \ref{gmineqns} is
solved numerically to obtain the value of $\gamma^{\prime}_{\rm min,
  fs(rs)}$. In the case of $\gamma^{\prime}_{\rm max} \gg
\gamma^{\prime}_{\rm min}$ and $q^{\prime} > 2$, the above expressions
reduce to
\begin{equation}
\label{gminlimit}
\gamma^{\prime}_{\rm min, fs(rs)} = 1837 \frac{q^{\prime} -
  2}{q^{\prime} - 1} \frac{\varepsilon^{\prime}_{\rm
    e}}{\zeta^{\prime}_{\rm e}} (\Gamma^{\prime}_{\rm fs(rs)} - 1)~.
\end{equation} 

The value of $Q^{\prime \rm inj}_{\rm 0, fs(rs)} (t^{\prime})$ is
obtained by assuming that a fraction of the kinetic energy stored in
the shocked plasma is transferred, per unit time, to non-thermal
electrons \citep{bd2010}. This yields
\begin{equation}
\label{linjeqn}
\frac{\varepsilon^{\prime}_{\rm e} U^{\prime}_{\rm
    fs(rs)}}{t^{\prime}_{\rm cr, fs(rs)}} =
\int\limits_{\gamma^{\prime}_{\rm min, fs(rs)}}^{\gamma^{\prime}_{\rm
    max, fs(rs)}} Q_{\rm e, fs(rs)}^{\prime \rm
  inj}(\gamma^{\prime}_{\rm fs(rs)}, \rm t^{\prime})
\gamma^{\prime}_{\rm fs(rs)} m_e c^2 d\gamma^{\prime}_{\rm fs(rs)}~.
\end{equation}

Using Eqns. \ref{linjeqn} and \ref{Qinjeqn}, we obtain the expression
for $Q^{\prime \rm inj}_{\rm 0, fs(rs)} (t^\prime)$ as
\begin{eqnarray}
\label{qoeqn}
Q^{\prime \rm inj}_{\rm 0, fs(rs)}(t^{\prime}) =
\left\{ \begin{array}{ll} \frac{\varepsilon^{\prime}_{\rm e}
    U^{\prime}_{\rm fs(rs)} (2 - q^{\prime})}{\rm m_e c^2
    t^{\prime}_{\rm cr, fs(rs)} (\gamma^{\prime (2 - q^{\prime})}_{\rm
      max, fs(rs)} - \gamma^{\prime (2 - q^{\prime})}_{\rm min,
      fs(rs)})} & \textrm{if $q^{\prime} \neq 2$}\\ 
\\
\frac{\varepsilon^{\prime}_{\rm e} U^{\prime}_{\rm fs(rs)}}{\rm m_e
  c^2 t^{\prime}_{\rm cr, fs(rs)} ln(\gamma^{\prime}_{\rm max,
    fs(rs)}/ \gamma^{\prime}_{\rm min, fs(rs)})} & \textrm{if
  $q^{\prime} = 2$}~.
\end{array} \right.
\end{eqnarray}

\section{\label{timescale}Photon Escape Timescale}

For a cylindrical geometry, the volume and angle-averaged photon
escape timescale is not the same as that of a spherical one. The
average photon escape timescale is used in the numerical computation
of the evolution of the particle and photon population inside the
emission region. Here, we derive a semi-analytical expression of the
escape timescale and assume the cylindrical emission region to be
Compton thin.

We consider three possible directions by which a photon can escape
from a cylindrical region: forward, sideways and backward. All
quantities in this section and thereafter refer to the comoving frame
only, unless otherwise indicated, hence the prime notation is no
longer used.

\begin{figure}
\plotone{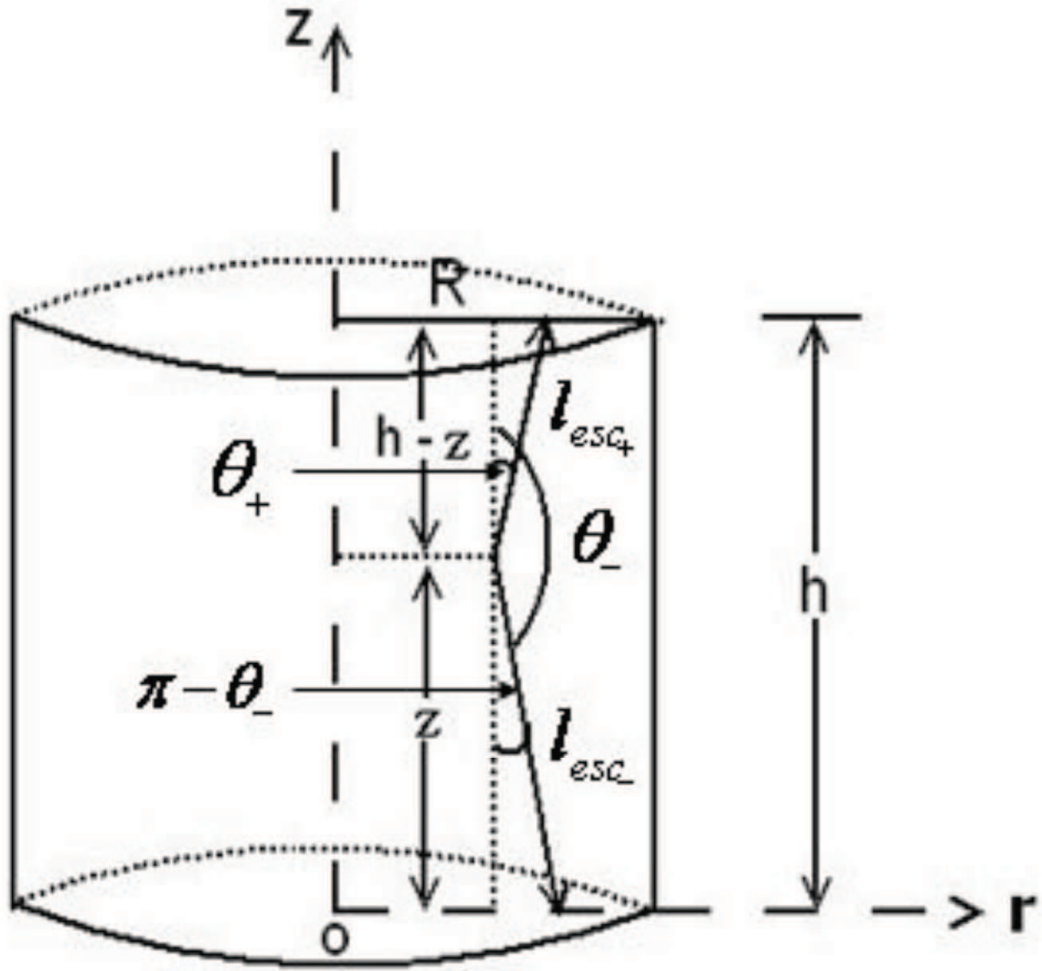}
\caption{Illustration of the three possible directions of escape for a
  photon from a cylindrical region. The quantities $l_{esc+}$ and
  $l_{esc-}$, respectively, refer to the escape path lengths for a
  photon in forward and backward directions, and $\theta_+$ and $\pi -
  \theta_{-}$ are the corresponding angles.}
\label{escape_dir}
\end{figure}

Figure \ref{escape_dir} depicts the angle considerations for the three
possible directions of escape for a photon from a cylindrical region
of height (width) h and radius R. As stated earlier, the z-axis is
along the axis of the cylinder. The volume-averaged photon escape
timescale can then be written as

\begin{equation}
\label{escleqn}
{\langle{t_{\rm ph,esc}}\rangle}_{\rm V} = \frac{1}{V_{\rm cyl}}
\int\limits_{0}^{R}\int\limits_{0}^{h}\int\limits_{0}^{2\pi}
\langle{t_{\rm ph,esc}\left(r, z\right)}\rangle rd\phi\,dz\,dr~,
\end{equation}
where the angle-averaged escape timescale is written as

\begin{equation}
\label{escteqn}
\langle{t_{\rm ph,esc}(r, z)}\rangle = \frac{1}{4 \pi c}
\int\limits_{0}^{2\pi}\int\limits_{-1}^{+1} l_{\rm esc} (\mu, \Phi; r,
z) d\mu\,d\Phi\
\end{equation}
Here, $l_{\rm esc}$ is the distance required to escape from the region
in any direction, and $\mu = \cos \theta$. We have solved the above
integral semi-analytically to obtain the final expression for the
volume and angle-averaged photon escape timescale for a cylindrical
region, which is given in Eqn. \ref{esctimeeqn}. The intermediate
steps needed to obtain the final expression are given in Appendix
\ref{photesctime} of this paper.

\begin{eqnarray}
\label{esctimeeqn}
\langle{t_{\rm ph,esc,V}}\rangle = \frac{h}{4c} - \frac{h}{2c} \ln(a)
+ \frac{h}{2\pi c} \int\limits_0^{2\pi}\int\limits_0^1\int\limits_0^1
xy \ln\left(k^2+(ay)^2\right) dy\,dx\,d\phi + \nonumber\\ 
\frac{R}{\pi c}\int\limits_0^{2\pi}\int\limits_0^1 xk
\arcsin\left(\frac{a}{\sqrt{k^2 + a^2}}\right) dx\,d\phi - \nonumber\\
\frac{R^2}{2\pi hc} \int\limits_0^{2\pi}\int\limits_0^1 x k^2
\ln\left(1 + \frac{a^2}{k^2}\right) dx\,d\phi~,
\end{eqnarray}
where $a = h/R$ and $k = \sqrt{1 - (x \sin\phi)^2} - x\cos\phi$.

Using the geometry of a cylinder, the volume-averaged probability of
escape for a photon in the forward direction is $P_{\rm fwd} = \pi
R^{2}/(2 \pi R (R+h))$. The volume-averaged probability for a photon
to go in the backward direction is the same as its counterpart in the
forward direction, hence $P_{\rm back} = P_{\rm fwd}$. Similarly, the
volume-averaged probability of escape in the sideways direction is
$P_{\rm side} = 2 \pi R h/(2 \pi R (R+h)) = h/(R+h)$. These
probabilities have been used to calculate the appropriate
volume-averaged photon escape rates in the three directions, as
described in the next section.

\section{\label{method}Numerical Method}

The emission region parameters and the average photon escape timescale
form the necessary set of quantities required to calculate radiative
energy loss rates and photon emissivities. They are needed to follow
the evolution of electron and photon populations inside the emission
region in a self-consistent and time-dependent manner along the length
of the region. Since the same set of equations is applicable to both
the forward and reverse emission regions, we do not distinguish
between the various quantities used in the equations.

We consider instantaneous acceleration of relativistic particles
behind the shock front. The accelerated particles are injected into
the region with a single-power-law distribution described by
Eqn. \ref{Qinjeqn}. The acceleration of particles and the longitudinal
expansion of the emission region continues until the shocks reach
their respective boundaries of the merged shell. As the shocks
propagate in their region, energetic electrons continue to be produced
at the shock front. Owing to energy losses, only less energetic
electrons will be found at progressively larger distances from the
shock fronts. This creates an energy gradient within the emission
region.

Once the acceleration sets in, particles begin to lose their energy
via synchrotron and SSC processes. The time-dependent evolution of
the electron and photon populations, in each of the emission regions,
is followed by using the equations

\begin{equation}
\label{neevoleqn}
{\partial n_{e} (\gamma, t) \over \partial t} = -{\partial \over
\partial \gamma} \left[\left({d\gamma \over dt}\right)_{loss} n_{e}
(\gamma, t)\right] + Q_{e} (\gamma, t) - \frac{n_{e} (\gamma, t)}
{t_{e,esc}}
\end{equation}

and
\begin{equation}
\label{npevoleqn}
{\partial n_{ph} (\epsilon, t) \over \partial t} = \dot n_{ph,em}
(\epsilon, t) - \dot n_{ph,abs} (\epsilon, t) - \frac{n_{ph}
(\epsilon, t)}{t_{ph,esc}}~.
\end{equation}
Here, $(d\gamma/dt)_{\rm loss}$ is the electron energy loss rate, due
to synchrotron and SSC emission, $Q_{e} (\gamma, t)$ is the sum of
external injection and intrinsic $\gamma - \gamma$ pair production
rate, and $t_{\rm e,esc} = {\eta}R/c$ is the electron escape time
scale with $\eta$ being the escape parameter. The quantities $\dot
n_{\rm ph,em} (\epsilon, t)$ and $\dot n_{\rm ph,abs} (\epsilon, t)$
are the photon emission and absorption rates corresponding,
respectively, to the electrons' radiative losses and gains, $\epsilon
= h \nu /m_{\rm e}c^{2}$ is the dimensionless photon energy, and
$t_{\rm ph,esc}$ is the volume and angle-averaged photon escape
timescale for a cylindrical emission region
(Eqn. \ref{esctimeeqn}). In order to obtain the temporal evolution of
the electron distribution in an emission region, equation
\ref{neevoleqn} is discretized and rearranged in such a way that it
forms a tridiagonal matrix, which is then solved numerically
\citep[]{cg1999, pr1992}.

Our current approach simulates the early phase of $\gamma$-ray
production. During this phase, the radiative cooling dominates over
the adiabatic cooling and the emission region remains optically thick
out to $\lesssim$ GHz radio frequencies. As a result, the simulated
radio flux is well below the observed radio data. We do not model the
phase in which the jet components gradually become transparent to
radio frequencies because that would introduce several additional and
poorly constrained parameters.

The synchrotron loss rate is calculated from \citep{rl1979}

\begin{equation}
\label{gdoteqn}
\dot \gamma_{\rm syn} = -(4/3)c \sigma_{\rm T} \frac{B^{2}
\gamma^{2}}{8 \pi m_{\rm e} c^{2}}~,
\end{equation}
where $\sigma_{\rm T} = 6.65 \times 10^{-25}~ \rm cm^{2}$ is the
Thomson cross-section, and B is obtained from Eqn. \ref{ubeqn}. The
synchrotron photon production rate per unit volume in the energy
interval $[\epsilon, \epsilon + d{\epsilon}]$ is calculated using the
formula
\citep{cs1986}
\begin{equation}
\label{syndeneqn}
\dot n_{\rm syn}(\epsilon) = \frac{\sqrt{3} e^{3} B}{2 \pi h^{2} \nu}
\int\limits_{1}^{\infty} R(x) n_{\rm e}(\gamma) d{\gamma}~,
\end{equation}
where $x = \frac{4 \pi m_e c \nu}{3 e B \gamma^{2}}$. A numerically
simplified version of the synchrotron emissivity (described in
Appendix \ref{synfit}) has been used in our simulations to save CPU
time. 

We have incorporated the effects of SSA on the synchrotron spectrum in
our calculation. The SSA optical depth is calculated using the formula
\citep{rl1979}

\begin{equation}
\label{ssaeqn}
\tau_{\rm SSA} = \frac{ -l_{\rm ph,esc} \sqrt{3} e^{3} B}{16 \pi^{2}
\left(m_{\rm e} c \nu\right)^{2}} \int\limits_{1}^{\infty} R(x)
\gamma^{2} \frac{d}{d \gamma} (\frac{n_{\rm e}(\gamma)}{\gamma^2})
d{\gamma}~,
\end{equation}
where $l_{\rm ph,esc} = t_{\rm ph,esc} ~c$ is the mean path length
traversed by a photon escaping from its point of origin inside a
cylindrical region. The synchrotron emission from a cylindrical region
is then obtained using the expression

\begin{equation}
\label{synndoteqn}
\dot N_{\rm syn}^{\rm esc}(\epsilon) = \dot n_{\rm syn}(\epsilon)
\frac{(1 - \exp(-\tau_{\rm SSA}))}{\tau_{\rm SSA}} \pi R^2 h~.
\end{equation}

The electron energy loss rate due to scattering an isotropic,
monochromatic radiation field of photon energy $\epsilon$ and density
$\rm n_{ph}$~($\rm cm^{-3}$), including the effects of the
Klein-Nishina scattering cross section, is calculated using equation
32 of \cite{bms1997}. The photon production rate per unit volume of
isotropic SSC emission in the region is calculated by \citep[]{jf1968,
  bms1997}

\begin{equation}
\label{sscphoteqn}
\dot n_{\rm SSC}(\epsilon_{\rm s}, \Omega_{\rm s}) = \frac{1}{4 \pi}
\int\limits_{1}^{\infty} d \gamma n_{\rm e}(\gamma)
\int\limits_{0}^{\infty} d \epsilon n_{\rm ph}(\epsilon)
g(\epsilon_{\rm s}, \epsilon, \gamma)~,
\end{equation}
where $\epsilon_{\rm s}$ is the energy of the scattered photon, and
$n_{\rm ph}(\epsilon)$ is the radiation field at photon energy
$\epsilon$ available for SSC scattering in the region. The quantity
$g(\epsilon_{\rm s}, \epsilon, \gamma)$ is a function that describes
the probability of Compton scattering, and is given by \citep{jf1968}

\begin{eqnarray}
\label{geqn}
g(\epsilon_{\rm s}, \epsilon, \gamma) = \left\{ \begin{array}{ll}
   \frac{3c \sigma_{\rm T}}{16 \gamma^{4} \epsilon} \left(\frac{4 
\gamma^{2} \epsilon_{\rm s}}{\epsilon} - 1\right) & \textrm{for 
$\frac{\epsilon}{4 \gamma^{2}} \leq \epsilon_{\rm s} \leq \epsilon$}\\
\\ 
   \frac{3c \sigma_{\rm T}}{4 \gamma^{2} \epsilon} \left[2q \ln (q) + 
(1+2q)(1-q) + \frac{ \left(4 \epsilon \gamma q\right)^{2} (1 - q)}
{2 \left(1 + 4 \epsilon \gamma q\right)}\right] & \textrm{for 
$\epsilon \leq \epsilon_{\rm s} \leq \frac{4 \epsilon \gamma^{2}}{1 + 
4 \epsilon \gamma}$}~,
\end{array} \right.
\end{eqnarray}
where $q = \epsilon_{\rm s}/{4 \epsilon \gamma^{2} (1 - {\epsilon_{\rm
      s}/ \gamma})}$. The radiation field available for SSC scattering
includes all contributions from the photon field of the previous time
step. This accounts for SSC scattering to arbitrarily higher
orders. The optical depth for a $\gamma$-ray photon of energy
$\epsilon_1$ due to $\gamma-\gamma$ absorption is calculated using the
formula \citep{bms1997}

\begin{equation}
\label{gamgameqn}
\tau_{\gamma\gamma}(\epsilon_1) = 2 \pi l_{\rm ph,esc}
\int\limits_{-1}^{+1} d{\mu} (1 - \mu) \int\limits_{\frac{2}
{\epsilon_{1}(1- \mu)}}^{\infty} d\epsilon \sigma_{\gamma\gamma}
(\epsilon_1, \epsilon, \mu) n_{\rm ph}(\epsilon, \Omega)~,
\end{equation}
where $n_{\rm ph}$ is the photon density present in the emission
region that provides the photon field for $\gamma - \gamma$
absorption, and $\sigma_{\gamma\gamma}$ is the pair production
cross-section. The subsequent pair production rate is obtained using
the exact and analytic solution, given in Eqn. (26) of
\cite{bs1997}. The contribution from pair production is added to
Eqn. (\ref{neevoleqn}) at every time step of the simulation. Using the
radiative transfer equation, the high-energy emission that is able to
escape from the region can be calculated as

\begin{equation}
\label{sscndoteqn}
\dot N_{\rm SSC}^{\rm esc}(\epsilon_{\rm s}, \Omega_{\rm s}) = \dot
n_{\rm SSC}(\epsilon_{\rm s}, \Omega_{\rm s}) \frac{(1 -
\exp(-\tau_{\gamma\gamma}))}{\tau_{\gamma\gamma}} \pi R^2 h~.
\end{equation}
The temporal evolution of the photon population inside the emission
region, for synchrotron and SSC emission, is followed using equation
\ref{npevoleqn}. The time step used in the simulations is a fraction
of the minimum time scale among the relevant ones (cooling, electron
and photon escape, injection, and shock crossing for as long as the
shocks are in the region) from both regions. The simulation time step
is taken to be common for both emission regions.

\subsection{\label{zones}The Slice Scheme}

In order to reproduce the observed spectral variability patterns of a
blazar, it is important to consider the inhomogeneity in the particle
as well as photon density throughout the emission region. Observations
in the optical and higher energy bands indicate that the acceleration
and/or cooling timescales can be shorter than the light crossing time
for the region \citep{cg1999}. The highly energetic electrons that
have been accelerated freshly at the shock front evolve on a cooling
timescale shorter than the light crossing time, farther from the shock
front those ultra-high energy electrons have had time to cool
significantly. This induces an energy gradient in the electron
population within the region and creates an inhomogeneity in the
particle as well as photon energy density throughout the emission
region. As a result, the observer sees a combination of various
spectra being produced in different parts of the region \citep[see
  also][]{smm2004, gr2008}.

\begin{figure}
\plotone{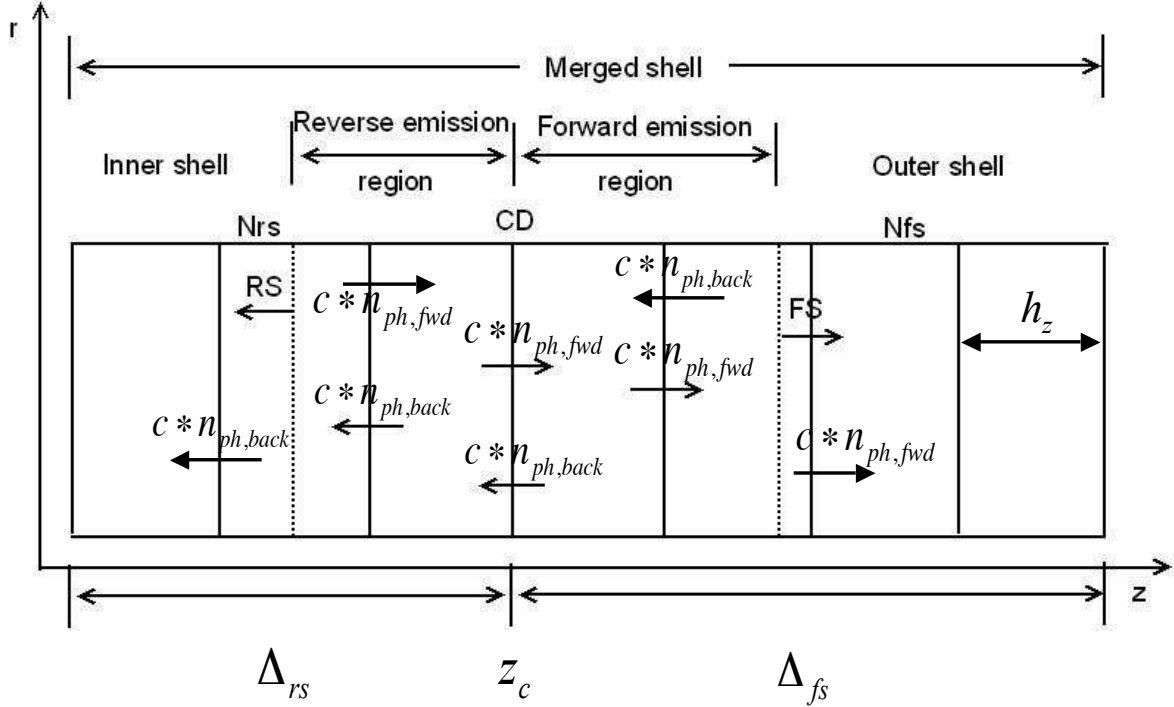}
\caption{Illustration of the slice scheme used to calculate the
  integrated radiation spectrum resulting from the emission regions
  shown in Fig. \ref{emission_region}. The entire forward and reverse
  emission regions are divided into $N_{\rm fs}$ and $N_{\rm rs}$
  slices, each of width $h_{z}$. The quantities $c \times n_{\rm ph,
    fwd}, c \times n_{\rm ph, back}, and c \times n_{\rm ph, side}$
  are the photon fluxes (photons per unit time per unit area of the
  slice) in the forward, backward, and sideways directions delineating
  the transfer of photons in between the slices throughout the forward
  and reverse emission regions.}
\label{zone_dia}
\end{figure}

In our model, we have incorporated the inhomogeneity in the particle
and photon energy densities by dividing both emission regions into
multiple slices, each of width $h_{\rm z}$ and radius R. As shown in
Figure \ref{zone_dia}, for simplicity we start with a jet that is
devoid of relativistic particles. The first population of
ultrarelativistic particles is injected by the shocks very close to
the CD.

The slice closest to the CD in the forward (reverse) emission region
will have the forward (reverse) shock propagating through it first. As
the shock advances, it injects particles into the slice with a
power-law distribution that is dictated by the shock parameters. The
normalization factor, $Q^{\rm inj}_{0}$ is calculated for every slice
using equation \ref{qoeqn} and replacing $U_{\rm fs(rs)}$ by $U_{\rm
  fs(rs), zn} = \frac{U_{\rm fs(rs)}}{N_{\rm fs(rs)}}$ and $t_{\rm cr,
  fs(rs)}$ by $t_{\rm cr, fs(rs), zn} = \frac{h_{\rm z}}{\beta_{\rm
    fs(rs)} c}$. The quantities described in \S \ref{method} are
calculated for each slice. As the shock enters a new slice, it
gradually populates that slice and the entire process of injection,
acceleration and cooling is repeated. For sake of brevity, we refer to
this phase of a slice as the acceleration phase in the following
text. The injection and acceleration stops in the previous slice,
which the shock has just left, and only the cooling continues. We
refer to this phase of the slice as the cooling phase.

The photon density of a slice, along with $P_{\rm fwd}$, $P_{\rm
  back}$, and $P_{\rm side}$ (see \S \ref{timescale}), are used to
calculate photon escape rates in all three directions, according to
\begin{equation}
\label{ndotupeqn}
\frac{dn_{\rm ph, fwd}(\epsilon, \Omega)}{dt} = \frac{n_{\rm
    ph}(\epsilon, \Omega)}{t_{\rm ph, esc}} P_{\rm fwd}
\end{equation}
and
\begin{equation}
\label{ndotsideeqn}
\frac{dn_{\rm ph, side}(\epsilon, \Omega)}{dt} = \frac{n_{\rm
    ph}(\epsilon, \Omega)}{t_{\rm ph, esc}} P_{\rm side}~,
\end{equation}
with $dn_{\rm ph, back}(\epsilon, \Omega) = dn_{\rm ph, fwd}(\epsilon,
\Omega)$. The sideways propagating photon density becomes part of the
observed spectrum originating from that slice. The forward and
backward propagating photon densities are used to provide feedback to
the neighbouring slices. The density of photons escaping from the
region is subtracted from the total density of that slice for the
current time step and the remainder is used for calculations in the
next time step. We point out here that even if the volume and
angle-averaged photon escape time scale for a slice turns out to be
greater than the simulation time step, on average one would still
expect some photons to escape out of the slice, as soon as they are
created, if they happen to be produced close to the boundary. In that
respect, our assumption that some percentage of photons do escape out
of the region is justified and can be used to provide feedback to
adjacent slices without having to track their movement individually at
every time step.

The process of providing photon feedback from a particular slice to
its adjacent slices, throughout the emission region (forward and
reverse), at every time step accounts for the inhomogeneity in the
photon density in the jet. In order to accurately calculate the SSC
process, the non-local retarded-time SSC losses due to photons
produced in other parts of the source need to be taken into
account. These are very important for SSC dominated sources, such as
TeV blazars, and thus cannot be ignored \citep{gr2008}. Our
multi-slice scheme with radiation feedback, lets us inherently address
this issue without having to calculate the photon density rates at
retarded times \citep[see also][]{smm2004, gr2008, bd2010}, along the
length of the emission region. We note that our approach does not
separately address the retarded-time photon field in the radial
direction as calculating the SSC process in 2D is computationlly more
intensive and beyond the scope of this work.

\section{\label{delays}Light Travel Time}

Since photons travel with a finite speed the time delays should be
incorporated in the calculation of arrival time of the radiation in
the observer's frame. Different distances are covered by photons
originating from different parts of the source but reaching the
observer at the same time. In the case of blazars, the axis of the jet
makes an angle $\theta_{obs}^*$ with the observer's line of
sight. Since the emission region moves towards us, we need to take
into account only the radiation that is emitted in the direction of
$\theta_{\rm obs}^*$. Hence to calculate the time delay of the emitted
photons, we consider the sideways direction for both emission regions,
and the forward direction for only the forward emission region. In
this section, starred quantities refer to the observer's frame,
whereas primed quantities indicate the comoving frame of the emission
region.

In case of the sideways direction, the time delay equation for the
$i^{\rm th}$ slice, of width $h^{\prime}_{\rm z}$, in the emission
region is given by

\begin{equation}
\label{tdelside}
\Delta t^{\prime}_{\rm side, a}(i_{\rm a}) = \frac{((N_{\rm tot} -
N_{\rm b} - i_{\rm a} - 1) h^{\prime}_{\rm z, a} + h^\prime_{\rm z, b}
N_{\rm b})\cos \theta^{\prime}}{c}~.
\end{equation}
Here $N_{\rm tot} = N_{\rm fs} + N_{\rm rs}$ is the total number of
slices in the entire emission region, and $\theta^{\prime}$ is the jet
viewing angle given by
\begin{equation}
\label{muprime}
\cos \theta^{\prime} = \frac{\cos \theta^*_{\rm obs} - \beta_{\rm
    sh}}{1 - \beta_{\rm sh} \cos \theta^*_{\rm obs}}~.
\end{equation}
The subscript `a' stands for the fs (forward) or rs (reverse) emission
region. The value of $N_{\rm b} = 0$ when the time delay for the
forward region is calculated and $N_{\rm b} = N_{\rm fs}$ when the
same is calculated for the reverse region.

The distance calculation using Eqn \ref{tdelside} becomes effective
only after the shocks leave a given slice. Until then, the precise
location of the shock fronts is used to calculate the time delay in
the sideways direction:
\begin{equation}
\label{fsdisteqn}
l^{\prime}_{\rm fs} = N_{\rm fs} h^{\prime}_{\rm z, fs} - c
\beta_{\rm fs} t^{\prime}
\end{equation}
and
\begin{equation}
\label{rsdisteqn}
l^{\prime}_{\rm rs} = N_{\rm fs} h^{\prime}_{\rm z, fs} + c
\beta_{\rm rs} t^{\prime}~,
\end{equation}
where $t^{\prime}$ is the total simulation time elapsed at the current
time step. The above two equations incorporate the fact that the space
in front of the shocks is still devoid of ultra-relativistic particles
and hence is inactive. In the forward direction, the time delay from
the forward emission region is calculated as

\begin{equation}
\label{tdelupfs}
\Delta t^{\prime}_{\rm up, fs} = \frac{l^{\prime}_{\rm fs} \cos
\theta^\prime}{c}~.
\end{equation}

We carry out the forward time-delay calculation for as long as the
shock is inside the forward emission region. Once the shock leaves the
region, only the forward radiation originating from the slice closest
to the observer's line of sight is observed. The forward radiation
from other slices provides feedback to adjacent slices.

All time delays are calculated with respect to the closest slice to the
observer. In addition to the above mentioned time delays, the beaming
of the radiation as observed in the observer's frame also needs to be
included. This effect is incorporated using the Doppler boosting
factor, $D = [\Gamma_{\rm sh}(1 - \beta_{\rm sh} \cos\theta^*_{\rm
    obs})]^{-1}$, that connects the comoving frame of the emission
region to the observer's frame. The total time-delay in the reception
of photons from either direction, in the observer's frame, for the
$i^{th}$ slice of an emission region is given by
\begin{equation}
\label{totdelayeqn}
t^{* ~\rm obs}_{\rm side, fs(rs); up, fs} = \frac{1 + Z}{D}
\left(t^\prime + \Delta t^{\prime}_{\rm side, fs(rs); ~up, fs}(\rm
i)\right)~,
\end{equation} 
where Z is the redshift of the source.

The frequencies have been transformed to the observer's frame by
multiplying by D, whereas the energy fluxes per unit frequency have
been transformed by multiplying by $D^{3}$ such that $(\nu F_{\nu})^*$
in Jy Hz is given by
\begin{equation}
\label{nufnueqn}
(\nu F_{\nu})^* = \frac{D^4}{1 + Z} \frac{\dot N^{\prime}_{\rm ph} h
  \nu^{\prime 2}}{4 \pi d^{*~2}_{\rm L}}~.
\end{equation}
Here $\dot N^{\prime}_{\rm ph}$ is the sum of the comoving photon
escape rates in the forward and/or sideways direction, from both
emission regions, at the current time step, and $d^{*}_{\rm L}$ is the
luminosity distance to the source.

In our model, we can accomodate the case of $\cos \theta^{*}_{\rm obs}
< \beta_{\rm sh}$. This would imply that in the comoving frame of the
emission region, the observer is located outside the cone of
superluminal motion and is looking at the region from behind. As a
result, the backside of the region would be visible to the observer
sooner than the front side. Thus, in this case, the observer would see
the sideways and backward emission of the reverse region before seeing
the sideways emission from the forward emission region. Since the
emission in the direction of $\theta^{*}_{\rm obs}$ needs to be
considered, no radiation in the forward direction from the forward
emission region would be visible to the observer.

\section{\label{study}Parameter Study}

The combination of various input parameters dictates the overall
evolution of the radiation spectrum and lightcurves in a
simulation. This makes it important to understand the effects of
varying such parameters in order to reproduce the observed features of
a source. These parameters can be broadly categorized into shock,
emission region, and jet parameters. In what follows, the unprimed
quantities refer to the AGN frame, primed quantities to the comoving
frame, and starred quantities indicate the observer's frame. The shock
parameters that affect the overall evolution of the SED are the
kinetic luminosity, $L_{\rm w}$, the total duration of the ejection
event, $t_{\rm w}$, the mass of the outer shell, $M_{\rm o}$, and the
widths and BLFs of the inner and outer shells, $\Delta_{\rm i}$ \&
$\Gamma_{\rm i}$ and $\Delta_{\rm o}$ \& $\Gamma_{\rm o}$. The
emission region parameters that have an impact on the evolution of the
radiation spectrum are the equipartition parameters,
$\varepsilon^{\prime}_{\rm e}$ \& $\varepsilon^{\prime}_{\rm B}$ (see
\S \ref{emitparam}), the particle acceleration fraction,
$\zeta^{\prime}_{\rm e}$, the electron acceleration rate parameter,
$\alpha^{\prime}$, and the particle injection index, $q^{\prime}$.
Finally, the jet parameters that play a role in the evolution of the
spectrum are the radius of the slices and the jet, $R^{\prime}$, and the
orientation angle of the jet, $\theta^{*}_{\rm obs}$.

We have carried out approximately 25 simulations to test and study the
effects of varying the values of each of these parameters on the
time-averaged simulated SEDs and their corresponding lightcurves. For
all our simulations, the flux values are calculated for the frequency
range $\nu^{\prime} = (10^{8} - 10^{26})$ Hz and electron energy
distribution (EED) range $\gamma^{\prime} = 10 - 10^{8}$ with both
ranges divided into 150 grid points. The entire emission region is
divided into 100 slices with 50 slices in the forward and 50 in the
reverse shock region. We constrain the values of some of the
parameters analytically to make sure that unphysical values are not
used in the simulations. The acceleration time scale of the highest
energy electron should be less than their corresponding synchrotron
cooling time scale. The following condition is used to place an upper
limit on the value of $\gamma^{\prime}_{\rm max}$ for particles in the
both forward and reverse shock emission regions:
\begin{equation}
\label{gmaxcondition}
\gamma^{\prime}_{\rm max} \leq \sqrt{\frac{3e}{B^{\prime} \sigma_{\rm
      T}}}~,
\end{equation}
where $e = 4.8 \times 10^{-10}$ esu is the electron's charge in cgs
units. The value of $\gamma^{\prime}_{\rm max}$ should not exceed the
maximum value of the EED range, mentioned above. In case of
$\gamma^{\prime}_{\rm min}$, the value should not fall below the
lowest value of the EED range. As pointed out by \cite{mi2004}, the
Larmor radius, $r^{\prime}_{\rm L}$, of the fastest moving
(highest-energy) electron should be smaller than the slice width so
that the magnetic field strength exceeds a certain minimum value and
the shock acceleration can take place. This condition is used to make
sure that the number of slices for both regions is selected in such a
way that
\begin{equation}
\label{rleqn}
\frac{r^{\prime}_{\rm L}}{h^{\prime}_{\rm z}} = \frac{m_e
  c^2}{eB^{\prime}} \frac{\sqrt{\gamma^{\prime~2}_{\rm max} -
    1}}{h^{\prime}_{\rm z}} < 1~.
\end{equation}

In order to carry out the parameter study, the value of each of the
shock, emission region, and jet parameters is varied twice. We study
the effects of variation in terms of changes in the flux level,
spectral hardness (SH) of the simulated SED, and simulated lightcurves
for different photon energies. Table \ref{basesetlist} shows the
values of the base set (run 1) parameters used to obtain the baseline
model.

\begin{deluxetable}{ccc}
\tabletypesize{\scriptsize}
\tablecaption{Parameter list of run 1 used to obtain the baseline
  model. \label{basesetlist}}
\tablewidth{0pt}
\tablehead{
\colhead{Parameter} & \colhead{Symbol} & \colhead{Value}
}
\startdata
Kinetic Luminosity & $L_w$ & $10^{47}$~erg/s\\
Event Duration & $t_w$ & $10^{7}$~s\\
Outer Shell Mass & $M_o$ & $4 \times 10^{31}$~g\\
Inner Shell BLF & $\Gamma_i$ & 25\\
Outer Shell BLF & $\Gamma_o$ & 10\\
Inner Shell Width & $\Delta_i$ & $3 \times 10^{15}$~cm\\
Outer Shell Width & $\Delta_o$ & $6 \times 10^{15}$~cm\\
Electron Energy Equipartition Parameter & $\varepsilon^{\prime}_e$ &
$5 \times 10^{-1}$\\
Magnetic Energy Equipartition Parameter & $\varepsilon^{\prime}_B$ &
$2 \times 10^{-3}$\\
Fraction of Accelerated Electrons & $\zeta^{\prime}_e$ & $2 \times
10^{-2}$\\
Acceleration Timescale Parameter & $\alpha^{\prime}$ & $8 \times
10^{-6}$\\
Particle Injection Index & $q^{\prime}$ & 3.4\\
Slice/Jet Radius & $R^{\prime}_z$ & $3 \times 10^{16}$~cm\\
Observer Frame Observing Angle & $\theta^{*}_{\rm obs}$ & $3.15
\deg$\\
Redshift & $z^{*}$ & 0.306\\
\enddata
\end{deluxetable}

Table \ref{paramlist} shows the values of each of the parameters that
are varied in the rest of the simulations. We study the effect on the
simulated SED and lightcurves with respect to that of the baseline
model.

\begin{deluxetable}{cc}
\tabletypesize{\scriptsize}
\tablecaption{Parameter list for other simulations. \label{paramlist}}
\tablewidth{0pt}
\tablehead{
\colhead{Run \#} & \colhead{Parameter Value}
}
\startdata
2 & $L_w = 5 \times 10^{47}$~ergs/s\\
3 & $L_w = 1 \times 10^{46}$~ergs/s\\
4 & $M_o = 1 \times 10^{32}$~g\\
5 & $\Gamma_i = 18$\\
6 & $\Gamma_i = 30$\\
7 & $\Gamma_o = 5$\\
8 & $\Gamma_o = 14$\\
9 & $\Delta_i = 3 \times 10^{16} ~\&~ \Delta_o = 6 \times 10^{16}$~cm\\
10 & $\Delta_i = 6 \times 10^{14} ~\&~ \Delta_o = 9 \times 10^{14}$~cm\\
11 & $\varepsilon^{\prime}_e = 9.5 \times 10^{-1}$\\
12 & $\varepsilon^{\prime}_e = 1 \times 10^{-2}$\\
13 & $\varepsilon^{\prime}_B = 1 \times 10^{-2}$\\
14 & $\varepsilon^{\prime}_B = 2 \times 10^{-5}$\\
15 & $\zeta^{\prime}_e = 9.5 \times 10^{-1}$\\
16 & $\zeta^{\prime}_e = 2 \times 10^{-3}$\\
17 & $\alpha^{\prime} = 6 \times 10^{-5}$\\
18 & $\alpha^{\prime} = 8 \times 10^{-8}$\\
19 & $q^{\prime} = 4.6$\\
20 & $q^{\prime} = 2.0$\\
21 & $R^{\prime}_z = 3 \times 10^{17}$~cm\\
22 & $R^{\prime}_z = 1.4 \times 10^{16}$~cm\\
23 & $\theta^{*}_{\rm obs} = 3.8 \deg$\\
24 & $\theta^{*}_{\rm obs} = 1.15 \deg$\\
25 & $\theta^{*}_{\rm obs} = 10.0 \deg$\\
\enddata
\end{deluxetable}

Simulations 2 and 3 involve changing the value of $M_o$ as well in
order to satisfy Eqn. \ref{lweqn} and obtain a non-negative value of
$M_i$, where $M_i = \frac{L_w t_w - (M_o \Gamma_o c^2)}{\Gamma_i
  c^2}$. On the other hand, changing the value of $M_o$ \textit{only}
could be carried out just once because changing it in the other
direction resulted in $M_i > M_o$, while we have assumed that the
inner shell is the faster moving one and thus should have a relatively
lower mass than the slower moving outer shell.

\section{\label{outcome}Results}

\begin{figure}
\plottwo{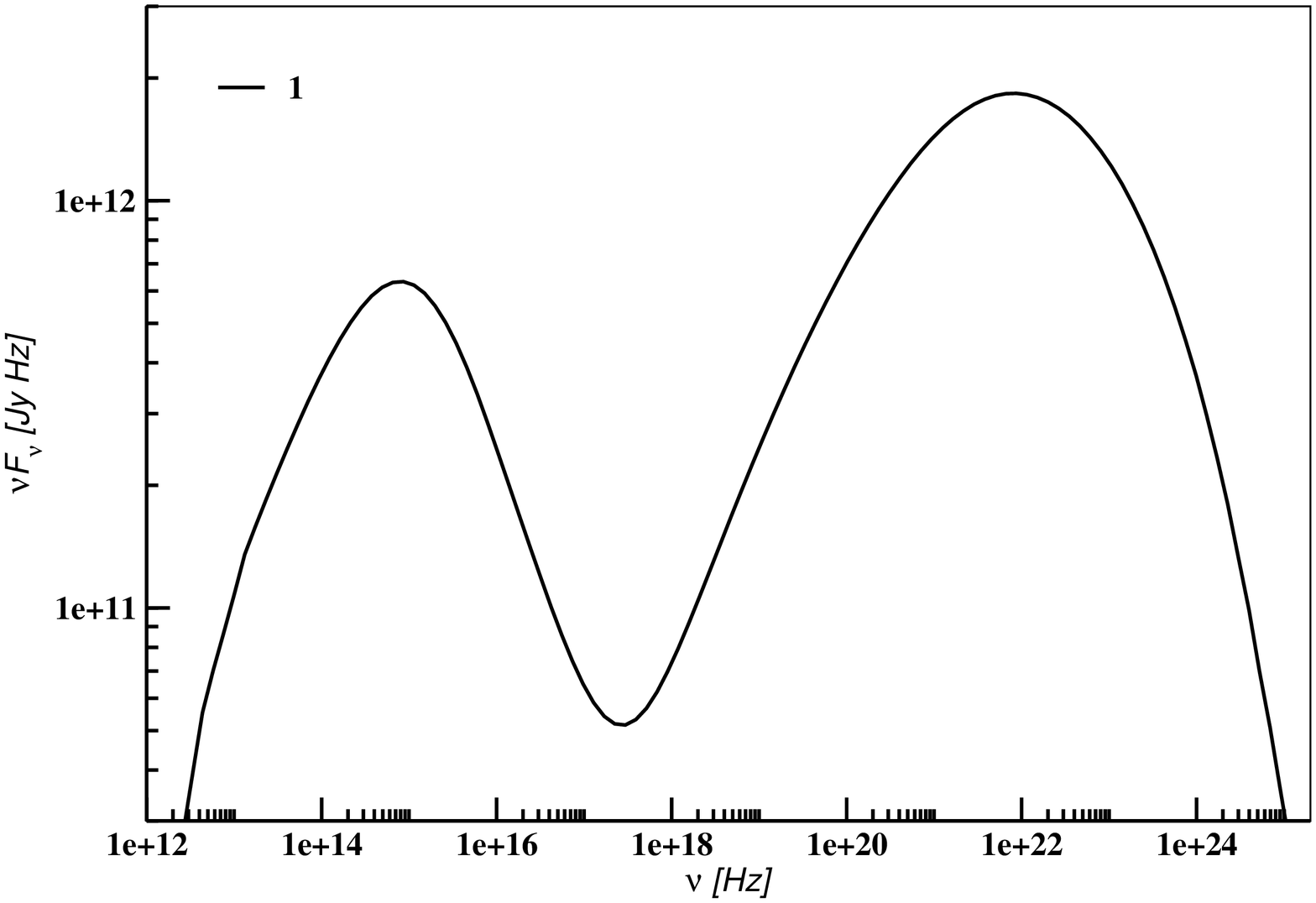}{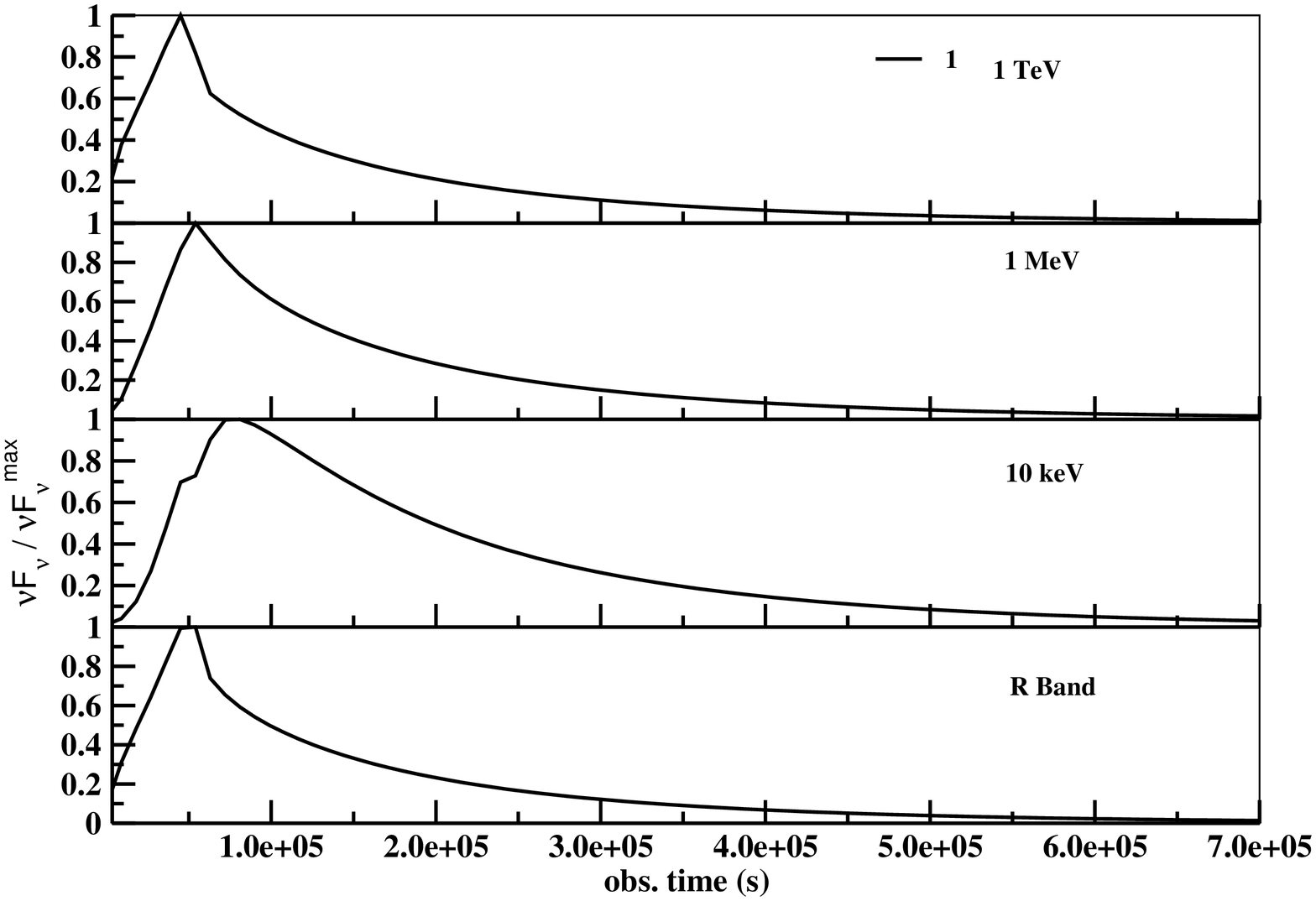}
\caption{Simulated time-integrated SED (left) and lightcurves (right)
  of our baseline model obtained using the input parameters listed in
  Table \ref{paramlist}. The panels on the right show the lightcurves
  that have been evaluated at various energies; from bottom to top, R
  band, 10 keV, 1 Mev, and 1 TeV.}
\label{1sedlc}
\end{figure}

Figure \ref{1sedlc} shows the resultant time-integrated SED and
lightcurves of the baseline model for a generic blazar source over an
8-day period. The input parameters chosen for the base set result in a
value for the BLF of the emission region as $\Gamma_{\rm sh} = 14.9$,
with the values of BLF of the forward and reverse shocks being
$\Gamma^{\prime}_{\rm fs} = 1.08$ and $\Gamma^{\prime}_{\rm rs} =
1.14$. A magnetic field value of $B^{\prime} = 2.51$ G is obtained for
both the forward and reverse emission regions. The values of
$\gamma^{\prime}_{\rm min}$ and $\gamma^{\prime}_{\rm max}$ resulting
from the input parameters, and evaluated using equations
\ref{gmineqns} \& \ref{gmaxeqn}, are obtained as $\gamma^{\prime}_{\rm
  min,fs} = 2.18 \times 10^{3}$, $\gamma^{\prime}_{\rm min,rs} = 3.74
\times 10^{3}$, and $\gamma^{\prime}_{\rm max;fs,rs} = 8.31 \times
10^{4}$. The derived total width of the forward and reverse emission
region is $\Delta^{\prime}_{\rm fs} = 8.19 \times 10^{15}$ cm and
$\Delta^{\prime}_{\rm rs} = 9.93 \times 10^{15}$ cm, with the shock
crossing time for each of the emission regions being $t^{\prime}_{\rm
  cr, fs} = 7.20 \times 10^{5}$ s and $t^{\prime}_{\rm cr, rs} = 6.94
\times 10^{5}$ s.

The time-integrated SED of the base set shows that the synchrotron
flux peaks in the optical-near UV at $\nu^{*}_{\rm syn} = 8.59 \times
10^{14}$ Hz, whereas the high-energy SSC component peaks in the MeV
regime at $\nu^{*}_{\rm SSC} = 8.72 \times 10^{21}$ Hz. The
synchrotron to SSC transition takes place in the hard X-rays at
$\nu^{*}_{\rm turn} = 2.96 \times 10^{17}$ Hz, implying that
synchrotron photons extend into the soft X-ray regime for this generic
blazar source. The derived Compton dominance factor (CDF), defined as
CDF = $\nu F^{*~\rm SSC, peak}_{\nu} / \nu F^{*~\rm syn, peak}_{\nu}$,
is 2.90. The photon spectral index in the X-ray (2 - 10 keV) range is
found to be $\alpha^{*}_{\rm 2-10 keV} = 0.549$, whereas in the Fermi
range (calculated at 10 GeV) it is $ \alpha^{*}_{\rm 10 GeV} = 2.08$.

The right side of Figure \ref{1sedlc} shows the simulated lightcurves
of the baseline model for the energy bands in the optical (R), X-ray
(10 keV), high-energy (1 MeV), and VHE regime (1 TeV). The flux value
for each energy band in all the graphs has been normalized with
respect to the maximum flux value of that energy band. 

The lightcurves indicate that the optical flux continues to rise as
long as the shocks are present in the emission region. It peaks at
$t^{*}_{\rm peak} = 54$ ks. Once the shocks break out of the system,
cooling dominates and the pulse begins to decline steadily. Thus, the
duration for which the optical pulse lasts at its half maximum (FWHM)
is 0.91 days in the observer's frame. Optical synchrotron photons are
produced from higher energy electrons, which cool on a time scale
shorter than the dynamical time scale of photons ($t^{\prime}_{\rm
  dyn} = l^{\prime}_{\rm esc} / c$) for a particular slice. Here
$l^{\prime}_{\rm esc}$ is the comoving average escape length for a
photon originating anywhere in a slice. As a result, the distribution
of such electrons remains steady during the acceleration phase of a
slice. This implies that the total optical flux includes a
contribution from the accelerating slices and the observer sees an
increase in the flux due to an increase in the volume of the emission
region. Once the shocks exit the respective emission regions and the
cooling phase dominates the region, the optical flux decreases
immediately. This makes the rising and decaying phases of the pulse
almost equal leading to a quasi-symmetrical profile of the optical
pulse.

On the other hand, the X-ray flux continues to rise even after the
shocks exit the emission region. This happens because of two reasons:
the 10 keV photons are produced from upscattering of lower energy
(near IR) synchrotron photons off lower energy electrons. Such
electrons remain in the system for a longer time due to slow
cooling. As a result, the rise in the corresponding intensity is
slow. Also, since no more electrons are added into the system during
the cooling phase there is a continued increase of late-arriving
photons at the sites of scattering due to the internal light crossing
time effect. As a result, initially we receive the emission from the
distribution that starts to cool in the back slices while the front
slices are still accelerating. Once the cooling phase takes over the
emission region, an important contribution comes from the front slices
as the back slices have a higher number of cooled off lowest energy
electrons. As a result, for a SSC dominated radiation of $\nu \sim
10^{16}$ Hz, the pulse experiences delayed peaking and decays with a
long and slow tail compared to its rise. This explains why the 10 keV
pulse has an asymmetrical profile and peaks last at $t^{*}_{\rm peak}
= 81$ ks and persists for the longest time compared to the pulses at
other frequencies, with a FWHM of 1.8 days (in the observer's
frame). Dividing the emission region into various slices increases the
spatial resolution of the region. As a result of this enhanced
resolution, when the shocks exit the system and pulses of higher
synchrotron and SSC frequencies die due to the reason explained above,
the 10 keV pulse, which builds up slowly over time bears the signature
of this exit. The kink seen in the lightcurve of the 10 keV photons is
the signature of shocks completely breaking out of the system with the
forward shock - the last shock to come out of the region - exiting the
forward emission region at $t^{*}_{\rm cr} = 5.27 \times 10^{4}$ s.

The 1 MeV photons are a result of Compton upscattering of higher
energy optical photons off low energy electrons, making the pulse peak
at the same time as R-band photons at $t^{*}_{\rm peak} = 54$ ks. At
all energies, the pulse rises sharply and declines comparatively
gradually. This implies that, for the reasons explained in the case of
the R-band pulse profile, the profile of the 1 MeV pulse is also
quasi-symmetrical.

The very high energy (VHE) photons (1 TeV) result from the SSC
scattering of higher energy (soft X-rays) synchrotron photons off
higher energy electrons. As these electrons cool at a very fast rate
compared to the electrons emitting at optical frequencies, the VHE
photons are produced only during the acceleration phase. This is why
the corresponding pulse has a sharper rise and peaks sooner than its
lower energy counterparts, at $t^{*}_{\rm peak} = 45$ ks. The pulse
undergoes a sharp drop in the 1 TeV energy regime exactly at the time
the shocks completely leave the system, thereby having the shortest
pulse duration with a FWHM of 0.81 days and a nearly symmetrical
profile.

For all simulations, including the baseline model, the average SED has
been evaluated over an integration time of 9000 s. This facilitates
comparison with observations, which usually require extended exposure
times. Table \ref{parameffect} lists the effects of changing the
values of the input parameters (according to Table \ref{paramlist}) on
the flux, the peak synchrotron and SSC frequencies ($\nu^{*}_{\rm syn}
~\&~ \nu^{*}_{\rm SSC}$, respectively), the trough frequency
($\nu^{*}_{\rm turn}$) indicating the transition from the synchrotron
to SSC component, the SH represented by energy spectral indices
$\alpha^{*}$ in the 2 -- 10~keV range and at 10~GeV, and the CDF of
the baseline model's time-integrated SED.

\begin{deluxetable}{ccccccccc}
\tabletypesize{\scriptsize}
\setlength{\tabcolsep}{0.02in}
\tablecaption{Parameter study of shock, emission-region, and jet
  parameters on the time-integrated SED. \label{parameffect}}
\tablewidth{0pt}
\tablehead{
\colhead{Input Parameter} & \colhead{Value} & \colhead{$\nu
  F_{\nu}^{*}$} & \colhead{$\nu^{*}_{\rm syn}$ [Hz]} &
\colhead{$\nu^{*}_{\rm turn}$ [Hz]} & \colhead {$\nu^{*}_{\rm SSC}$ }
& \colhead{$CDF^{*}$} & \colhead{$SH^{*}$ (2-10 keV)} &
\colhead{$SH^{*}$ (10 GeV)}
}
\startdata
Baseline & & & $8.59 \times 10^{14}$ & $2.96 \times 10^{17}$ & 8.72
$\times 10^{21}$ & 2.90 & $\alpha_{\rm 2-10keV} = 0.549$ &
$\alpha_{\rm 10GeV} = 2.08$\\
$L_w$ & $\uparrow$ & $\uparrow$ & $\uparrow$ & $\uparrow$ & $\uparrow$
& 3.09 & $\downarrow \alpha_{\rm 2-10keV} = 0.682$ & $\uparrow
\alpha_{\rm 10GeV} = 2.04$\\
      & $\downarrow$ & $\downarrow$ & $\downarrow$ & $\uparrow$ &
$\downarrow$ & 2.20 & $\uparrow \alpha_{\rm 2-10keV} = 0.444$ &
$\downarrow \alpha_{\rm 10GeV} = 2.09$\\
$M_o$ & $\uparrow$ & $\downarrow$ & $\uparrow$ & $\uparrow$ &
$\uparrow$ & 4.84 & $\downarrow \alpha_{\rm 2-10keV} = 0.847$ &
$\uparrow \alpha_{\rm 10GeV} = 1.75$\\
$\Gamma_i$ & $\uparrow$ & $\uparrow$ & $\uparrow$ & $\uparrow$ &
$\uparrow$ & 2.86 & $\downarrow \alpha_{\rm 2-10keV} = 0.824$ &
$\uparrow \alpha_{\rm 10GeV} = 1.92$\\
           & $\downarrow$ & $\downarrow$ & $\downarrow$ & $\downarrow$
& $\downarrow$ & 2.65 & $\uparrow \alpha_{\rm 2-10keV} = 0.441$ &
$\downarrow \alpha_{\rm 10GeV} = 2.23$\\
$\Gamma_o$ & $\uparrow$ & $\downarrow$ & $\downarrow$ & $\downarrow$ &
$\downarrow$ & 2.34 & $\uparrow \alpha_{\rm 2-10keV} = 0.377$ &
$\downarrow \alpha_{\rm 10GeV} = 2.36$\\
           & $\downarrow$ & $\uparrow$ & $\uparrow$ & $\uparrow$ &
$\uparrow$ & 2.68 & $\downarrow \alpha_{\rm 2-10keV} = 1.71$ &
$\uparrow \alpha_{\rm 10GeV} = 1.34$\\
$\Delta_i \& \Delta_o$ & $\uparrow$ & $\downarrow$ & $\downarrow$ &
$-$ & $\downarrow$ & 2.45 & $\uparrow \alpha_{\rm 2-10keV} = 0.348$ &
$\uparrow \alpha_{\rm 10GeV} = 2.01$\\
                       & $\downarrow$ & $\uparrow$ & $\uparrow$ & $-$
& $\uparrow$ & 3.04 & $\downarrow \alpha_{\rm 2-10keV} = 0.679$ &
$\downarrow \alpha_{\rm 10GeV} = 2.10$\\
$\varepsilon_e$ & $\uparrow$ & $\uparrow$ & $\uparrow$ & $\uparrow$ &
$\uparrow$ & 4.73 & $\downarrow \alpha_{\rm 2-10keV} = 0.993$ &
$\uparrow \alpha_{\rm 10GeV} = 1.84$\\
                & $\downarrow$ & $\downarrow$ & $\downarrow$ &
$\downarrow$ & $\downarrow$ & 0.0386 & $\downarrow \alpha_{\rm
  2-10keV} = 1.44$ & $\downarrow \alpha_{\rm 10GeV} = 2.39$\\
$\varepsilon_B$ & $\uparrow$ & $\uparrow$ & $\uparrow$ & $\uparrow$ &
$\uparrow$ & 0.826 & $\downarrow \alpha_{\rm 2-10keV} = 0.962$ &
$\uparrow \alpha_{\rm 10GeV} = 2.07$\\
                & $\downarrow$ & $\downarrow$ & $\downarrow$ &
$\downarrow$ & $\downarrow$ & 33.3 & $\uparrow \alpha_{\rm 2-10keV} =
0.087$ & $\uparrow \alpha_{\rm 10GeV} = 1.95$\\
$\zeta_e$ & $\uparrow$ & $\downarrow$ & $\downarrow$ & $\downarrow$ &
$\downarrow$ & 1.86 & $\downarrow \alpha_{\rm 2-10keV} = 0.986$ &
$\downarrow \alpha_{\rm 10GeV} = 3.45$\\
          & $\downarrow$ & $\uparrow$ & $\uparrow$ & $\uparrow$ &
$\uparrow$ & 1.66 & $\downarrow \alpha_{\rm 2-10keV} = 1.71$ &
$\uparrow \alpha_{\rm 10GeV} = 0.844$\\
$\alpha$ & $\uparrow$ & $\uparrow$ & $-$ & $\downarrow$ & $-$ & 2.90 &
$-$ & $\uparrow \alpha_{\rm 10GeV} = 2.07$\\
         & $\downarrow$ & $\downarrow$ & $\uparrow$ & $\uparrow$ &
$\uparrow$ & 3.71 & $\uparrow \alpha_{\rm 2-10keV} = 0.540$ &
$\downarrow \alpha_{\rm 10GeV} = 2.19$\\
$q$ & $\uparrow$ & $\uparrow$ & $-$ & $\downarrow$ & $\downarrow$ &
3.37 & $\uparrow \alpha_{\rm 2-10keV} = 0.340$ & $\downarrow
\alpha_{\rm 10GeV} = 2.67$\\
    & $\downarrow$ & $\downarrow$ & $\uparrow$ & $\uparrow$ &
$\uparrow$ & 2.65 & $\downarrow \alpha_{\rm 2-10keV} = 1.39$ &
$\uparrow \alpha_{\rm 10GeV} = 1.40$\\
$R_z$ & $\uparrow$ & $\downarrow$ & $\downarrow$ & $\downarrow$ &
$\downarrow$ & 1.81 & $\uparrow \alpha_{\rm 2-10keV} = 0.228$ &
$\downarrow \alpha_{\rm 10GeV} = 2.11$\\
      & $\downarrow$ & $\uparrow$ & $\uparrow$ & $\uparrow$ &
$\uparrow$ & 3.07 & $\downarrow \alpha_{\rm 2-10keV} = 0.669$ &
$\uparrow \alpha_{\rm 10GeV} = 2.04$\\
$\theta_{\rm obs}$ & $\uparrow$ & $\downarrow$ & $\downarrow$ &
$\downarrow$ & $\downarrow$ & 2.92 & $\uparrow \alpha_{\rm 2-10keV} =
0.525$ & $-$\\
                   & $\downarrow$ & $\uparrow$ & $\uparrow$ &
$\uparrow$ & $\uparrow$ & 2.86 & $\downarrow \alpha_{\rm 2-10keV} =
0.662$ & $\uparrow \alpha_{\rm 10GeV} = 1.87$\\
outside beaming cone & $\uparrow$ & $\downarrow$ & $\downarrow$ &
$\downarrow$ & $\downarrow$ & 2.86 & $\uparrow \alpha_{\rm 2-10keV} =
0.449$ & $\downarrow \alpha_{\rm 10GeV} = 2.24$\\
\enddata 
\tablecomments{The slope of the low-energy component of the
  time-integrated SED obtained from run 1 in the X-ray range (2-10
  keV) is $\alpha^{*}_{\rm 2-10keV} = 0.549$ and that of the
  high-energy component in the Fermi range (10 GeV) is
  $\alpha^{*}_{\rm 10GeV} = 2.08$. The time-integrated low-energy
  component of the baseline model peaks in the optical-UV range at
  $\nu^{*}_{\rm syn} = 8.59 \times 10^{14}$ Hz. The high-energy
  component peaks in the MeV range at $\nu^{*}_{\rm SSC} = 8.72 \times
  10^{21}$ with the transition from synchrotron to SSC component
  taking place in the X-rays at $\nu^{*}_{\rm turn} = 2.96 \times
  10^{17}$. The $\uparrow$ implies an increase in the value of a
  particular quantity, $\downarrow$ refers to a decrease in the value,
  and $-$ stands for no change in the value. The abbreviation CDF
  stands for Compton Dominance Factor, defined in \S \ref{outcome},
  and SH stands for Spectral Hardness, defined in \S \ref{study}.}
\end{deluxetable}

Table \ref{lceffect} parametrizes the simulated lightcurves in terms
of the time of the last shock to break out of the region, $t^*_{\rm
  cr}$, the time at which the pulse at a given energy peaks, $t^*_{\rm
  peak}$, and the FWHM (indicated as $W^*$ in the table) of the pulse
at a given energy. 

\begin{deluxetable}{ccccccccccc}
\tabletypesize{\scriptsize}
\setlength{\tabcolsep}{0.02in}
\tablecaption{Parameter study of shock, emission-region, and jet
  parameters on simulated lightcurves. \label{lceffect}}
\tablewidth{0pt}
\tablehead{
\colhead{} & \colhead{} & \colhead{} & \multicolumn{2}{c}{R-Band} &
\multicolumn{2}{c}{10 keV} & \multicolumn{2}{c}{1 MeV} &
\multicolumn{2}{c}{1 TeV}\\
\cline{4-5} \cline{6-7} \cline{8-9} \cline{10-11}\\
\colhead{Input Parameter} & \colhead{Value} & \colhead{$t^*_{\rm cr}$
  [days]} & \colhead{$t^*_{\rm peak}$ [ks]} & \colhead{$W^*$ [days]} &
\colhead{$t^*_{\rm peak}$ [ks]} & \colhead{$W^*$ [days]} &
\colhead{$t^*_{\rm peak}$ [ks]} & \colhead{$W^*$ [days]} &
\colhead{$t^*_{\rm peak}$ [ks]} & \colhead{$W^*$ [days]}
}
\startdata
Baseline & & 0.61 & 54 & 0.91 & 81 & 1.8 & 54 & 1.1 & 45 & 0.81\\
$L_w$ & $\uparrow$ & 0.61 & 45 & 0.60 & 54 & 1.1 & 45 & 0.78 & 45 &
0.56\\
      & $\downarrow$ & 0.61 & 45 & 1.1 & 180 & 3.2 & 99 & 2.3 & 45 &
0.56\\
$M_o$ & $\uparrow$ & 1.6 & 63 & 1.6 & 162 & 2.1 & 144 & 2.2 & 54 &
1.5\\
$\Gamma_i$ & $\uparrow$ & 0.55 & 45 & 0.78 & 72 & 1.8 & 54 & 1.1 & 45
& 0.95\\
           & $\downarrow$ & 0.83 & 72 & 0.97 & 108 & 1.9 & 72 & 1.3 &
63 & 0.85\\
$\Gamma_o$ & $\uparrow$ & 1.3 & 72 & 1.5 & 144 & 2.3 & 99 & 1.7 & 72 &
1.2\\
           & $\downarrow$ & 0.37 & 18 & 0.34 & 36 & 1.6 & 18 & 0.68 &
18 & 1.4\\
$\Delta_i \& \Delta_o$ & $\uparrow$ & 6.1 & 522 & 5.5 & 504 & 4.4 &
495 & 5.0 & 522 & 5.5\\
                       & $\downarrow$ & 0.11 & 9 & 0.27 & 18 & 0.97 &
9 & 0.40 & 0 & 0.26\\
$\varepsilon^{\prime}_e$ & $\uparrow$ & 0.61 & 54 & 0.85 & 72 & 1.8 &
54 & 1.1 & 45 & 0.76\\
                & $\downarrow$ & 0.61 & 45 & 0.61 & 45 & 2.2 & 144 &
3.5 & 45 & 0.50\\
$\varepsilon^{\prime}_B$ & $\uparrow$ & 0.61 & 45 & 0.56 & 45 & 1.3 &
45 & 0.80 & 45 & 0.54\\
                & $\downarrow$ & 0.61 & 63 & 1.39 & 162 & 2.44 & 99 &
1.7 & 54 & 1.2\\
$\zeta^{\prime}_e$ & $\uparrow$ & 0.61 & 180 & 3.3 & 162 & 3.4 & 117 &
2.4 & 45 & 0.88\\
          & $\downarrow$ & 0.61 & 45 & 0.64 & 45 & 0.54 & 45 & 1.7 &
45 & 0.54\\
$\alpha^{\prime}$ & $\uparrow$ & 0.61 & 54 & 0.91 & 81 & 1.8 & 54 &
1.1 & 45 & 0.81\\
         & $\downarrow$ & 0.61 & 45 & 0.62 & 72 & 1.9 & 45 & 0.94 & 45
& 0.58\\
$q^{\prime}$ & $\uparrow$ & 0.61 & 54 & 0.91 & 72 & 1.8 & 54 & 1.0 &
45 & 0.88\\
    & $\downarrow$ & 0.61 & 45 & 0.88 & 45 & 1.1 & 54 & 1.3 & 45 &
0.63\\
$R^{\prime}_z$ & $\uparrow$ & 0.61 & 117 & 7.4 & 1332 & 28 & 540 & 18
& 54 & 0.60\\
      & $\downarrow$ & 0.61 & 45 & 0.55 & 45 & 0.68 & 45 & 0.59 & 45 &
0.55\\
$\theta^*_{\rm obs}$ & $\uparrow$ & 0.72 & 54 & 1.1 & 90 & 2.1 & 63 &
1.5 & 54 & 0.83\\
                   & $\downarrow$ & 0.40 & 36 & 0.70 & 54 & 1.3 & 36 &
0.77 & 27 & 0.81\\
outside beaming cone & $\uparrow$ & 2.8 & 234 & 1.9 & 432 & 5.7 & 252
& 5.1 & 234 & 2.0\\
\enddata 
\tablecomments{The lightcurves of the baseline model are parametrized
  to study the effects of variation of various input parameters,
  listed in Table \ref{paramlist}. We parametrize these lightcurves
  according to the shock crossing time, $t^*_{\rm cr}$, time of
  peaking of a pulse at a given energy, $t^*_{\rm peak}$, and full
  width at half maximum (FWHM), indicated as $W^*$, of a pulse at a
  particular energy.}
\end{deluxetable}

The simulated time-integrated SEDs and the corresponding lightcurves
are affected in several ways as input parameters are varied. The
variation directly changes the values of some of the physical
quantities like $\Gamma^{\prime}_{\rm fs}$, $\Gamma^{\prime}_{\rm
  rs}$, relative velocity of the shells, internal energy of the
shocks, magnetic field, $\gamma^{\prime}_{\rm min} ~\&~
\gamma^{\prime}_{\rm max}$, particle density, and size of the region,
which affect the overall flux, SH, and CDF of the spectrum. This also
directly alters the time at which a pulse at a given energy peaks, as
well as its FWHM. We describe below some of the dominant effects
observed on the simulated SED and lightcurves according to the changes
in the values of the above mentioned physical quantities.

A) An increase in the internal energy of the shocks boosts the flux
level of the spectrum.

B) An increase in the density of injected electrons increases the
overall flux of the broadband spectrum. It also increases the photon
number density in the region. This implies rapid radiative cooling for
electrons due to synchrotron and SSC processes, and a higher value of
CDF due to an increased level of SSC flux. In terms of lightcurve
profiles, this translates into acceleration and cooling setting in
sooner, thereby making the pulse at all energies peak sooner and last
for a shorter time compared to the case where the acceleration and
cooling set in more gradually.

C) An increase in the value of relative velocity of the two shells,
along with the value of their individual shock propagation Lorentz
factors ($\Gamma^{\prime}_{\rm fs} ~\&~\Gamma^{\prime}_{\rm rs}$),
increases the acceleration efficiency of the collision. As a result,
the overall flux of the spectrum rises. In terms of lightcurve
profiles, due to efficient acceleration the particles are energized
sooner and start to cool down faster. Thus, the higher the
acceleration efficiency, the faster is the particle acceleration and
the subsequent cooling. Hence, pulses at all energies peak sooner and
last for a shorter duration of time, except for the VHE pulse. The
FWHM of the VHE pulse lasts for a longer time, compared to that of the
baseline model, because of efficient acceleration that energizes more
particles to higher energies. As a result, high-energy electrons
capable of producing TeV electrons are available for a longer time,
and the FWHM of the TeV pulse increases.

D) An increase in the value of $\gamma^{\prime}_{\rm min}$ implies
more higher energy electrons being injected into the system. The
resultant spectrum shifts toward higher frequencies. The shift in peak
($\nu^*_{\rm syn}$ \& $\nu^*_{\rm SSC}$) and trough ($\nu^*_{\rm
  turn}$) frequencies makes the spectrum softer in the X-ray (2 - 10
keV) but harder in the Fermi range (10 GeV). Usually the X-ray range
forms the lower energy part of the SSC spectrum while the Fermi range
is around or beyond the SSC peak. However, if $\nu^*_{\rm sy}$
increases, the high-frequency end of the synchrotron component may
extend into the 2 -- 10~keV regime. As a result, the corresponding
value of $\alpha^*_{\rm 2-10keV}$ increases, indicating a softer X-ray
range spectrum. A shift of $\nu^*_{\rm SSC}$ to higher frequencies
will harden the Fermi spectrum.

E) A much lower value of $\gamma^{\prime}_{\rm min}$ implies the
presence of very low energy electrons in the system. This shifts the
spectrum leftwards, toward lower frequencies, and higher order SSC
components become prominent. This happens because the Thomson depth of
the region increases and Klein-Nishina effects no longer suppress the
higher-order SSC components. In terms of lightcurve profiles, this
implies that acceleration and hence cooling set in very gradually for
low energy electrons. On the other hand, higher energy electrons do
not last for a long time in the system. As a result, photons resulting
directly from these electrons are produced very early on and last for
a very short time.

F) The magnetic field in the emission region is responsible for
synchrotron radiation. An increase in the value of the magnetic field
decreases the synchrotron cooling time scale, which makes the
synchrotron mechanism the primary mode of radiative cooling for the
electrons. As a result, the synchrotron component of the spectrum
becomes dominant while the photon flux due to SSC scattering
decreases, thus reducing the resultant CDF value. The overall flux of
the spectrum increases, but due to excessive radiative cooling the
spectrum becomes softer, as it comes from a cooled population of
electrons. In terms of lightcurve profiles, the shorter synchrotron
cooling time scales lead to shorter pulses compared to the baseline
model.

G) A larger emission region implies a larger comoving volume, but
lower values of the magnetic field, shocks' internal energies, and
density of injected electrons in the region. As a result, the flux and
CDF of the spectrum decrease following the explanation given in (A)
and (B). In terms of lightcurve profiles, since the acceleration and
radiative cooling set in gradually, the pulse at a given energy peaks
later and has a higher FWHM relative to that of the baseline model.

H) An increase in the value of $\gamma^{\prime}_{\rm max}$ without any
change in the magnetic field value implies a shorter synchrotron
cooling timescale for the highest-energy electrons. This results in
more synchrotron photons in the region, thereby boosting the flux
level of the SSC emission.

Subsections \ref{shocksed} - \ref{jetsed} briefly discuss the effects
of varying various input parameters on the time-averaged simulated
spectra and lightcurves of a generic blazar source.

\subsection{\label{shocksed}Shock Parameters}

\begin{figure}
\plottwo{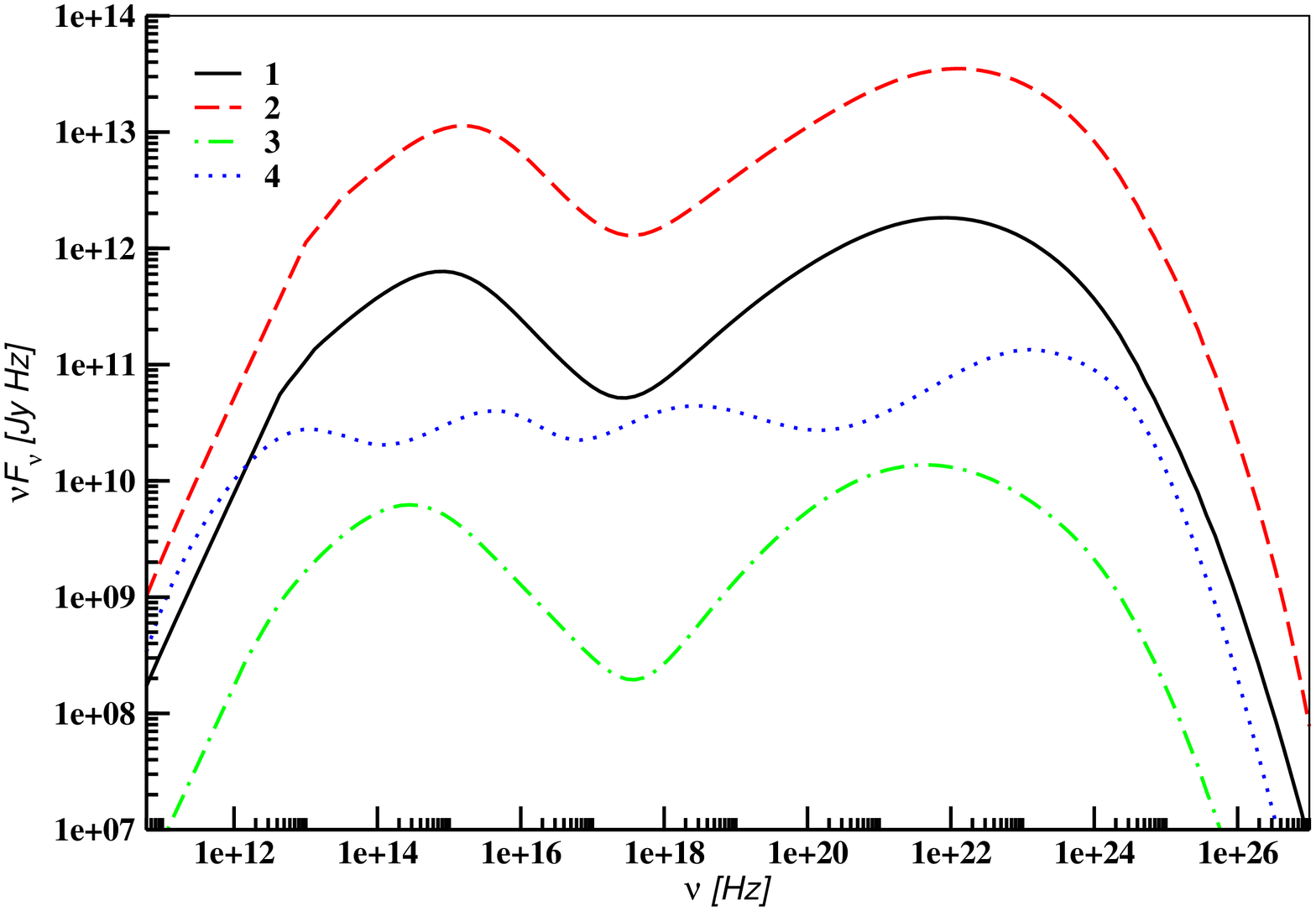}{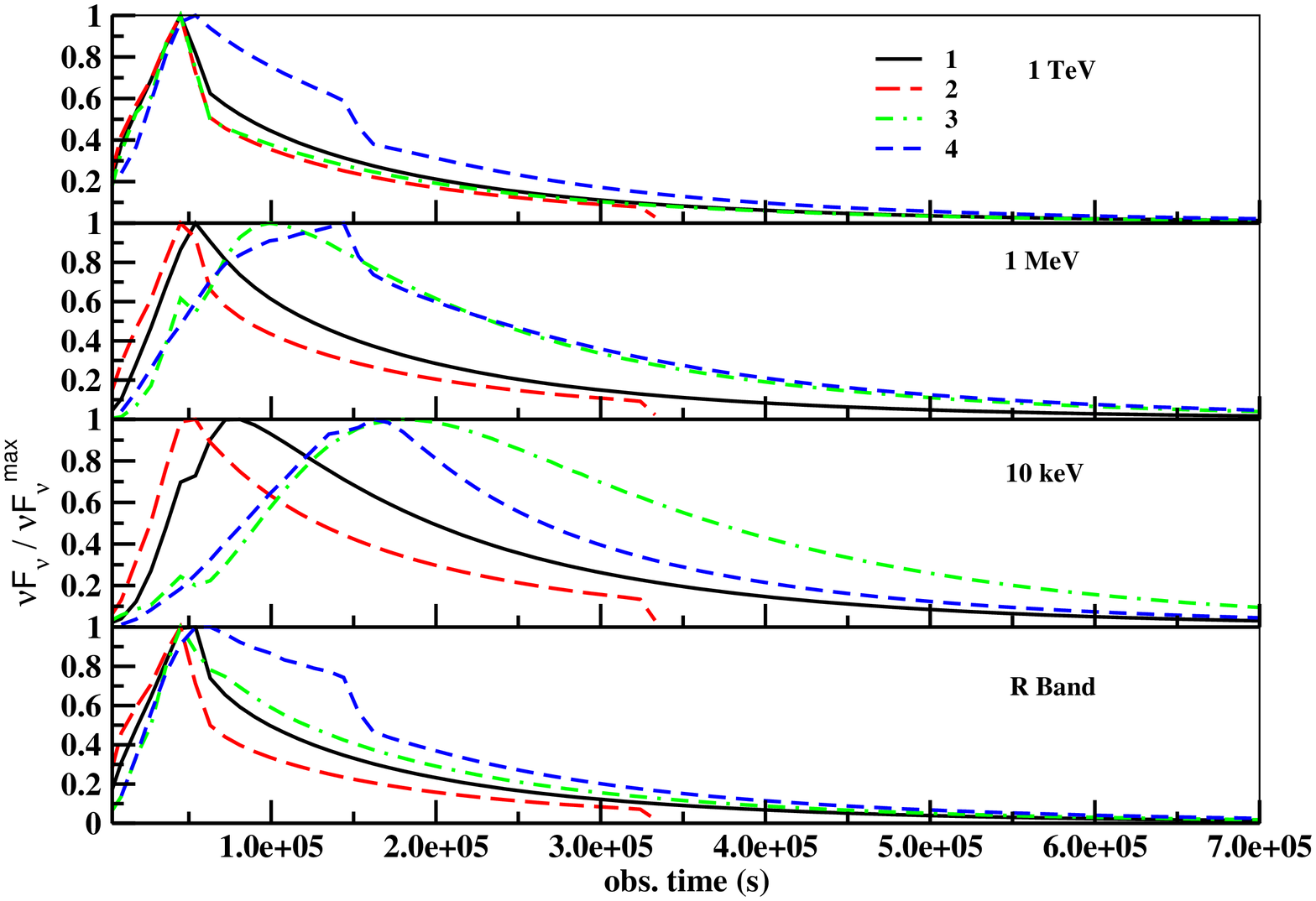}
\caption{Simulated time-integrated SED and lightcurves of a generic
  blazar source illustrating the effects of varying shock parameters,
  $L_{\rm w}$ (runs 2 \& 3), $M_{\rm o}$ (run 4) according to Table
  \ref{paramlist}.}
\label{234sedlc}
\end{figure}

\begin{figure}
\plottwo{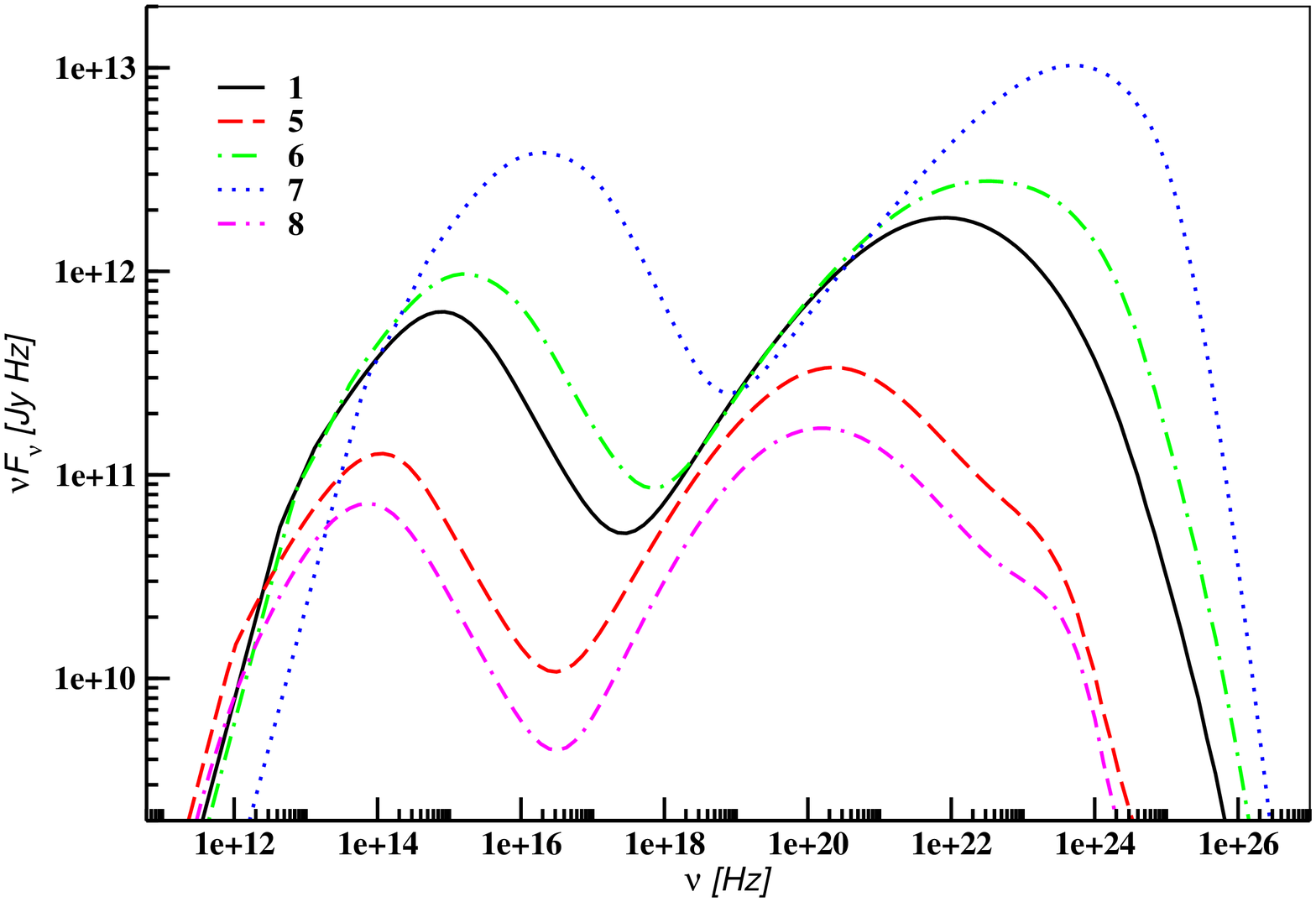}{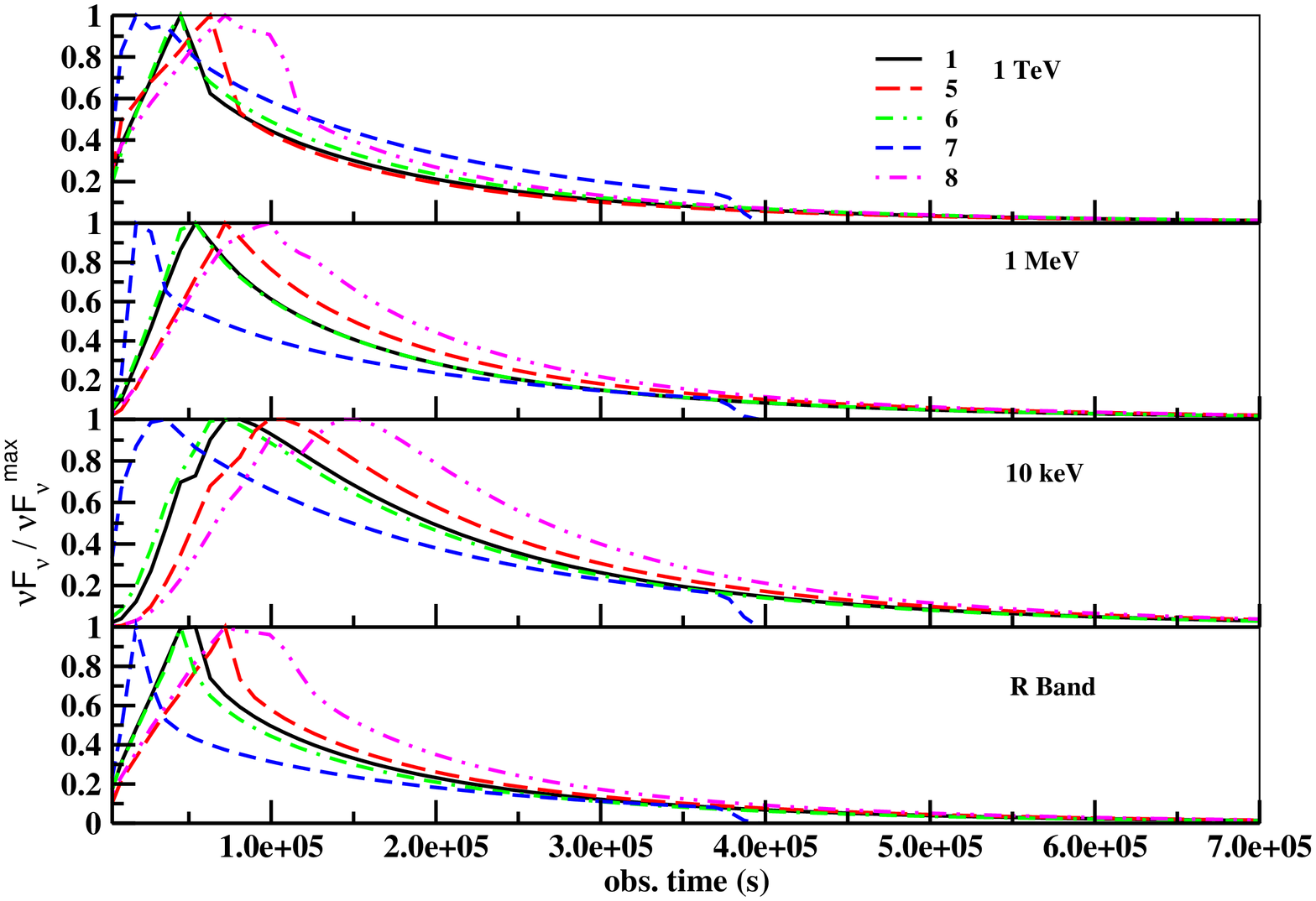}
\caption{Simulated time-integrated SED and lightcurves of a generic
  blazar source illustrating the effects of varying shock parameters
  $\Gamma_{\rm i}$ (runs 5 \& 6) and $\Gamma_{\rm o}$ (runs 7 \& 8)
  according to Table \ref{paramlist}.}
\label{5678sedlc}
\end{figure}

\begin{figure}
\plottwo{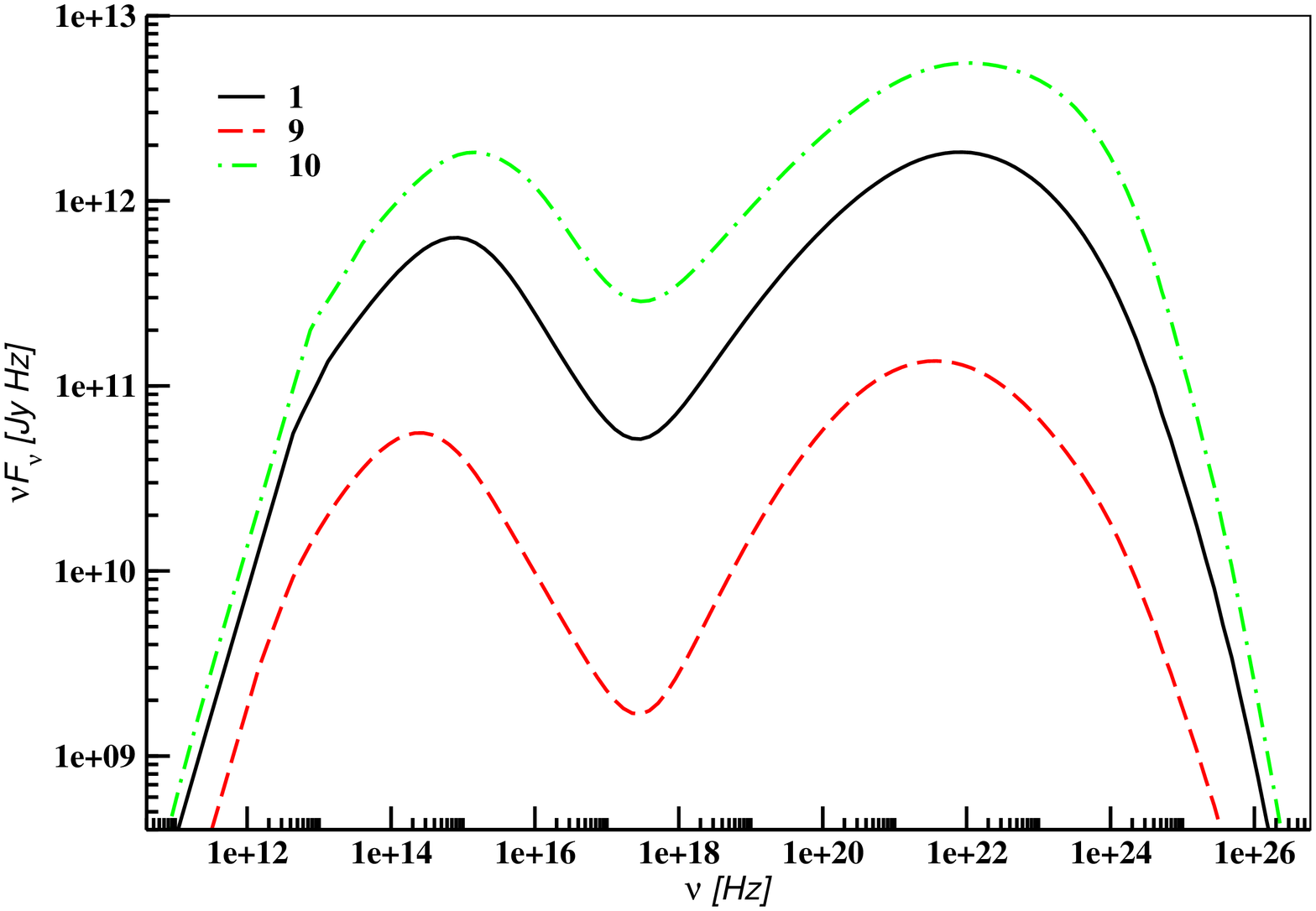}{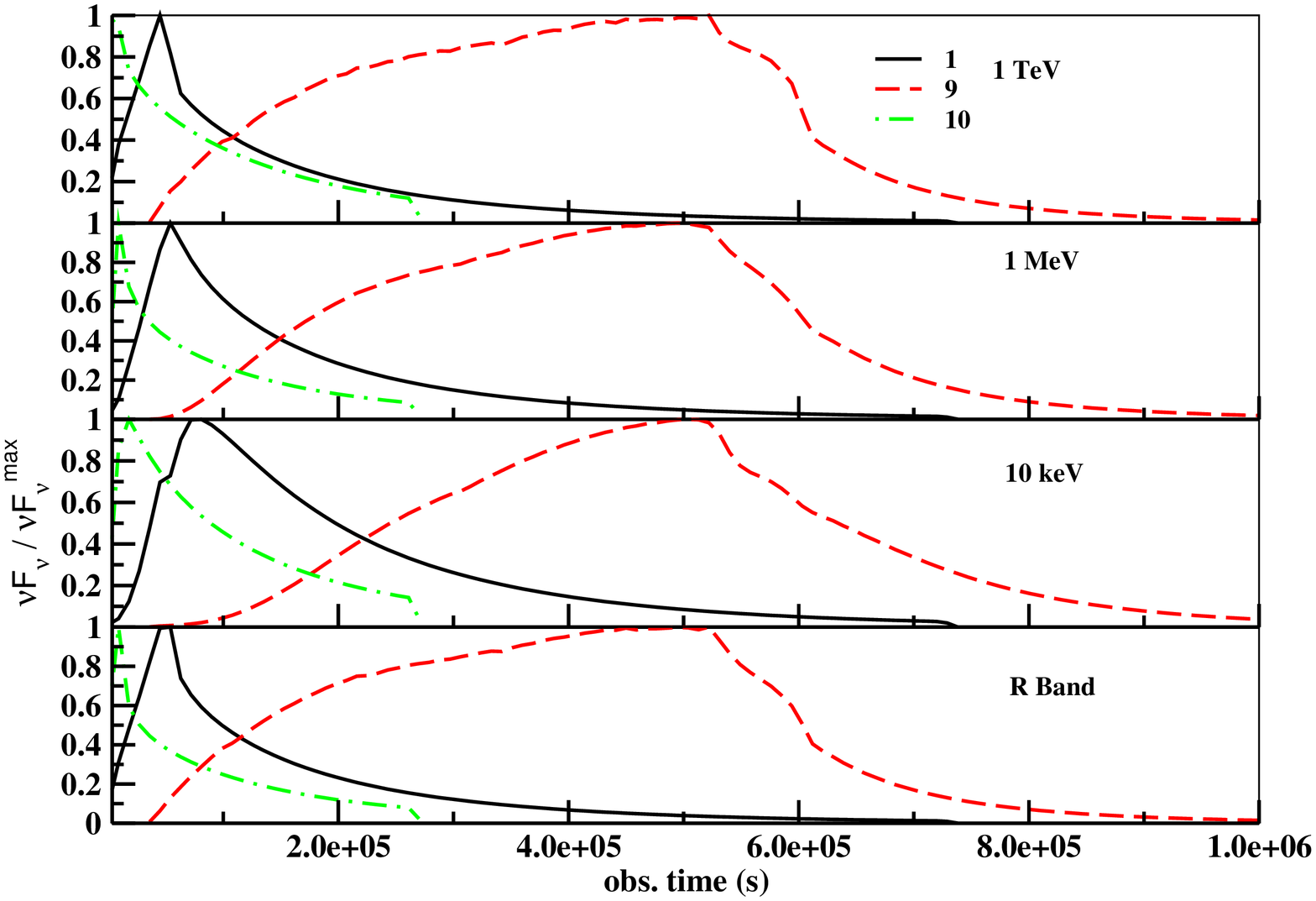}
\caption{Simulated time-integrated SED and lightcurves of a generic
  blazar source illustrating the effects of varying shock parameters
  $\Delta_{\rm i}$ \& $\Delta_{\rm o}$ (runs 9 \& 10) according to
  Table \ref{paramlist}.}
\label{910sedlc}
\end{figure}

Figures \ref{234sedlc} - \ref{910sedlc} depict the effects of varying
shock parameters ($L_{\rm w}$, $M_{\rm o}$, $\Gamma_{\rm i}$,
$\Gamma_{\rm o}$, and $\Delta_{\rm i}$ \& $\Delta_{\rm o}$) on the
simulated time-integrated SED (left side) and lightcurves (right side)
of the baseline model (run 1).

Increasing $L_{\rm w}$ (run 2) increases the value of the internal
energy of the shocks, the magnetic field, $\gamma^{\prime}_{\rm min}$,
and the density of the injected electrons in both emission
regions. Due to the reasons cited in (A), (B), and (D) the overall
flux of the spectrum rises (see Fig. \ref{234sedlc}) and the SH
decreases in the X-ray range but increases in the Fermi range (see
Table \ref{parameffect}). The CDF value increases marginally due to
the reason cited in (B). On the other hand, decreasing $L_{\rm w}$
(run 3) from its original value results in a spectrally harder
spectrum in the X-rays and a softer one in the Fermi range, along with
an overall lower flux of the broadband time-integrated SED and a
reduced CDF value.

Making the outer shell heavier ($\uparrow M_o$, run 4) makes the shell
denser while making the inner shell lighter and rarer. This
combination of densities decreases the values of $\Gamma_{\rm sh}$ and
$\Gamma^{\prime}_{\rm fs}$. The shocks' internal energies and the
magnetic field values decrease along with the density of electrons
injected into the forward emission region. This results in an overall
decrease of the flux throughout the spectrum according to the reasons
cited in (A) and (B). The $\gamma^{\prime}_{\rm min, fs}$ value
decreases, whereas $\gamma^{\prime}_{\rm min, rs}$ increases due to an
increase in $\Gamma^{\prime}_{\rm rs}$. Consequently, a mixed
population of lower and higher energy electrons arises in the emission
region during the cooling phase, giving rise to a bumpy
time-integrated SED. This also leads to a comparatively longer
acceleration and cooling phase in the emission region. Since the X-ray
range falls in the 3rd hump of the spectrum, which forms the lower
energy part of the SSC component, the spectrum is softer (see Table
\ref{parameffect}) in this case. On the other hand, an increase in
$\gamma^{\prime}_{\rm min, rs}$ results in the shifting of $\nu^*_{\rm
  syn}$, $\nu^*_{\rm turn}$, and $\nu^*_{\rm SSC}$ to higher
frequencies. As a result, the spectrum is harder in the Fermi range,
as explained in (D) and indicated by the value of $\alpha^*_{\rm
  10GeV}$ in Table \ref{parameffect}. An increase in
$\Gamma^{\prime}_{\rm rs}$ decreases the comoving width of the reverse
emission region, thereby decreasing the comoving volume and increasing
the comoving number density of electrons in the region. This leads to
an increased contribution of SSC photons from the reverse emission
region, thus slightly increasing the CDF value.

The lightcurves of run 2 ($L_{\rm w} \uparrow$), 3 ($L_{\rm w}
\downarrow$), and 4 ($M_{\rm o} \uparrow$; right side of
Fig. \ref{234sedlc}) indicate a similar pattern as that of run 1. The
same reasons used to explain the time of peaking of a pulse and its
FWHM, at a given energy, for run 1 in \S \ref{outcome} can be applied
to runs 2, 3, and 4 to understand their pulse profiles. In case of run
2, pulses at all energies peak sooner compared to run 1, except for
the TeV pulse. They also last for a shorter time due to the reasons
cited in (B). The opposite holds true for run 3, although the optical
pulse peaks sooner while the TeV pulse has a smaller FWHM (see Table
\ref{lceffect}) compared to run 1. This happens because the R-band
synchrotron photons and the TeV Compton upscattered photons are now a
result of relatively higher energy electrons, in comparison to run 1.
In the case of run 4, the pulse peaks later and lasts longer at all
energies compared to that of run 1 (see Table \ref{lceffect}) due to
the reasons mentioned in (B) for the case of lower particle density in
the region.

A higher value of $\Gamma_i$ (run 6) and a lower value of $\Gamma_o$
(run 7) increase the relative velocity of the two shells along with
the value of their individual shock propagation Lorentz factors
($\Gamma^{\prime}_{\rm fs,rs}$), internal shock energies, magnetic
field, $\gamma^{\prime}_{\rm min}$, and density of higher energy
electrons injected into the region. The overall flux of the spectrum
increases (see Fig. \ref{5678sedlc}) due to the reasons cited in (A),
(B), and most importantly (C). In both cases, the synchrotron spectrum
extends into the X-rays: for run 6 $\nu^*_{\rm turn} \sim 6.86 \times
10^{17}$ Hz whereas for run 7 the transition from synchrotron to SSC
component occurs beyond 10 keV. As a result, the spectrum is softer in
the X-ray range and harder in the Fermi range (see Table
\ref{lceffect}) due to reasons given in (D). The magnetic field value
for these runs is high enough to affect the CDF value according to the
reason cited in (F).

The lightcurves in Fig. \ref{5678sedlc} have similar profiles as that
of run 1, with the same explanation as given in the lightcurve profile
discussion of run 1 in \S \ref{outcome}. The pulses of runs 6
($\Gamma_i \uparrow$) and 7 ($\Gamma_o \downarrow$) behave according
to the explanation given in (C) at all energies (see Table
\ref{lceffect}).

As shown in Figure \ref{910sedlc} and Table \ref{parameffect},
increasing the shell widths (run 9) decreases the flux and CDF of the
spectrum following the explanation given in (G). The values of the
peak frequencies of both radiation components decrease relative to run
1, thus increasing the SH in the X-ray range as explained in (D) but
for the case of a low value of $\gamma^{\prime}_{\rm min}$. On the
other hand, the location of the trough frequency does not change in
this case. As a result, the SH for the VHE regime increases as well,
as shown in Table \ref{parameffect}.

As can be seen from Figure \ref{910sedlc}, the lightcurves from run 9
($\Delta_i ~\&~ \Delta_o \uparrow$) are much more extended than those
of run 1. This is because run 9 corresponds to larger widths of the
colliding shells, thereby resulting in a wider emission region. As a
result, the time taken for the shocks to propagate through the region
is longer. Hence, pulses at all energies follow a profile as described
in (G). In the case of run 9, the forward shock is the last to break
out of the system, at $t^*_{\rm cr} \sim 5.27 \times 10^{5}$ s, while
the reverse shock exits the reverse emission region at $\sim 5.08
\times 10^5$ s.

\subsection{\label{emitsed}Emission Region Parameters}

\begin{figure}
\plottwo{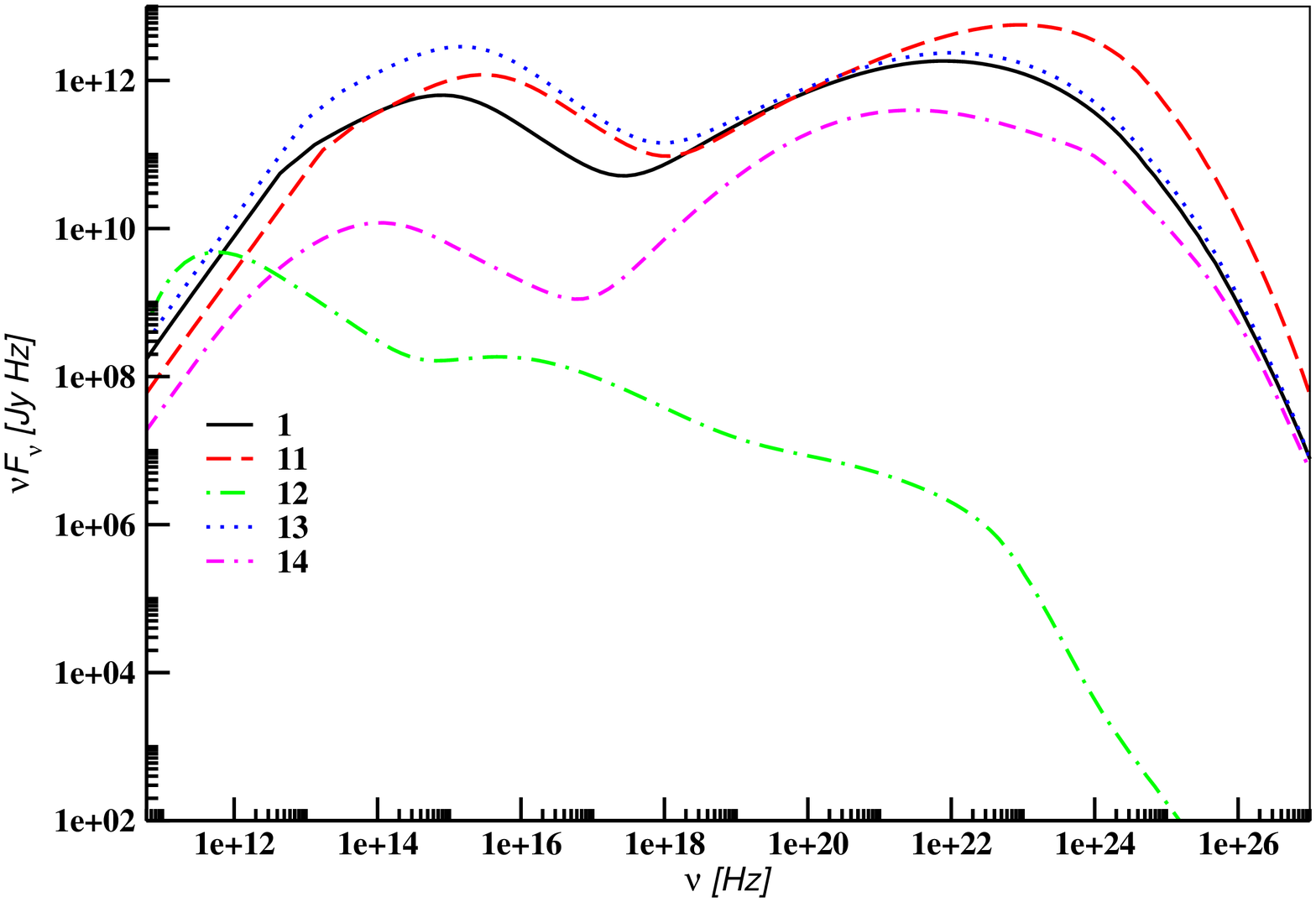}{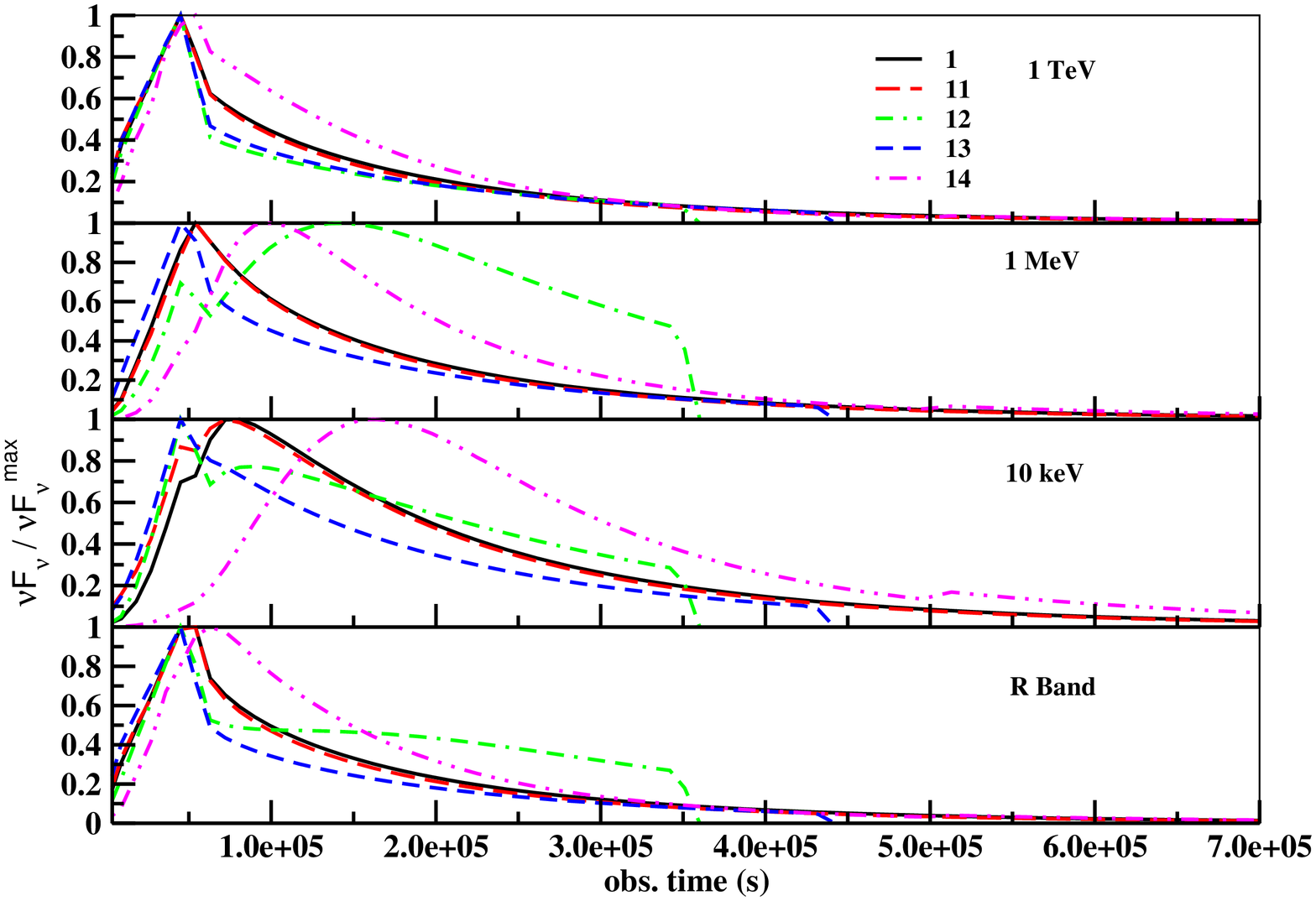}
\caption{Simulated time-integrated SED and lightcurves of a generic
  blazar source illustrating the effects of varying emission region
  parameters $\varepsilon^{\prime}_{\rm e}$ (runs 11 \& 12), and
  $\varepsilon^{\prime}_{\rm B}$ (runs 13 \& 14) according to Table
  \ref{paramlist}.}
\label{11121314sedlc}
\end{figure}

As can be seen from Fig. \ref{11121314sedlc}, increasing the value of
$\varepsilon^{\prime}_e$ (run 11) increases $\gamma^{\prime}_{\rm
  min}$ and the density of injected electrons in the region. The
resultant spectrum and its SH in the X-ray and Fermi ranges, along
with the value of the CDF (see Table \ref{parameffect}), behave
according to the patterns described in (B) and (D). In the case of
decreasing $\epsilon^{\prime}_e$ (run 12), the synchrotron component
peaks in the sub-mm range ($\nu^*_{\rm syn} \sim 6.21 \times 10^{11}$
Hz) with the transition from the synchrotron to the SSC component
taking place in the optical ($\nu^*_{\rm turn} \sim 6.50 \times
10^{14}$ Hz). The MeV photons are a result of higher order SSC
scattering. The spectrum becomes softer in both X-ray and Fermi range
(see Table \ref{parameffect}) because the entire spectrum shifts
toward low frequencies, due to an extremely low value of
$\gamma^{\prime}_{\rm min}$.

Increasing $\varepsilon^{\prime}_B$ (run 13) increases the magnetic
field while marginally increasing $\gamma^{\prime}_{\rm min}$ and the
normalization of the injected electron spectrum. The spectral flux,
its SH, and the corresponding CDF value undergo the same changes as
cited in (F), and (D).

As can be seen from Fig. \ref{11121314sedlc} and Table
\ref{parameffect}, the lightcurve profiles of run 11
($\varepsilon^{\prime}_e \uparrow$) follow a similar pattern to those
of run 1 at all energies. In the case of run 12
($\varepsilon^{\prime}_e \downarrow$), since R-band photons lie in the
highest energy part of the synchrotron component (see
Fig. \ref{11121314sedlc}), the high energy electrons responsible for
these photons last for a short time in the system. As a result, the
corresponding pulse peaks sooner with a smaller FWHM relative to run
1. On the other hand, the X-ray photons form the higher energy part of
the 1st order SSC component and are the result of scattering of high
energy synchrotron photons off low energy electrons. Thus, the
corresponding pulse peaks sooner relative to run 1 but has maximum
contribution from cooling, and therefore a larger FWHM. The MeV
photons form the lower energy part of the higher order SSC component
and result from the higher order SSC scattering of SSC photons off
lower energy electrons. Thus, the pulse at MeV energies peaks later
and lasts for a longer time than that of run 1. The TeV photons are a
result of higher order SSC scattering off high energy electrons. Since
such electrons last for a very short time in the system, the pulse has
a fairly equal contribution from both acceleration and cooling, making
it approximately symmetric about its maximum (see
Fig. \ref{11121314sedlc}). As can be seen from
Fig. \ref{11121314sedlc} and Table \ref{lceffect}, the pulses in run
13 ($\varepsilon^{\prime}_B \uparrow$) follow the same pattern as
explained in (F).

\begin{figure}
\plottwo{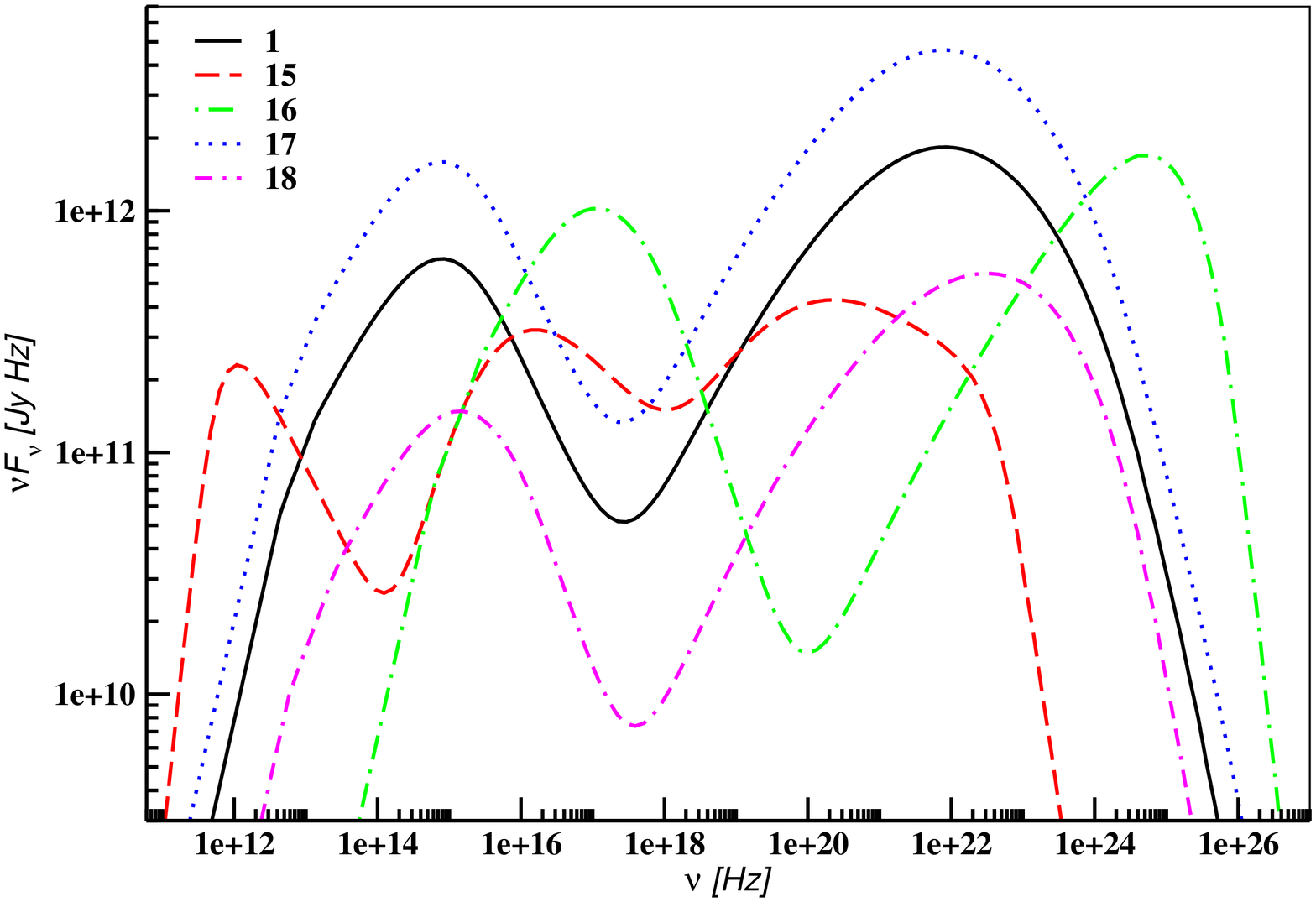}{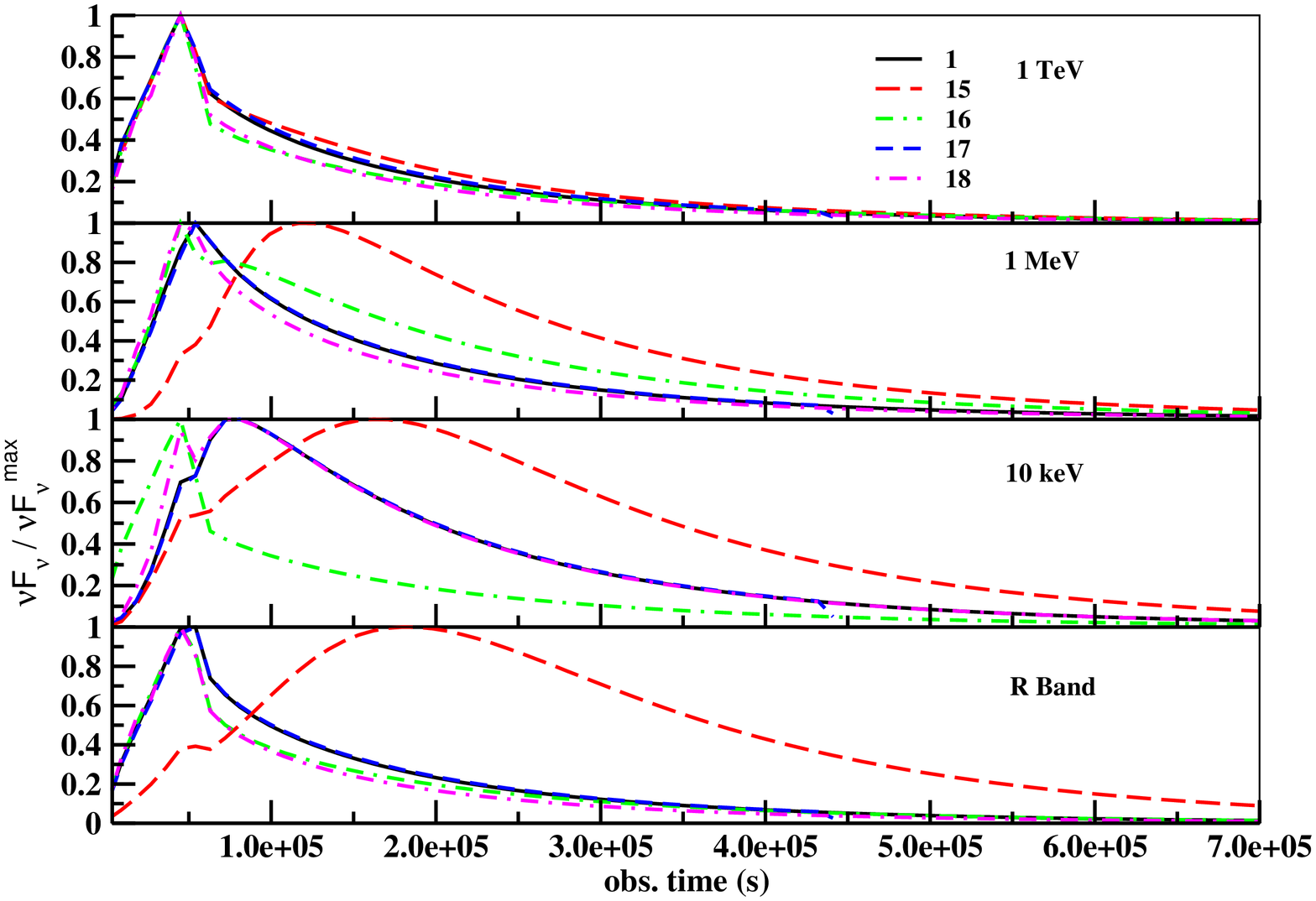}
\caption{Simulated time-integrated SED and lightcurves of a generic
  blazar source illustrating the effects of varying emission region
  parameters, $\zeta^{\prime}_{\rm e}$ (runs 15 \& 16), and
  $\alpha^{\prime}$ (runs 17 \& 18) according to Table
  \ref{paramlist}.}
\label{15161718sedlc}
\end{figure}

As shown in Figure \ref{15161718sedlc} and Table \ref{parameffect},
increasing $\zeta^{\prime}_e$ (run 15) increases the fraction of
accelerated electrons in the region while decreasing the value of
$\gamma^{\prime}_{\rm min}$ and the density of injected high-energy
electrons. The effects on the spectrum can be explained using reasons
given in (E), and (B) for the case of lower density of electrons. As
can be seen from the figure, the synchrotron component for this case
peaks in the IR instead of the optical, as in run 1. Thus, the optical
photons form the lower energy part of the SSC component and are
produced from the SSC scattering of much lower energy synchrotron
(radio) photons off the lowest energy electrons. Hence, the higher
order SSC component peaks at $\sim$ 1 MeV and is a result of
upscattering of first-order SSC photons off low energy electrons. On
the other hand, the TeV photons are the result of higher order SSC
scattering of SSC photons off the highest-energy electrons. The
spectrum in the X-ray range becomes softer (see Table
\ref{parameffect}) because, as can be seen from the figure, the
transition between the SSC and the higher order SSC components lies in
the 2-10 keV energy range. On the other hand, the spectrum in the
Fermi range becomes softer due to the shift in $\nu^*_{\rm SSC}$ to
lower frequency.

As can be seen from Fig. \ref{15161718sedlc} and Table
\ref{parameffect}, increasing the value of $\alpha^{\prime}$ (runs 17)
increases the value of $\gamma^{\prime}_{\rm max}$ while mildly
decreasing the value of $\gamma^{\prime}_{\rm min}$ and the
normalization of the injected electron spectrum. As a result, the SH
of the spectrum and the CDF value are hardly affected (see Table
\ref{parameffect}), but the overall flux of the spectrum increases due
to the reasons given in (H).

As can be seen from the right side of Figure \ref{15161718sedlc},
pulse profiles of all runs at all energies follow the same trend as
described in \S \ref{outcome} for run 1. In the case of run 15
($\zeta^{\prime}_e \uparrow$), since the particle density of the
region decreases the $t^*_{\rm peak}$ and FWHM of a pulse at a given
energy follow the same explanation as given in (B) for that case. As
can be seen from Fig. \ref{15161718sedlc} and Table \ref{lceffect},
the pulse profiles of run 17 ($\alpha^{\prime} \uparrow$) are highly
similar at all energies to that of run 1, as are their $t^*_{\rm
  peak}$ and FWHM values.

Increasing the value of $q^{\prime}$ (run 19) increases
$\gamma^{\prime}_{\rm min}$ slightly along with the normalization of
the injected electron spectrum, while keeping the magnetic field value
the same. Thus, the overall flux level of the SED rises along with the
value of the CDF, as explained in (B). Since a softer population of
electrons is injected into the region, the peak and trough frequencies
shift toward lower frequencies, even though the value of
$\gamma^{\prime}_{\rm min}$ increases slightly. As a result, the SH in
the X-ray and Fermi ranges are affected according to (D), but for the
case of a lower value of $\gamma^{\prime}_{\rm min}$.

\begin{figure}
\plottwo{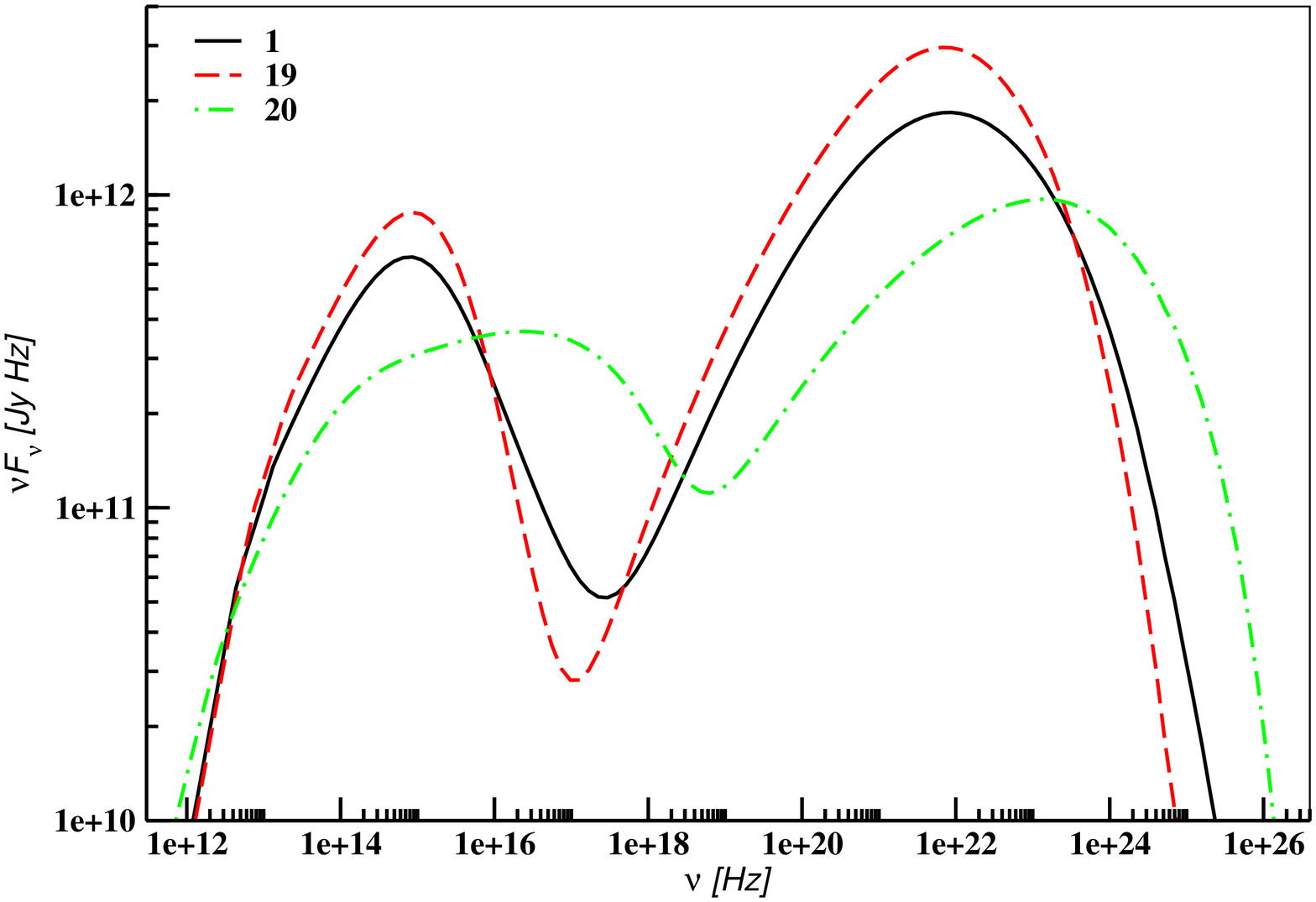}{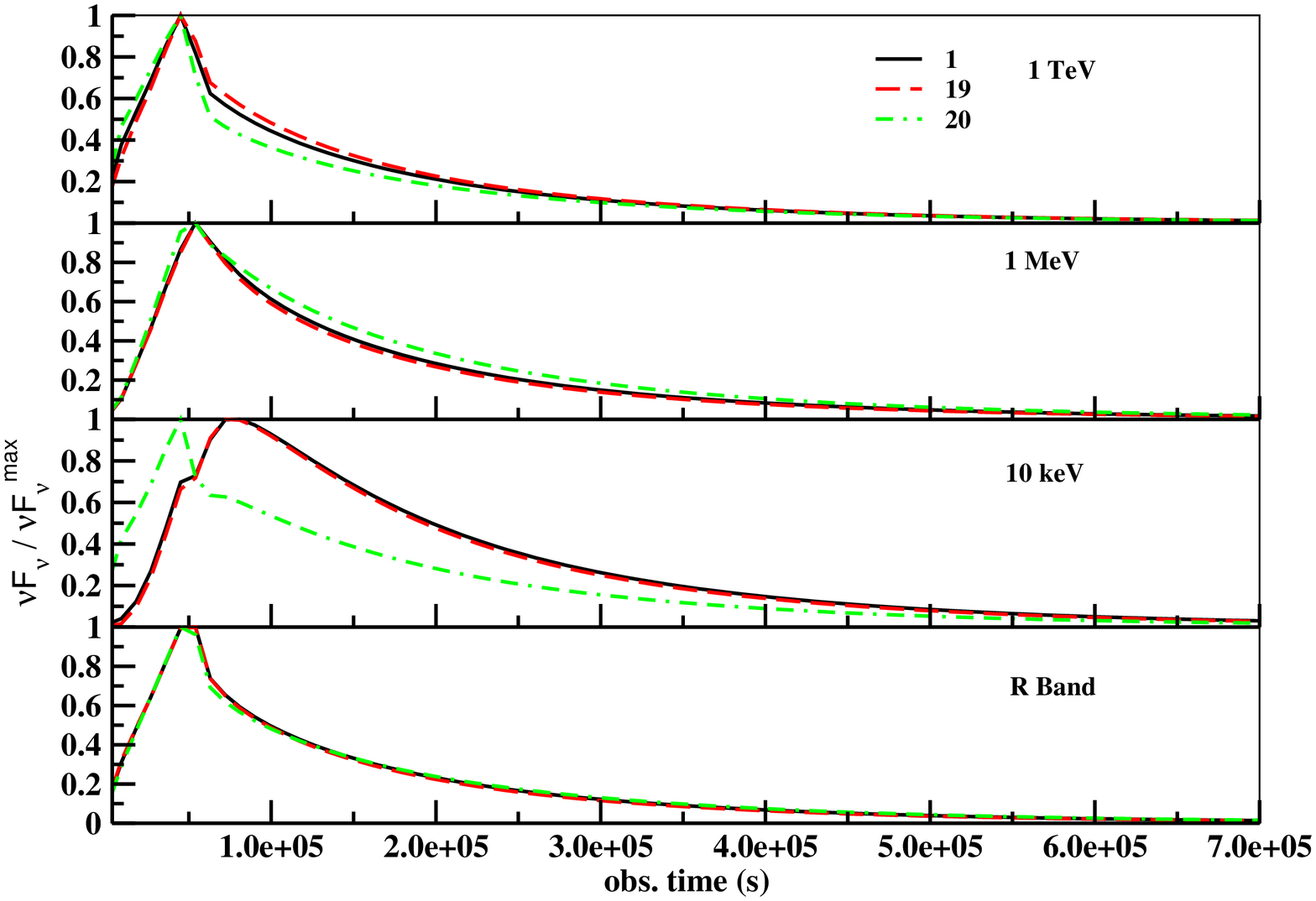}
\caption{Simulated time-integrated SED and lightcurves of a generic
  blazar source illustrating the effects of varying emission region
  parameter $q^{\prime}$ (runs 19 \& 20) according to Table
  \ref{paramlist}.}
\label{1920sedlc}
\end{figure}

The lightcurves from run 20 ($q^{\prime} \downarrow$) (Figure
\ref{1920sedlc}) show that a harder energy distribution of the
electron population results in a sharper rise and a steeper decline of
the pulse. Since the parent population of electrons is more energetic
than that of runs 1 (base set) \& 19 ($q^{\prime} \uparrow$), the
electrons lose their energy faster and more energetic synchrotron
photons (hard X-rays; Figure \ref{1920sedlc}) are produced, although
only for a short time. As a result, the rise in the corresponding
pulses at X-ray and optical energies is sooner and the FWHM shorter
compared to those of runs 1 \& 19. For the blazar source under
consideration, since TeV photons are a result of SSC scattering of
higher energy (soft X-rays) photons off higher energy electrons, they
are produced simultaneously with the lower energy (optical)
synchrotron photons but with the shortest FWHM (see Table
\ref{lceffect}). The TeV photons would lag behind the soft X-ray
synchrotron photons by a time of the order of the cooling time scale
for these synchrotron photons. The MeV pulse peaks the last and lasts
the longest relative to that of run 1 \& 19 because for this case,
they are a result of SSC scattering of lower energy (optical)
synchrotron photons off low energy electrons.

\subsection{\label{jetsed}Jet Parameters} 

\begin{figure}
\plotone{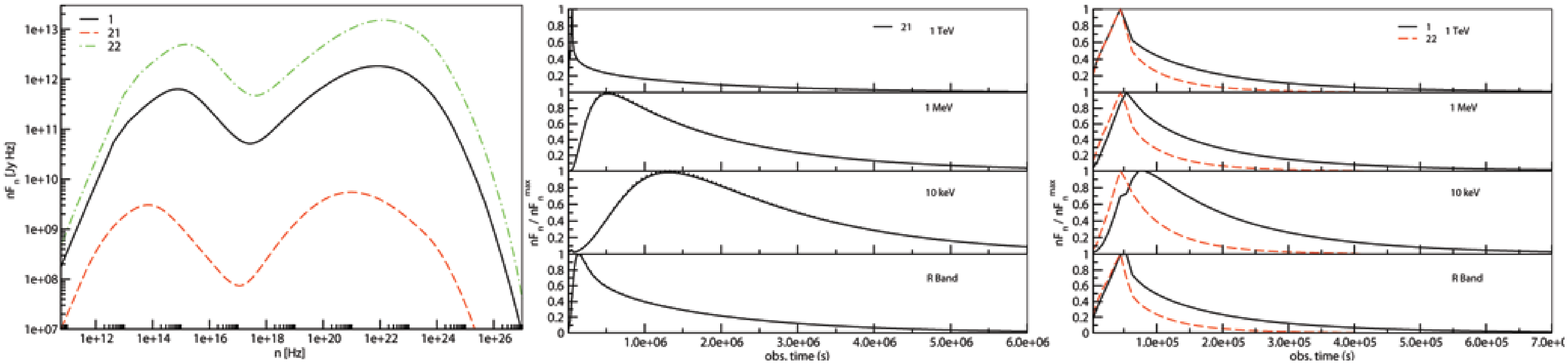}
\caption{Simulated time-integrated SED and lightcurves of a generic
  blazar source illustrating the effects of varying jet parameter
  $R^{\prime}_{\rm z}$ (runs 21 \& 22) according to Table
  \ref{paramlist}.}
\label{2122sedlc}
\end{figure}

As can be seen from Fig. \ref{2122sedlc}, increasing $R^{\prime}_z$
(run 21) implies a larger emission region. This decreases the overall
flux of the SED and the CDF value (see Table \ref{parameffect})
following the reasons given in (G). Since the value of
$\gamma^{\prime}_{\rm min}$ decreases slightly in this case, the peak
and trough frequencies, along with the SH, in both energy ranges are
affected in the opposite way to what is explained in (D).

For run 21 ($R^{\prime}_z \uparrow$), the pulse profiles at all
energies (except at TeV) are much more extended relative to run 1 (see
Table \ref{lceffect}) due to the reasons explained in (G). Since the
width of the emission region for both runs 21 \& 22 ($R^{\prime}_z
\downarrow$) remains the same, the shock crossing time does not
change. As a result, the TeV pulse lasts only for as long as the
shocks are present in the system. Thus, the FWHM of the pulse is
almost the same as the shock crossing time ($t^*_{\rm cr} \sim 0.61$
days, in the observer's frame) of the forward shock.

\begin{figure}
\epsscale{1.0}\plottwo{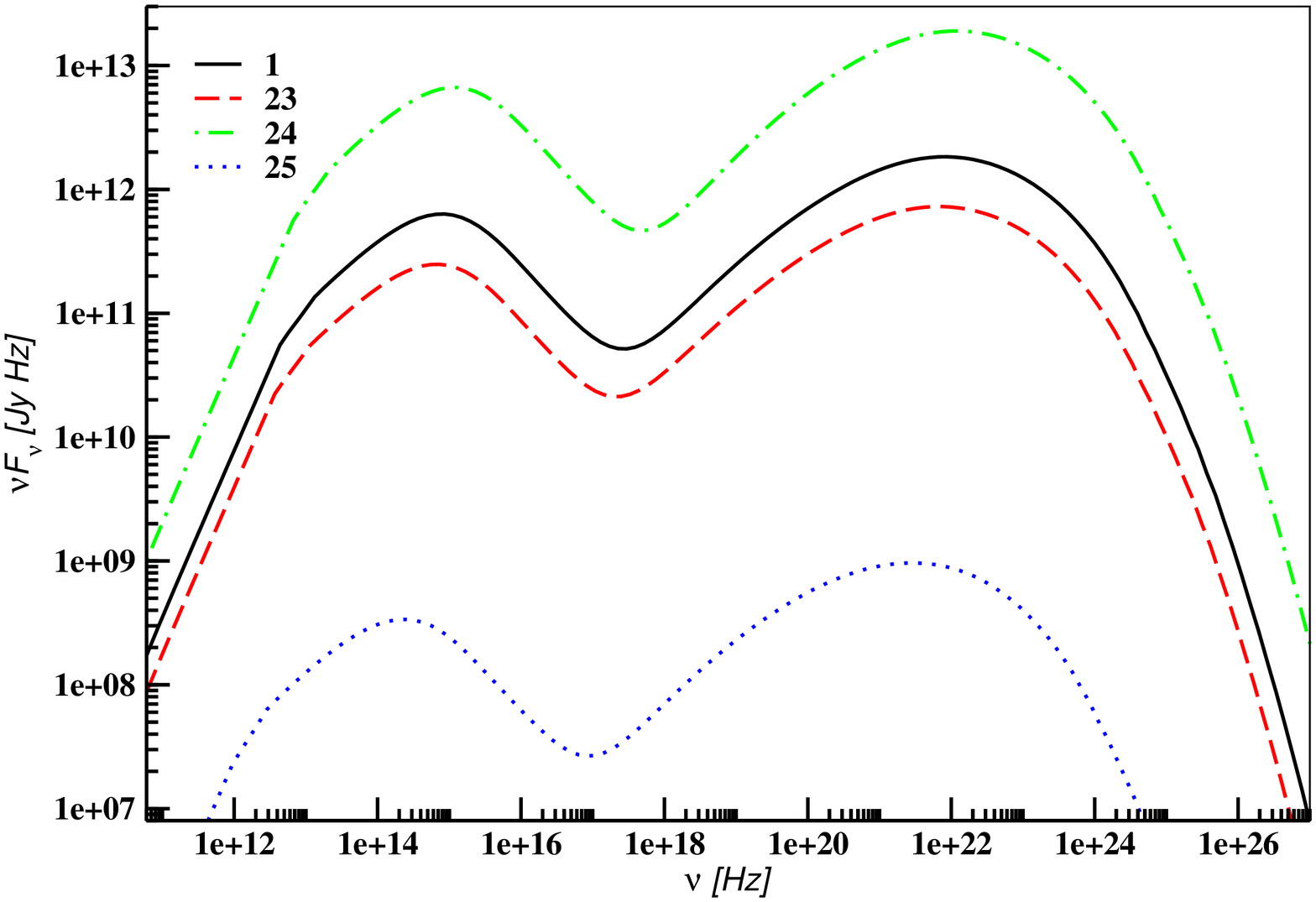}{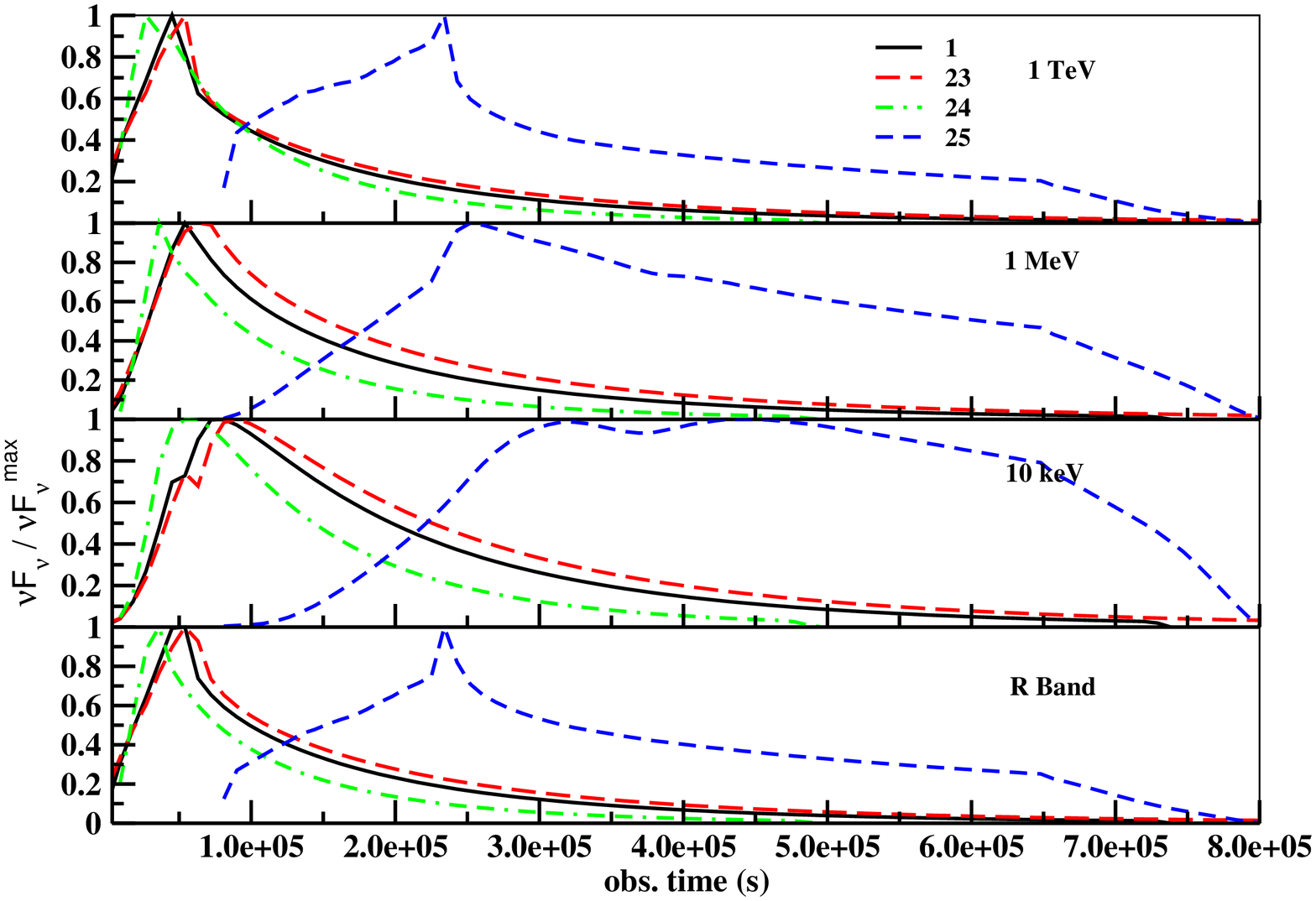}
\caption{Simulated time-integrated SED and lightcurves of a generic
  blazar source illustrating the effects of varying jet parameter
  $\theta^*_{\rm obs}$ (runs 23, 24 \& 25) according to Table
  \ref{paramlist}.}
\label{232425sedlc}
\end{figure}

As can be seen from Figure \ref{232425sedlc} and Table
\ref{parameffect}, decreasing the value of $\theta^*_{\rm obs}$ (run
24) increases the overall flux of the spectrum without changing the
values of magnetic field, the shocks' internal energies,
$\gamma^{\prime}_{\rm min}$, $\gamma^{\prime}_{\rm max}$, and the
particle density. Since the jet is now closely aligned with our line
of sight, the overall radiation is boosted more strongly in our
direction, thereby increasing the overall flux. Due to a lower value
of $\theta^*_{\rm obs}$, the frequencies when transformed into the
observer's frame assume higher values. As a result, even though the
synchrotron and SSC components peak at the same location in the
comoving frame, in the observer's frame the values of $\nu^*_{\rm
  syn}$, $\nu^*_{\rm turn}$, and $\nu^*_{\rm SSC}$ shift to higher
frequencies. This does not affect the CDF value, as can be seen from
Table \ref{parameffect} for all three runs 23, 24, and 25. However, in
the case of run 24, this shift of the spectrum towards higher
frequencies causes the synchrotron component in the observer's frame
to extend more into the X-rays, with the transition from synchrotron
to SSC component taking place at $\nu^*_{\rm turn} \sim 4.53 \times
10^{17}$ Hz. As a result, the spectrum in the 2-10 keV range appears
to be softer as the synchrotron component begins to contribute to the
soft X-ray flux. On the other hand, the SSC component in the
observer's frame peaks at $\nu^*_{\rm SSC} \sim 1.34 \times 10^{22}$
Hz, with the entire hump extending further into the Fermi range. This
makes the spectrum in the 10 GeV range appear harder, as can be seen
from the value of $\alpha^*_{\rm 10GeV}$ in Table \ref{parameffect}.

We consider two cases of a larger value of $\theta^*_{\rm obs}$. The
first one is run 23, where the value of $\theta^*_{\rm obs}$ increases
such that $\cos \theta^*_{\rm obs_{23}} > \beta_{\rm sh}$ and the line
of sight is still within the cone of maximum superluminal motion. The
second one is run 25 where the value of $\theta^*_{\rm obs}$ is high
enough to put the line of sight outside the beaming cone. In the
comoving frame, this would translate into the observer looking at the
emission region from behind, as a result of which the backside of the
region would become visible to the observer before the front side
does. As shown in Fig. \ref{232425sedlc} and Table \ref{parameffect},
for both runs 23 \& 25 the opposite trend, compared to run 24, is
observed. Since the value of $\theta_{\rm obs}$ in run 25 is higher
than that of run 23, the effect on the overall flux, the frequency
shift, and the SH of the spectrum is even more dramatic compared to
that of run 23.

The lightcurves of runs 23 ($\theta^*_{\rm obs} \uparrow$), 24
($\theta^*_{\rm obs} \downarrow$), and 25 ($\theta^*_{\rm obs}
\uparrow$, outside beaming cone) are all shifted in time with respect
to those of run 1 (see Fig. \ref{232425sedlc}). In the case of run 24,
since the value of $\theta^*_{\rm obs}$ is smaller, the required time
for photons at all energies to reach from the far side to the near
side, of the jet emission region, is reduced. As a result, the pulses
at all energies peak sooner and last for a shorter duration of time
compared to those for run 1 (see Table \ref{lceffect}). Due to a
smaller $\theta^*_{\rm obs}$, the shock crossing time of the forward
shock for run 24, in the observer's frame, turns out to be shorter
than that for run 1. As a result, the kink that is present in the
lightcurve of 10 keV photons for run 1, indicating the time of exit of
the shock from the system, is smeared out to give a smoother pulse
profile at that energy. The exact opposite effect can be seen in the
pulse profiles at all energies of run 23. In the case of run 25, since
the line of sight, in the comoving frame of the emission region,
passes through the region from behind, the light travel time increases
by a significant amount. This stretches the entire pulse profiles at
all energies compared to runs 1, 23, \& 24. Once again, the value of
$t^*_{\rm peak}$ for R-band and TeV photons almost marks the time when
the shock breaks out of the system for all the three runs.

\section{\label{conc}Summary And Conclusions}

We have developed a 1D multi-slice internal shock model with radiation
feedback scheme to simulate the radiative signatures of blazars in a
time-dependent manner. We consider a cylindrical geometry for the
emission region and semi-analytically compute the volume and
angle-averaged photon escape time scale for it. The inclusion of
radiation feedback through a multi-slice scheme makes our model a
powerful tool to address the issue of inhomogeneity in the photon and
particle density throughout the emission region. Under this scheme,
the cylindrical emission region is divided into multiple slices and
the subsequent time-dependent radiation transfer, within each slice
and in between slices, is calculated by using the appropriate photon
escape rates of the slices. Our scheme inherently addresses the
non-locality and time-delay effects due to internal light-travel time
while calculating the SSC emission of a blazar source. By increasing
the spatial resolution along the length of the emitting region and
considering the contribution of all slices in the emission region, our
model mimicks an inhomogenous source where the observer sees photons
from a mixed population of electrons of various ages. Such a scenario
is important for cases where the electron population evolves on time
scales shorter than the dynamical time. Thus our approach facilitates
the simulation of high-energy variability observed in blazars,
especially the SSC dominated TeV sources, more accurately. Our model
includes synchrotron and SSC emission for evaluating the
self-consistent time-dependent radiation transfer along the length of
the emission region. We account for the light travel time delays, in
our model, to evaluate the radiative output in the observer's frame.

We consider a single inelastic collision between two shells of plasma
of different mass, velocity, and energy. The collision results in the
formation of a forward and a reverse shock, internal to the jet. The
relativistic shocks are separated by a contact discontinuity (CD). We
use the relativistic hydrodynamic Rankine-Hugoniot jump conditions to
obtain the characteristics of the shock dynamics. The bulk Lorentz
factor (BLF) of the emission region in the lab frame is evaluated
using the equality of pressures of the shocked fluids across the CD.

The emission region parameters, such as $\gamma_{\rm max}$,
$\gamma_{\rm min}$, the normalization of the injected electron
spectrum, and the magnetic field, \textit{B}, are obtained using the
BLFs of the shocks and the width of the emission region in the
comoving frame of the shocked fluid. The emission region parameters
are then used to evaluate the radiation spectrum resulting from the
collision for a cylindrical emission region. A numerical
simplification of the synchrotron emission formula has been used in
order to speed up the computing time. Synchrotron spectra calculated
with this numerical simplification match those calculated with the
exact expression to a relative accuracy of better than 1 $\%$.

The code has been tested to satisfy various analytical constraining
conditions. We have carried out 25 simulations to perform an extensive
parameter study of the code, in order to understand the effects of
varying various shock (wind luminosity, mass of the outer shell, BLF
and width of the two shells), emission-region (fraction of shock
energy stored in the electron and magnetic field energy density,
fraction of effectively accelerated electrons, acceleration timescale
parameter, and particle energy index), and jet parameters (radius of
the slice/jet, and observing angle inside and outside of the beaming
cone) on the resulting SED and spectral lightcurves of a generic
blazar source. The SED has been parametrized on the basis of the
change in the level of its overall flux, shift in its peak and trough
frequencies, change in its spectral hardness in the 2 - 10 keV and 10
GeV energy ranges, and the Compton dominance of the SSC component over
the synchrotron component. The lightcurves at optical, X-ray, MeV, and
TeV energy ranges have been parametrized on the basis of the time of
exit of the shock from the emission region, the time of the peak, and
the FWHM of a pulse at a given energy.

As demonstrated in \S \ref{study}, both SEDs and lightcurves give us
an idea about the dominant radiation mechanism responsible for
producing radiation of a particular frequency and the corresponding
time lags between various frequency bands. As shown in \S \ref{study},
our model can be exploited to reproduce long and short-term
variability of blazars based on the combination of shock,
emission-region, and jet parameters. The symmetric and asymmetric
profiles observed in the lightcurves of blazars are frequency
dependent. Lightcurves at higher synchrotron and SSC frequencies tend
to be more symmetric than their lower frequency counterparts
\citep[see e.g.,][]{cg1999, smm2004} for the reasons explained in \S
\ref{study}. As indicated in \S \ref{study}, such profiles of the
lightcurves can be reproduced using the right combination of shock,
emission-region, and jet parameters. The symmetric and asymmetric
profiles of the lightcurves are also dependent on the angle at which
the jet is being viewed (see Fig. \ref{232425sedlc}). As mentioned
below, since we are not carrying out a 2-D simulation of the jet
radiation processes, we are limited by the internal light crossing
time in the radial direction. As a result, the symmetric profiles that
appear from viewing the jet at non-zero angles \citep{smm2004} and are
dominated by light crossing time effect across the emission region are
not accounted for.

The particle injection spectra can be used to identify the
characteristic signatures of various particle acceleration
mechanisms. The particle acceleration at parallel shocks can give rise
to electron spectra following a power-law distribution with an
injection index of $q^{\prime} \sim$ 2.2 - 2.3 \citep[]{ac2001,
  ga1999}. Oblique shocks, on the other hand, are believed to produce
much softer injection spectral indices with $q^{\prime} > 2.3$
\citep[]{ob2002, no2004}. In contrast, 2nd order Fermi acceleration
might result in a harder injection index of the order of $q^{\prime}
\sim 1$ or beyond if the plasma is turbulent, behind the shock
front. Thus, the value of $q^{\prime}$ used for modeling the SED and
spectral lightcurves of a blazar source can give us insights on the
kind of shock propagating through its jet and the dominant mode of
acceleration present in the source. Our future modeling efforts will
help us gain a deeper understanding of the real data.

We would like to point out that for simplicity we have considered the
role of the magnetic field to be negligible in the dynamics of the
shock energetics. The structure of the mean large scale magnetic field
plays an important part in particle acceleration and shock
dynamics. We also emphasize that for making the problem at hand more
tractable we chose to have the particle injection spectral index, q,
as a free parameter. The value of q should really be the result of
shock dynamics and the underlying magnetic field structure. In our
model, the escape timescale for electrons from the emitting region is
a weakly constrained, free parameter. It involves a fudge factor that
parametrizes the randomness of the magnetic field direction. Hence a
more detailed model of the radiation transfer in blazar jets, dealing
with particle accleration, must address these physical parameters.

We note that our model, in its current form, does not address the
internal light travel time in the radial direction. The time
resolution of our model is limited by the light crossing time across
the cylinder. This feature is more relevant for carrying out a
comparison between synchrotron and SSC peak times at highest
frequencies, especially for the case where the jet is being viewed
face-on \citep{smm2004}. A 2-D multi-celled approach would be a more
appropriate tool to incorporate this aspect of frequency-dependent
time lags.

\acknowledgements 

We thank Prof. Alan Marscher for comments and discussions. We
acknowledge Dr. Justin Finke for useful suggestions. We thank the
anonymous referee for his/her helpful comments. The work of M.B. is
supported by NASA through Fermi Guest Investigator Grant NNX09AT82G
and Astrophysics Theory Program Grant NNX10AC79G. The work of M.J. is
partially supported through NRL BAA 76-03-01, contract no.
N00173-05-P-2004 (Ohio University) and National Science Foundation
grant AST-0907893 from Boston University.

\appendix
\section{\label{coeffeqns}Coefficients of $\Gamma_{\rm sh}$}

The coefficients of Eqn. \ref{gamsheqn} are as follows:

\begin{eqnarray}
\label{gamcoeffeqn}
a = \hat{\gamma}^2\left[\left(Y-1\right)^2 +
  4Y\left(n-2x^{2}m\right)\right],
\nonumber\\
b = 2\hat{\gamma}g\left[\left(Y+1\right)\left(\Gamma_{\rm o} +
  Y\Gamma_{\rm i}\right) - 2Yxpm\right],
\nonumber\\
c = 2\hat{\gamma}Y\left(2-3\hat{\gamma}\right)n +
2\hat{\gamma}\left(\hat{\gamma}-2\right)r - 2Yxm\left(g^2 -
4\hat{\gamma}^2 x\right) - 4Y\hat{\gamma}g +
\left(1-\hat{\gamma}^2\right)\left(1+Y^2\right),
\nonumber\\
d = 2g\left[Yp\left(1+\hat{\gamma}\left(x\left(1-2q\right) -
  1\right)\right) - \left(1+\hat{\gamma}\right)\left(\Gamma_{\rm o} +
  Y^{2}\Gamma_{rm i}\right) + \hat{\gamma}\left(\Gamma^{3}_{\rm o} +
  Y^{2}\Gamma^{3}_{\rm i}\right)\right],
\nonumber\\
e = \hat{\gamma}^{2}\left(\Gamma^{4}_{\rm o} + Y^{2}\Gamma^{4}_{\rm
  i}\right) + 2\hat{\gamma}Y\left(2 - gn - \hat{\gamma}\left(1 +
x^2\right)\right) + r\left(1 - \hat{\gamma}^2\right) - 2g^{2}Yxq -
2Y~.
\end{eqnarray}

Here, $x = \Gamma_{\rm o}\Gamma_{\rm i}$, $m = 1 - \beta_{\rm
  o}\beta_{\rm i}$, $q = \beta_{\rm o}\beta_{\rm i}$, $n =
\Gamma^{2}_{\rm o} + \Gamma^{2}_{\rm i}$, $p = \Gamma_{\rm o} +
\Gamma_{\rm i}$, $r = \Gamma^{2}_{\rm o} + Y^{2}\Gamma^{2}_{\rm i}$,
$g = 1 - \hat{\gamma}$, and $Y = \overline{\rho}_{\rm i} /
\overline{\rho}_{\rm o}$. All unprimed quantities refer to the AGN
frame and quantities with an overline refer to the unshocked fluid
frame.

\section{\label{photesctime} Photon Escape Time Calculation}

\begin{figure}[!ht]
\plotone{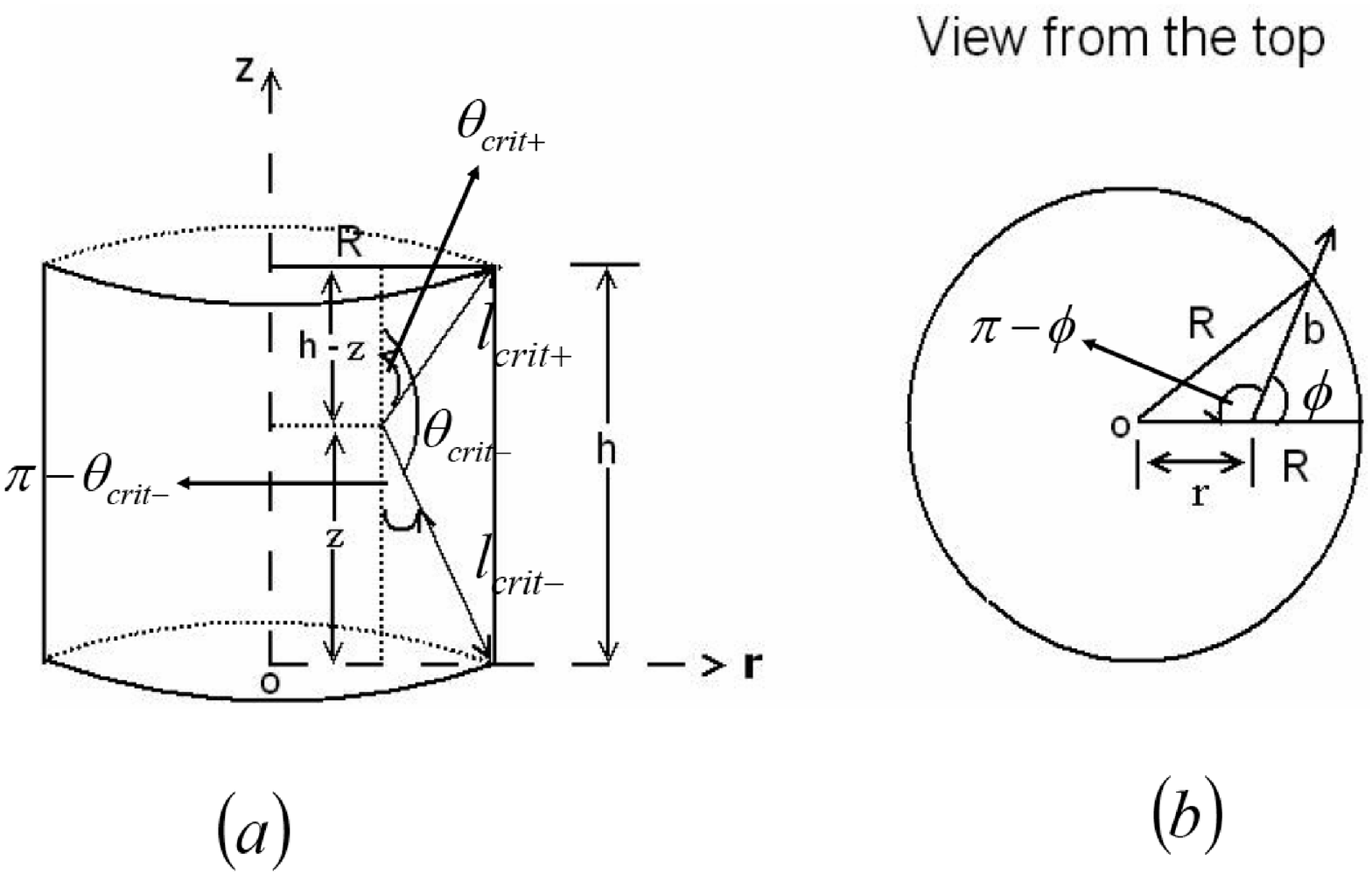}
\caption{In (a) we illustrate the three angles of escape along with
their corresponding escape lengths. In (b) we show the angle
consideration and the projection of the escape lengths on the
horizontal plane of the cylinder, when viewed from the top.}
\label{angle_depict}
\end{figure}

All quantities mentioned here refer to the comoving frame of the
emission region. The intermediate steps used to derive equation
(\ref{esctimeeqn}) are the following:

As shown in Figure \ref{angle_depict} (a), $l_{\rm crit+}$ is the
distance traveled by a photon that escaped from the edge of the region
in the forward direction, at an angle $\theta_{\rm crit+}$ with the
z-axis, and similarly for $l_{\rm crit-}$ at an angle $\pi -
\theta_{\rm crit-}$ with the z-axis; \textit{b} is the corresponding
projection of these lengths on the horizontal plane (Figure
\ref{angle_depict} (b)).

Using the law of cosines and solving the quadratic equation in
\textit{b}, and realizing that negative distances are unphysical, an
expression for \textit{b} can be obtained as

\begin{equation}
\label{beqn}
b = \sqrt{R^{2} - r^{2} \sin^{2} \phi} - r \cos \phi~.
\end{equation}
Now, referring to Figure \ref{angle_depict}, we can write

\begin{eqnarray}
\label{leqns}
\tan \theta_{\rm crit+} = \frac{b}{h - z},
\tan \theta_{\rm crit-} = \frac{-b}{z}, \nonumber\\
l_{\rm esc_{-}} = \frac{-z}{\mu_{-}} & \textrm{for
$-1 < \mu < \mu_{\rm crit-}$}, \nonumber\\
l_{\rm esc} = \frac{b}{\sqrt{1 - \mu^{2}}} & \textrm{for
$\mu_{crit-} < \mu < \mu_{\rm crit+}$}, \nonumber\\
l_{\rm esc_{+}} = \frac{h - z}{\mu_{+}} & \textrm{for
$\mu_{\rm crit+} < \mu < +1$}, \nonumber\\
\mu_{\rm crit+} = \frac{h-z}{\sqrt{b^{2} + \left(h-z\right)^{2}}},
\mu_{\rm crit-} = \frac{-z}{\sqrt{b^{2} + z^{2}}}~.
\end{eqnarray}

Substituting equations (\ref{leqns}) in equation (\ref{escteqn}) and
then equation (\ref{escteqn}) in (\ref{escleqn}), equation
(\ref{escleqn}) can be written as

\begin{eqnarray}
\label{1stteqn}
\langle{t_{\rm ph,esc, V}}\rangle = (\frac{1}{\pi R^{2} h})(\frac{1}{4 \pi c})
\int\limits_{0}^{2 \pi} \int\limits_{0}^{R} \int\limits_{0}^{h} \nonumber\\
\Big( \int\limits_{0}^{2 \pi } d{\Phi}
\left[\int\limits_{-1}^{\mu_{\rm crit-}} l_{\rm esc_{-}} d{\mu_{-}} +
\int\limits_{\mu_{\rm crit-}}^{\mu_{\rm crit+}} l_{\rm esc} d{\mu}
\int\limits_{\mu{\rm crit+}}^{+1} l_{\rm esc_{+}} d{\mu_{+}}\right] \Big)
d{\phi}\,rdr\,dz~,
\end{eqnarray}
which gives

\begin{eqnarray}
\label{2ndteqn}
\langle{t_{\rm ph,esc, V}}\rangle = (\frac{1}{\pi R^{2} h})(\frac{1}{4 \pi c})
\int\limits_{0}^{2 \pi} \int\limits_{0}^{R} \int\limits_{0}^{h} \nonumber\\
\Big( \int\limits_{0}^{2 \pi } d{\Phi}
\left[\int\limits_{-1}^{\mu_{\rm crit-}} \frac{-z}{\mu_{-}} d{\mu_{-}} +
\int\limits_{\mu_{\rm crit-}}^{\mu_{\rm crit+}} \frac{b}{\sqrt{1 - \mu^2}}
d \mu + \int\limits_{\mu{\rm crit+}}^{+1} \frac{h - z}{\mu_+} d \mu_+\right]
\Big) d{\phi}\,rdr\,dz~.
\end{eqnarray}

Carrying out integration over $\mu$, and realizing that the
integration of the right hand side expression over the angle $\Phi$
is going to give 2 $\pi$, the above equation can be written as

\begin{eqnarray}
\label{3rdteqn}
\langle{t_{\rm ph,esc, V}}\rangle = \frac{2}{R^2 h} \int\limits_{0}^{R}
\Big( \int\limits_{0}^{h} (\frac{1}{2c}[(z - h) \ln (h - z) - z \ln z] +
I_1 + I_2 + I_3) dz \Big) rdr~,
\end{eqnarray}
where

\begin{eqnarray}
\label{4thteqn}
I_1 = \frac{h - z}{8 \pi c} \int\limits_{0}^{2 \pi} \ln (b^2 + (h - z)^2)
d \phi, \nonumber\\
I_2 = \frac{z}{8 \pi c} \int\limits_{0}^{2 \pi} \ln (b^2 + z^2) d \phi,
\nonumber\\
I_3 = \frac{1}{4 \pi c} \int\limits_{0}^{2 \pi} b
[\arcsin(\frac{h - z}{\sqrt{b^2 + (h - z)^2}}) +
\arcsin(\frac{z}{\sqrt{b^2 + z^2}})] d\phi~.
\end{eqnarray}

The above three integrals can be solved semi-analytically to obtain
the final expression for the volume and angle-averaged photon escape
timescale for a cylindrical region given in Eqn. (\ref{esctimeeqn}).

\section{\label{synfit} Synchrotron Function Fitting}

All quantities mentioned here refer to the comoving frame of the
emission region. The function R(x), used in the calculation of the
synchrotron photon production density rate (see \S \ref{method},
Eqn. \ref{syndeneqn}), is given by \citep{cs1986}

\begin{equation}
\label{rxeqn}
R(x) = \frac{\pi x}{2} [W_{0, \frac{4}{3}}(x) W_{0, \frac{1}{3}}(x) -
W_{\frac{1}{2}, \frac{5}{6}}(x) W_{\frac{-1}{2}, \frac{5}{6}}(x)]~,
\end{equation}
where $x = \frac{4 \pi m_e c \nu}{3 e B \gamma^{2}}$ is the normalized
frequency, and $W_{\lambda, \mu}$(x) denotes Whittaker's function
\citep{as1970}.

In order to save CPU time while computing equation (\ref{syndeneqn})
(\S \ref{method}), we numerically approximate the function, R(x), as
follows:

\begin{equation}
\label{rxapprox}
R(x) = C_1 x^{p_1} e^{-x} - C_2 x^{-p_2} e^{-x}~,
\end{equation}
where $C_1 = 1.08895$, $C_2 = 2.35861 \times 10^{-3}$, $p_1 =
0.20949$, and $p_2 = 0.79051$. 

The asymptotic expansions of Whittaker functions \citep{cs1986} that
are used in the calculation of Eqn. \ref{syndeneqn}, for small ($x \ll
1$) and large ($x \gg 1$) values of x, are no longer required when
using the above approximation. Since R(x) has been approximated for
the expression given in Eqn. (\ref{rxeqn}), we normalize R(x) at lower
values of x, where R(x) $\propto x^{1/3}$. This is carried out to
obtain a good fit to the exact expression for the entire range of x
values.

\begin{figure}[!ht]
\plotone{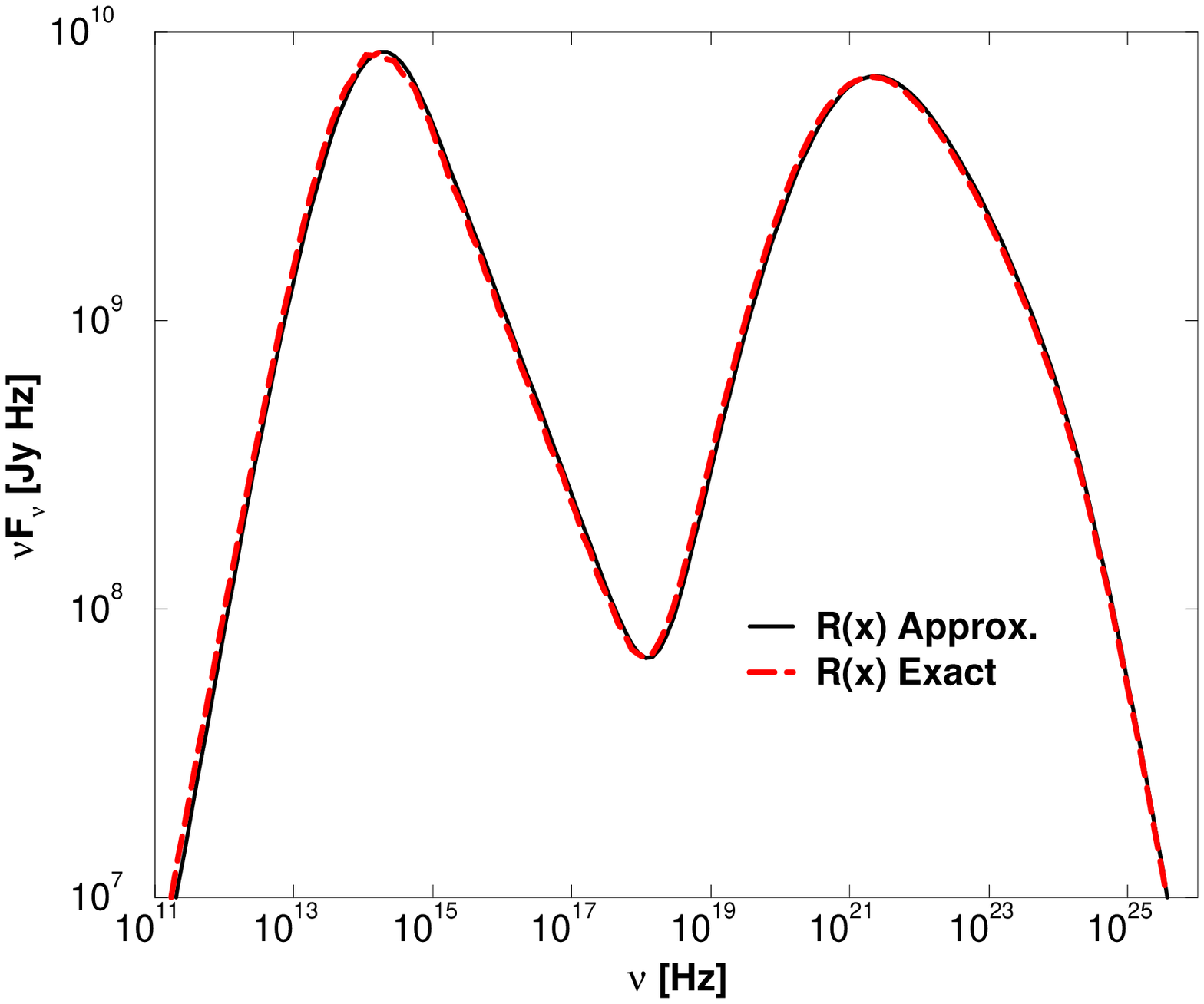}
\caption{Comparison of the exact expression and numerical
approximation of R(x). The fit is accurate to $\sim 0.5\%$ for the
entire frequency range.}
\label{rxfit}
\end{figure}

Figure \ref{rxfit} shows the comparison of simulated SEDs of a generic
blazar source, obtained from using the approximation and the exact
expression. As shown in the figure, the approximation is accurate to
better than $1\%$ over the entire frequency range.

\begin{thebibliography}{}

\bibitem[Achterberg et al.(2001)]{ac2001}
Achterberg, A., Gallant, Y. A., Kirk, J. G., \& Guthmann, A. W., 2001,
MNRAS, 328, 393

\bibitem[Abramowitz \& Stegun(1970)]{as1970}
Abramowitz, M., \& Stegun, I. A., 1970, Handbook of mathematical functions: 
with formulas, graphs, and mathematical tables, National Bureau of Standards

\bibitem[Aharonian et al.(2007)]{an2007}
Aharonian et al., 2007, ApJL, 664, 71

\bibitem[Albert et al.(2007)]{at2007}
Albert et al., 2007, ApJ, 669, 862

\bibitem[Blandford \& McKee(1976)]{bm1976}
Blandford, R. D., \& McKee, C. F., 1976, Phys. Fluids, 19, 1130 

\bibitem[Bloom \& Marscher(1996)]{bm1996}
Bloom, S. D., \& Marscher, A. P., 1996, ApJ, 461, 657

\bibitem[B\"{o}ttcher \& Bloom(2000)]{bb2000}
B\"{o}ttcher, M., \& Bloom, S. D., 2000, AJ, 119, 469

\bibitem[B\"{o}ttcher \& Chiang(2002)]{bc2002}
B\"{o}ttcher, M., \& Chiang, J., 2002, ApJ, 581, 127

\bibitem[B\"{o}ttcher \& Dermer(2010)]{bd2010}
B\"{o}ttcher, M., \& Dermer, C. D., 2010, ApJ, 711, 445

\bibitem[B\"{o}ttcher, Mause \& Schlickeiser(1997)]{bms1997}
B\"{o}ttcher, M., Mause, H., \& Schlickeiser, R., 1997, A\&A, 324, 395

\bibitem[B\"{o}ttcher \& Schlickeiser(1997)]{bs1997}
B\"{o}ttcher, M., \& Schlickeiser, R., 1997, A\&A, 325, 866

\bibitem[Catanese \& Weekes(1999)]{cw1999}
Catanese, M., \& Weekes, T. C., 1999, PASP, 111, 1193

\bibitem[Chiaberge \& Ghisellini(1999)]{cg1999}
Chiaberge, M., \& Ghisellini, G., 1999, MNRAS, 306, 551

\bibitem[Chiang \& Dermer(1999)]{cd1999}
Chiang, J., \& Dermer, C. D., 1999, ApJ, 512, 699

\bibitem[Costamante \& Ghisellini(2002)]{cg2002}
Costamante, L., \& Ghisellini, G., 2002, A\&A, 384, 56

\bibitem[Crusius \& Schlickeiser(1986)]{cs1986}
Crusius, A., \& Schlickeiser, R., 1986, A\&A, 164, 16

\bibitem[deJager \& Harding(1992)]{dh1992}
deJager, O. C., \& Harding, A. K., 1992, ApJ, 396, 161

\bibitem[Dermer \& Schlickeiser(1993)]{ds1993}
Dermer, C. D., \& Schlickeiser, R., 1993, ApJ, 416, 458 

\bibitem[Gaidos et al.(1996)]{ga1996}
Gaidos, J. A., et al., 1996, Nature, 383, 319

\bibitem[Gallant et al.(1999)]{ga1999} 
Gallant, Y. A., et al., 1999, A\&AS, 138, 549

\bibitem[Georganopoulos \& Marscher(1998)]{gm1998} 
Georganopoulos, M., \& Marscher, A. P., 1998, ApJ, 506, 621

\bibitem[Gradshteyn \& Ryzhik(1994)]{gr1994} 
Gradshteyn, I. S., \& Ryzhik, I. M., 1994, Table of integrals, series and 
products, New York: Academic Press

\bibitem[Graff et al.(2008)]{gr2008}
Graff, P. B., et al., 2008, ApJ, 689, 68

\bibitem[Jones(1968)]{jf1968}
Jones, F. C., 1968, Phys. Rev., 167, 1159

\bibitem[Jorstad et al.(2005)]{jo2005}
Jorstad, S. G., et al., 2005, AJ, 130, 1418

\bibitem[Joshi \& B\"{o}ttcher(2007)]{jb2007}
Joshi, M., \& B\"{o}ttcher, M., 2007, ApJ, 662, 884

\bibitem[Kirk et al.(1998)]{krm1998}
Kirk, J. G., Rieger, F. M., \& Mastichiadis, A., 1998, A\&A, 333, 452

\bibitem[Kobayashi et al.(1997)]{kps1997}
Kobayashi, M., Piran, T., \& Sari, R., 1997, ApJ, 490, 92

\bibitem[Longair(1994)]{lm1994}
Longair, M. S., 1994, High energy astrophysics. Vol.2: Stars, the
galaxy and the interstellar medium, Cambridge: Cambridge University
Press

\bibitem[Marscher \& Gear(1985)]{mg1985}
Marscher, A. P., \& Gear, W. K., 1985, ApJ, 298, 114

\bibitem[Maraschi, Ghisellini \& Celotti(1992)]{mgc1992}
Maraschi, L., Ghisellini, G., \& Celotti, A., 1992, ApJ, 397, L5 

\bibitem[Mimica et al.(2004)]{mi2004}
Mimica, P., et al., 2004, A\&A, 418, 947

\bibitem[Niemiec \& Ostrowski(2004)]{no2004}
Niemiec, J., \& Ostrowski, M., 2004, ApJ, 610, 851

\bibitem[Ostrowski \& Bednarz(2002)]{ob2002}
Ostrowski, M., \& Bednarz, J., 2002, A\&A, 394, 1141

\bibitem[Panaitescu \& M\'{e}sz\'{a}ros(1999)]{pm1999}
Panaitescu, A., \& M\'{e}sz\'{a}ros, P., 1999, ApJ, 526, 707

\bibitem[Press et al.(1992)]{pr1992}
Press W. H., et al., 1992, Numerical Recipes in C, 
Cambridge: Cambridge University Press

\bibitem[Rieger \& Duffy(2004)]{rd2004}
Rieger, F. M., \& Duffy, P., 2004, ApJ, 617, 155

\bibitem[Rybicki \& Lightman(1979)]{rl1979} 
Rybicki, G. B., \& Lightman, A. P., 1979, Radiative processes in
astrophysics, John Wiley \& Sons, New York
 
\bibitem[Sambruna et al.(2007)]{sa2007}
Sambruna, R. M., et al., 2007, ApJ, 670, 74

\bibitem[Sikora et al.(1994)]{sbr1994}
Sikora, M., Begelman, M. C., \& Rees, M. J., 1994, ApJ, 421, 153

\bibitem[Sikora et al.(2001)]{si2001}
Sikora, M., et al., 2001, ApJ, 554, 1

\bibitem[Sokolov et al.(2004)]{smm2004}
Sokolov, A., Marscher, A. P., \& McHardy, I. M., 2004, ApJ, 613, 725

\bibitem[Spada et al.(2001)]{sp2001}
Spada, M., et al., 2001, MNRAS, 325, 1559

\end{thebibliography}
\end{document}